\documentclass[acmtog,nonacm]{acmart}

\makeatletter
\@ifundefined{hyxmp@parse@acmart}{}{%
  \let\merm@hyxmp@parse@orig\hyxmp@parse@acmart
  \renewcommand*{\hyxmp@parse@acmart}{%
    \begingroup
      \let\@ACM@checkaffil\relax
      \long\def\email##1##2{}%
      \let\@authornotemark\relax
      \long\def\@@authornotemark##1{}%
      \merm@hyxmp@parse@orig
    \endgroup}%
}
\makeatother

\usepackage{booktabs}

\usepackage{wrapfig}
\usepackage{comment}
\usepackage[permil]{overpic}
\usepackage{subfigure}
\usepackage{xspace}
\usepackage{graphicx}
\usepackage{amsmath}
\usepackage{utfsym}
\usepackage{bm}
\usepackage{enumitem}
\usepackage{multirow}
\usepackage{threeparttable}

\usepackage[ruled]{algorithm2e}

\SetAlFnt{\small}
\SetAlCapFnt{\small}
\SetAlCapNameFnt{\small}
\SetAlCapHSkip{0pt}

\acmJournal{TOG}

\begin{document}

\title{MeRM: Large Mesh Reconstruction Model via Neural Geometry Tokens}

\author{Kai Ye}
\authornote{Equal contribution.}
\orcid{0009-0006-9691-1125}
\affiliation{
  \institution{Peking University}
  \city{Beijing}
  \country{China}
}
\email{ye\_kai@pku.edu.cn}

\author{Haoru Wang}
\authornotemark[1]
\orcid{0009-0006-6968-7287}
\affiliation{
  \institution{Peking University}
  \city{Beijing}
  \country{China}
}
\email{ou524u@gmail.com}

\author{Zhongyu Shi}
\orcid{0009-0004-1008-1809}
\affiliation{
  \institution{Peking University}
  \city{Beijing}
  \country{China}
}
\email{2300017734@stu.pku.edu.cn}

\author{Ding Liang}
\orcid{0000-0001-9774-4687}
\affiliation{
  \institution{VAST AI}
  \city{Beijing}
  \country{China}
}
\email{liangding1990@163.com}

\author{Yan-Pei Cao}
\orcid{0000-0002-0416-4374}
\affiliation{
  \institution{VAST AI}
  \city{Beijing}
  \country{China}
}
\email{caoyanpei@gmail.com}

\author{Wenzheng Chen}
\authornote{Equal advisory.}
\orcid{0009-0008-5623-1963}
\affiliation{
  \institution{Peking University}
  \city{Beijing}
  \country{China}
}
\email{wenzhengchen@pku.edu.cn}

\author{Baoquan Chen}
\authornotemark[2]
\orcid{0000-0003-4702-036X}
\affiliation{
  \institution{Peking University}
  \city{Beijing}
  \country{China}
}
\email{baoquan.chen@gmail.com}

\renewcommand{\shortauthors}{Ye et al.}

\newcommand{\name}{MeRM}

\newcommand{\highlight}[1]{\paragraph{#1}}

\makeatletter
\DeclareRobustCommand\onedot{\futurelet\@let@token\@onedot}
\def\@onedot{\ifx\@let@token.\else.\null\fi\xspace}

\def\eg{\emph{e.g}\onedot} \def\Eg{\emph{E.g}\onedot}
\def\ie{\emph{i.e}\onedot} \def\Ie{\emph{I.e}\onedot}
\def\cf{\emph{cf}\onedot} \def\Cf{\emph{Cf}\onedot}
\def\etc{\emph{etc}\onedot} \def\vs{\emph{vs}\onedot}
\def\wrt{w.r.t\onedot} \def\dof{d.o.f\onedot}
\def\iid{i.i.d\onedot} \def\wolog{w.l.o.g\onedot}
\def\etal{\emph{et al}\onedot}
\def\plucker{Plücker\xspace}
\makeatother

\begin{abstract}

High-fidelity surface reconstruction from sparse RGB images remains a significant challenge, where classical optimization-based approaches typically fail to resolve geometric ambiguity under insufficient epipolar constraints.
While recent feed-forward methods have achieved remarkable progress in directly regressing point clouds and 3D Gaussians from sparse inputs, extending such paradigms to explicit surface reconstruction often results in suboptimal geometric quality. We identify the primary obstacle in feed-forward surface reconstruction as an inherent structural misalignment between 2D visual inputs and 3D surface targets, which creates a learning bottleneck and leads to degraded performance.
In this work, we propose \textbf{\name}, a large-scale feed-forward Mesh Reconstruction Model that leverages a novel \textbf{Neural Geometry Token (NGT)} technique to address this challenge. Our key insight is that intermediate tokens of a multi-view Transformer—pretrained to predict dense depth maps, normal maps, and masks—can serve as a powerful geometric pre-aggregator to bridge the structural bottleneck. Benefiting from the highly informative geometric cues distilled by NGT, our {\name} delivers exceptional surface reconstruction quality while maintaining impressive training and inference efficiency. Extensive experiments demonstrate that our {\name} significantly outperforms prior feed-forward methods and even exceeds the performance of state-of-the-art optimization-based approaches on dense views, despite using only sparse inputs. We believe our approach establishes a novel framework for large-scale feed-forward mesh reconstruction, paving a new path for future research. Project page: \url{https://illusive-chase.github.io/MeRM-Homepage}.
\end{abstract}



\begin{teaserfigure}
    \centering
    \includegraphics[width=1.0\linewidth,trim={0cm 1cm 0cm 0cm},clip]{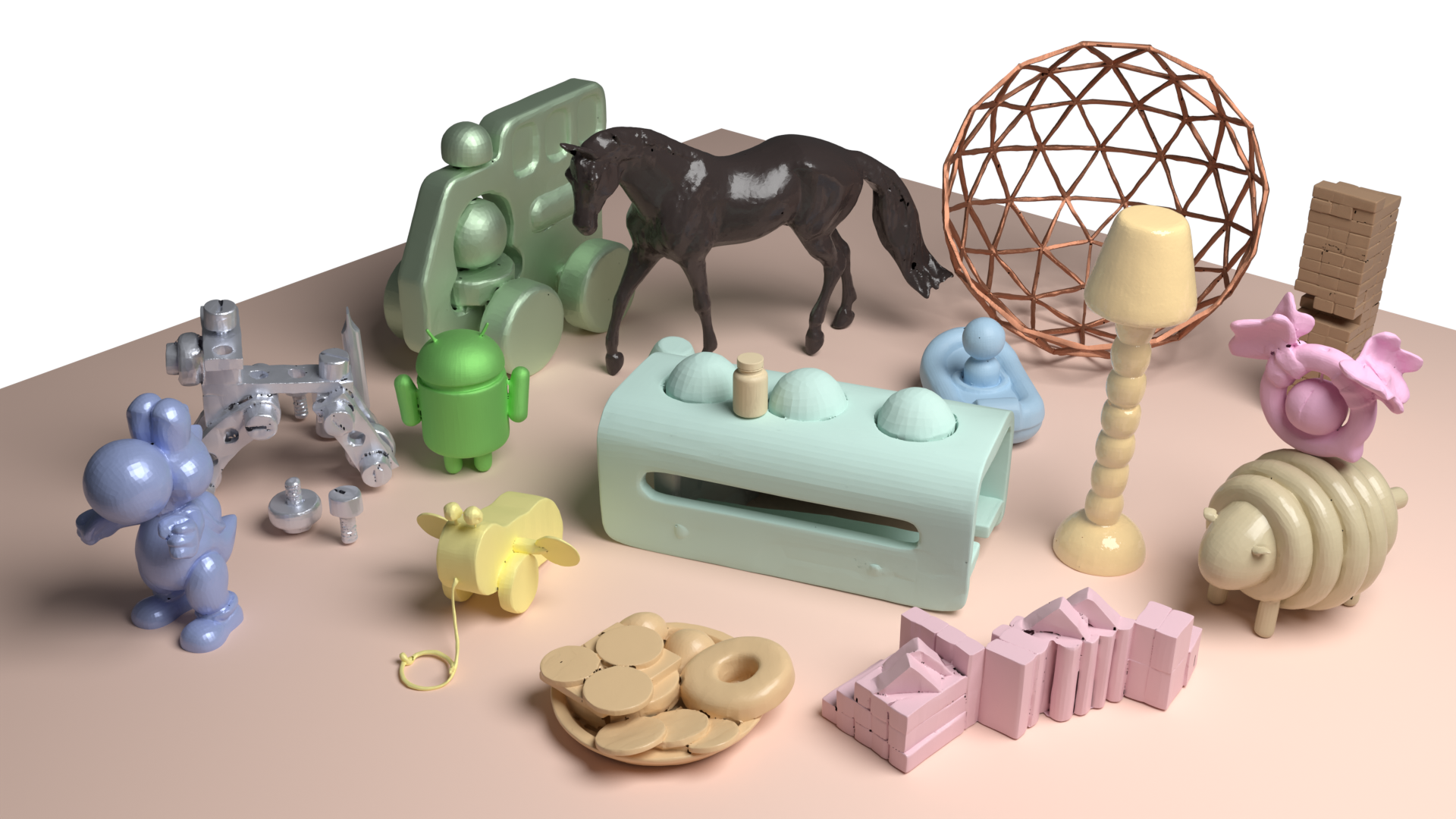}
    \Description{A gallery of reconstructed objects arranged on a pale platform, including toys, a horse, a wireframe sphere, a lamp, a toaster, and small household objects. The meshes show varied surface structures, including curved shells, thin rods, holes, ridges, and sharp-edged recesses. Most objects have solid pastel materials, while the horse retains a patterned appearance.}
\end{teaserfigure}

\maketitle

\section{Introduction}
\label{sec:intro}

High-fidelity surface reconstruction from RGB inputs has long been a central goal of computer graphics and 3D vision. Classical Multi-View Stereo~\cite{furukawa2015multi,schoenberger2016mvs} and recent differentiable-rendering pipelines~\cite{wang2021neus,li2023neuralangelo,huang20242d,yu2024gaussian} can recover impressive geometric detail, but require dense view coverage and costly per-scene optimization.
To lift these constraints, recent feed-forward frameworks~\cite{hong2023lrm,xu2024grm,wang2024dust3r,leroy2024grounding,wang2025vggt,wang2025pi,liu2025slam3r,zhang2024gs,smart2024splatt3r,ye2024no,xu2025depthsplat,charatan2024pixelsplat,chen2024mvsplat,jin2024lvsm,jiang2025rayzer,huang2025no} train large Transformers to regress pixel-aligned 3D primitives (\eg, point maps, pixelwise 3D Gaussians) directly from sparse RGB inputs. These methods achieve state-of-the-art geometric accuracy, significantly advancing the feed-forward 3D reconstruction.

Inspired by this success, a line of work~\cite{wei2024meshlrm,wang2025r,jiang2025dimer,ge2025prm,liu2024meshformer,luo2025ms3d} has attempted to transfer the feed-forward paradigm to mesh reconstruction. Despite massive training data and large model capacity, these attempts either yield blobby surfaces lacking fine-grained detail (MeshLRM~\cite{wei2024meshlrm}, R-MeshFusion~\cite{wang2025r}, DiMeR~\cite{jiang2025dimer}, PRM~\cite{ge2025prm}) or fall back to auxiliary normal-map inputs to compensate (MeshFormer~\cite{liu2024meshformer}, MS3D~\cite{luo2025ms3d}). This disparity raises a fundamental question: \emph{why does the feed-forward paradigm scale so gracefully for pixel-aligned regression, yet stall on mesh reconstruction?}

\begin{figure*}[t]
\centering
\includegraphics[width=\linewidth]{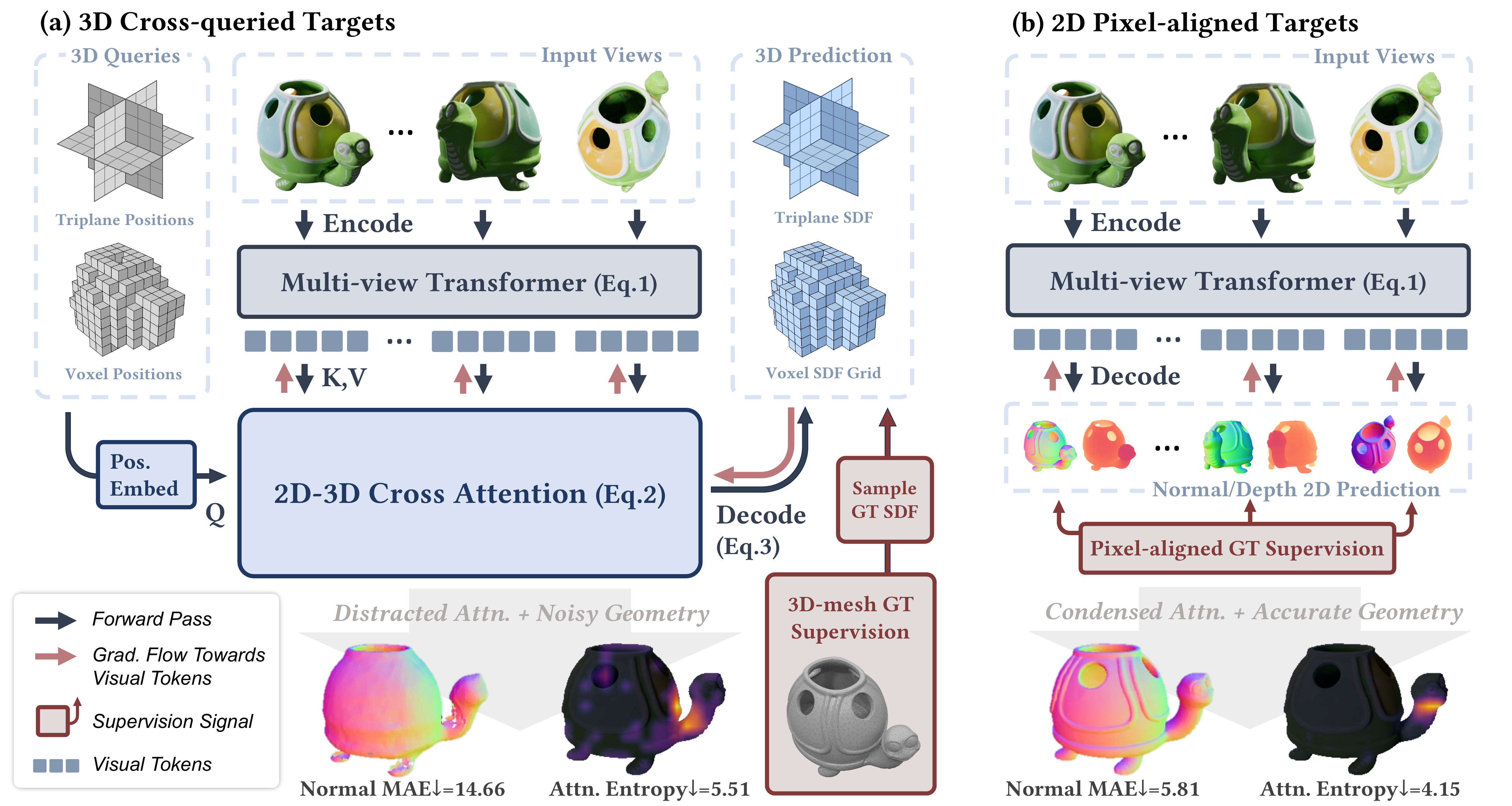}
\caption{\textbf{Structural bottleneck in feed-forward mesh regression.} (a)~Directly training the cross-queried pipeline (Eqs.~\ref{eq:ff_enc}--\ref{eq:ff_dec}) yields blobby surfaces lacking fine-grained detail. (b)~In sharp contrast, a model regressing 2D-aligned targets attains significantly higher geometric accuracy under matched data and capacity. We attribute such structural bottleneck to asymmetric \emph{gradient flow towards visual tokens}: 2D-aligned supervision drives dense gradients directly into every visual token, whereas the cross-queried design routes gradients only indirectly through 3D queries. Attention-weight visualization further confirms that accurate cross-view correspondences are implicitly learned under 2D-aligned supervision but fail to emerge in the cross-queried setting. See Sec.~\ref{sec:ffmr} for details.}
\label{fig:bottleneck}
\Description{Two side-by-side network diagrams compare supervision pathways using the same multi-view images of a green toy. On the left, 3D queries cross-attend to visual tokens before mesh decoding; supervision reaches the visual tokens indirectly through this readout. On the right, visual tokens decode directly into pixel-aligned normal and depth maps, providing a direct supervision pathway. The examples beneath the diagrams contrast a smoothed prediction and diffuse attention on the left with detailed geometry and concentrated attention on the right. Black arrows denote forward computation and red arrows denote gradient flow.}
\end{figure*}

We trace this failure to a \textbf{structural bottleneck} endemic to the \emph{cross-queried} architecture shared by existing feed-forward mesh regressors (Sec.~\ref{sec:ffmr}). In this design, a set of learnable 3D queries will cross-attend to the 2D inputs to aggregate geometric clues from RGB observations. However, the supervision signals are only applied to these 3D tokens; the informative 2D tokens receive gradient only \emph{indirectly}, through cross-attention. As a result, when the queries fail to attend to their geometrically corresponding tokens, the network struggles to correct the mismatch, and cross-attention could collapse into uninformative patterns. In sharp contrast, feed-forward models that regress \emph{pixel-aligned} targets~\cite{wang2024dust3r,wang2025vggt,jin2024lvsm} let every visual token simultaneously serve as query and information source, so gradients flow densely and directly into the visual stream (Fig.~\ref{fig:bottleneck}).

To this end, we propose \textbf{{\name}}, a large-scale feed-forward \underline{Me}sh \underline{R}econstruction \underline{M}odel
built on a novel \textbf{Neural Geometry Token (NGT)} representation to overcome the structural bottleneck. We first pretrain a multi-view Transformer to predict per-view depth, normal, and mask maps; its intermediate tokens are then exported as NGTs---tokens that remain 2D-aligned with the input pixels while already encoding cross-view-consistent geometric cues via dense 2D supervision signals. A downstream Transformer, \textit{NGTFormer}, \emph{reads out} these tokens into 3D queries via interleaved self-cross attention. As rich geometric information is pre-aggregated into NGTs, the reading out process is much easier to learn.

Extensive experiments show that \name{} reconstructs high-fidelity meshes from 4-8 RGB views in a single feed-forward pass, without auxiliary normal maps or fixed camera rigs. It significantly outperforms prior feed-forward mesh regressors and, despite operating on sparse inputs, even surpasses state-of-the-art optimization-based methods~\cite{huang20242d,yu2024gaussian,chen2024pgsr} that consume dense views. Ablations and probing analyses validate our hypothesis of the structural bottleneck and the effectiveness of our NGT design. We believe the proposed approach establishes a new, efficient feed-forward paradigm for mesh reconstruction.

\section{Related Work}
\label{sec:related}

\highlight{Feed-forward Multi-view Reconstruction.}
Recent feed-forward methods~\cite{hong2023lrm, wang2024dust3r, wang2025vggt, wang2025pi, zhang2024gs, xu2025depthsplat, jin2024lvsm, jiang2025rayzer} train end-to-end Transformer networks~\cite{vaswani2017attention} to regress pixel-aligned 3D targets (point maps, 3DGS primitives, \etc) from sparse multi-view RGB inputs, leveraging learned priors to support 2--8 sparse input views. Extending this paradigm to mesh regression has proven challenging: triplane-based methods~\cite{wei2024meshlrm, jiang2025dimer, ge2025prm} inherit the limitations of implicit fields, struggling with high-frequency details, while explicit voxel approaches~\cite{liu2024meshformer, luo2025ms3d} encounter difficulties in distilling geometric clues from RGB observations, often requiring additional normal map inputs.

\highlight{Native 3D Shape Generation.}
While existing feed-forward deterministic methods struggle to produce high-resolution meshes, native 3D generative methods~\cite{wu2024unique3d,zhang2024clay,xiang2025structured,xiang2025trellis2,he2025triposf,hunyuan3d2025hunyuan3d} have made striking progress in generating high-quality mesh assets. Powered by diffusion or flow-matching models~\cite{ho2020denoising,lipman2022flow} trained on large-scale 3D data, these methods learn to sample from a distribution over plausible shapes, producing meshes with significantly richer detail than existing reconstruction approaches---whether feed-forward~\cite{wei2024meshlrm} or optimization-based~\cite{yu2024gaussian}.

\highlight{Fidelity-aware 3D Generation.}
Despite their impressive mesh quality, 3D generative methods address fundamentally different needs than the faithful mesh reconstruction targeted by this work. With a training objective that rewards distributional coverage and perceptual quality, generative models produce plausible and diverse shapes but hallucinate geometry even on directly observed surfaces, making them unsuitable for tasks demanding geometric fidelity (\eg, scanning, simulation, measurement).
A recent line of work seeks to close this gap from within the generative paradigm, combining generative priors with reconstruction-oriented objectives. Two strategies have emerged. The first \emph{couples a reconstruction front-end to a generative back-end}: ReconViaGen~\cite{chang2025reconviagen} conditions a TRELLIS-style generative backend on multi-view reconstruction features and compensates for fidelity at inference time, and RnG~\cite{xiang2026rng} pursues a similar recon-conditioned formulation. The second enforces \emph{pixel alignment natively inside 3D generation}~\cite{li2026pixal3d}, back-projecting 2D features into the volume that a latent diffusion model denoises. Both strategies improve input fidelity, and---unlike earlier fidelity-aware work restricted to point clouds or 3DGS---the first now produces unstructured mesh surfaces directly. What they share is architectural: a generative predictor remains the sole producer of \emph{all} output geometry, so fidelity enters only as a soft signal that the sampler must rediscover through its conditioning channel, never as a hard constraint. Empirically the gap remains open (Sec.~\ref{sec:exp:recon_vs_gen}). \name{} occupies the complementary position: it is a \emph{strict-reconstruction} primitive that regresses observed geometry directly, with no diffusion sampler and no generative completion in the loop.

\highlight{Latent Representation for Mesh Modeling.}
Training deep networks to produce high-fidelity mesh surfaces is non-trivial. A fundamental challenge lies in the inherent mismatch between the structured latents and the unstructured meshes. Numerous efforts~\cite{cheng2023sdfusion,gupta20233dgen,li2023generalized,zheng2023locally,nta2023auto,hui2024makeashape,liu2024one2345,ren2024xcube,wu2024direct3d,lan2024ln3diff,Xiong_2025_SGP,meng2025lt3sd} have been made to fill this gap, proposing various latent surface representations ranging from SDF volumes to triplane embeddings.
Among these, recent approaches have achieved state-of-the-art progress. Vecset-based representations~\cite{zhang20233dshape,zhao2023michelangelo,zhang2024clay,hunyuan3d2025hunyuan3d} have demonstrated great compatibility with modern Transformer-based networks and exhibited promising compression ratio, while sparse voxel-based representations~\cite{xiang2025structured,he2025triposf,xiang2025trellis2} excel at recovering fine-grained details at the cost of tailored architecture dependency.

\begin{table}[b]
    \centering
    \caption{Design choices of $\mathcal{S}'$ in feed-forward mesh regression (Eq.~\ref{eq:ff_enc}--\ref{eq:ff_iso}).}
    \label{tab:ff_pipeline}
    \resizebox{\linewidth}{!}{
        \begin{tabular}{l|lll}
            \toprule
            \textbf{Method} & \textbf{\small{$\mathcal{S}'$ Parameterization}} & \textbf{\small{$\mathcal{S}'$ Representation}} & \textbf{\small{$\mathcal{S}'\rightarrow\mathcal{S}$ Extraction}} \\
            \midrule
            MeshLRM~\cite{wei2024meshlrm} & Triplane & Density Field & DiffMC~\cite{wei2025neumanifold} \\
            DiMeR~\cite{jiang2025dimer} & Triplane & SDF Field & FlexiCubes~\cite{shen2023flexicubes} \\
            PRM~\cite{ge2025prm} & Triplane & SDF Field & FlexiCubes~\cite{shen2023flexicubes} \\
            MeshFormer~\cite{liu2024meshformer} & Sparse Voxel & SDF Grid & DMC~\cite{lorensen1998marching} \\
            MS3D~\cite{luo2025ms3d} & Sparse Voxel & SDF Grid & DMC~\cite{lorensen1998marching} \\
            \midrule
            \name{} (Ours) & Sparse Voxel & Flexible Dual Grid & DC~\cite{ju2002dual} \\
            \bottomrule
        \end{tabular}
    }
\end{table}

\section{Structural Bottleneck in Feed-forward Mesh Regression}
\label{sec:ffmr}

\begin{figure*}
	\begin{center}
		\includegraphics[width=1.0\textwidth,trim={0cm 0.2cm 0cm 0.2cm},clip]{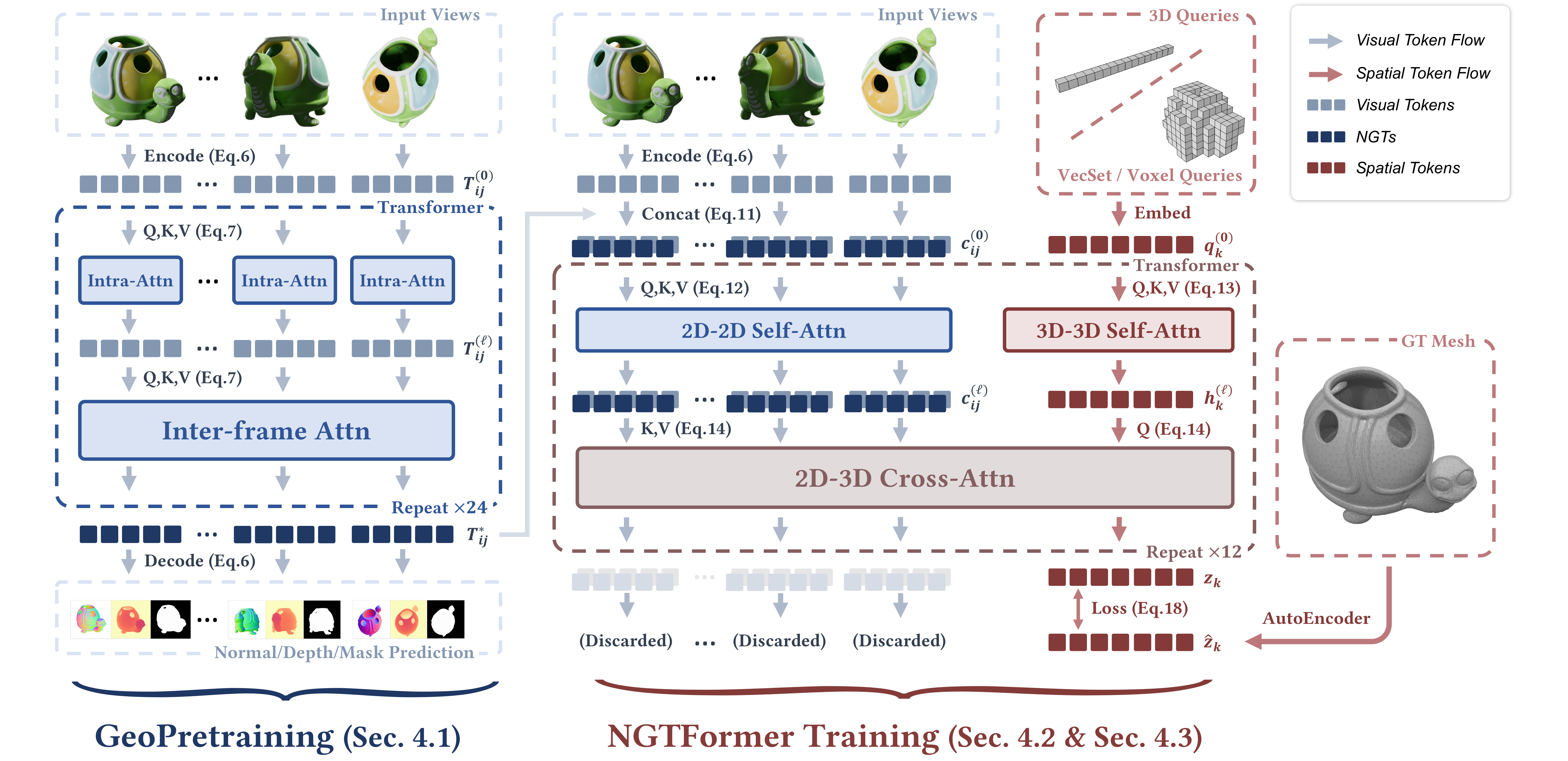}
	\end{center}
	\caption{\label{fig:pipeline} \textbf{{\name} Pipeline}. {\name} operates in a two-stage feed-forward manner: it first extracts \textit{Neural Geometry Tokens} (NGTs) via a multi-view Transformer pretrained on dense 2D geometric maps, and subsequently employs a Transformer (\textit{NGTFormer}) to aggregate these tokens into structured latents which can be decoded into high-fidelity mesh surfaces.}
    \Description{A two-stage architecture diagram. On the left, multi-view RGB images pass through alternating intra-frame and inter-frame attention blocks in GeoTransformer. Normal and depth prediction heads provide geometric pretraining. On the right, the pretrained encoder supplies Neural Geometry Tokens to NGTFormer, which combines image-aligned tokens with 3D spatial tokens using self-attention and cross-attention. Backend-specific branches predict sparse-voxel or vector-set latents, and a pretrained decoder converts these latents into a mesh. Auxiliary heads and latent losses indicate the training supervision paths.}
\end{figure*}

\highlight{Cross-queried Pipeline.}
Given calibrated multi-view images $\mathcal{I}=\{(I_i,C_i)\}_{i=1}^V$, feed-forward mesh regression aims to recover the mesh surface $\mathcal{S}$ in a single forward pass. Despite architectural variations, existing methods~\cite{wei2024meshlrm,jiang2025dimer,ge2025prm,liu2024meshformer,luo2025ms3d} share a PerceiverIO-style~\cite{jaegle2021perceiver} \textit{cross-queried} design (Fig.~\ref{fig:bottleneck}a) that bridges the structured visual tokens $\mathcal{E}$ and the unstructured mesh target $\mathcal{S}$:
\begin{align}
    \mathcal{E} &= \textrm{MultiViewTransformer}(\mathcal{I})\,,\label{eq:ff_enc}\\
    \bm{z}_k &= \textrm{CrossAttn}\!\left(\bm{q}_k,\;\mathcal{E}\right),\;\; k{=}1,\dots,K\,,\label{eq:ff_cross}\\
    \mathcal{S}' &= \textrm{DecoderNetwork}(\bm{z}_1,\dots,\bm{z}_K)\,,\label{eq:ff_dec}\\
    \mathcal{S} &= \textrm{MeshExtraction}(\mathcal{S}')\,,\label{eq:ff_iso}
\end{align}
where a multi-view Transformer encoder (Eq.~\ref{eq:ff_enc}) first processes RGB inputs $I_i$ augmented with camera-parameter embeddings~\cite{plucker1865xvii} into compact visual tokens $\mathcal{E}$. A set of learnable 3D queries $\{\bm{q}_k\}_{k=1}^K$---instantiated as triplane tokens~\cite{wei2024meshlrm,jiang2025dimer,ge2025prm} or sparse voxel features~\cite{liu2024meshformer,luo2025ms3d}---then aggregates geometric cues from $\mathcal{E}$ via cross-attention (Eq.~\ref{eq:ff_cross}). A decoder maps the aggregated queries $\{\bm{z}_k\}_{k=1}^K$ into a parameterized isosurface representation $\mathcal{S}'$ (Eq.~\ref{eq:ff_dec}), from which the mesh surface $\mathcal{S}$ is recovered via isosurfacing (Eq.~\ref{eq:ff_iso}). Table~\ref{tab:ff_pipeline} summarizes how existing methods instantiate $\mathcal{S}'$.

\highlight{Pixel-aligned Targets Are Much Easier.}
Although end-to-end trainable, the cross-queried pipeline consistently yields suboptimal results: as shown in Fig.~\ref{fig:bottleneck}a, an end-to-end model produces blobby surfaces lacking fine detail even with sufficient model capacity and training data. In sharp contrast, feed-forward methods that regress \emph{pixel-aligned} geometric targets---per-view depth, normal, and point maps~\cite{wang2024dust3r,wang2025vggt,jin2024lvsm}---readily attain high accuracy at large training scales, hinting at an inherent bottleneck of the cross-queried design. To make this contrast concrete, we additionally train a geometric model that regresses 2D pixel-aligned targets (Fig.~\ref{fig:bottleneck}b), bypassing any 3D primitive and directly outputting per-view depth and normal maps. Despite identical training data and far fewer parameters, this 2D-target model attains significantly higher geometric accuracy; attention analysis further reveals that accurate cross-view correspondences are implicitly learned. We report this comparison in detail in Sec.~\ref{sec:analysis:diag}.

\highlight{Structural Bottleneck in Mesh Regression.}
The success of 2D-aligned supervision exposes a deeper asymmetry between the query and information-source roles in cross-attention, which we identify as a \emph{structural bottleneck} caused by the misalignment between 2D inputs and 3D targets. Recovering 3D geometry from 2D RGB observations requires non-trivial information aggregation. In the 2D-aligned setting, every patch embedding simultaneously serves as the query $Q$ (where supervision is applied) and the information source $K,V$ (from which geometric clues are aggregated), enabling direct gradient flow into the visual tokens. In the cross-queried setting (Eq.~\ref{eq:ff_cross}), by contrast, the 3D queries $\bm{q}_k$ carry no intrinsic content yet receive the supervision signal, while gradients toward the informative 2D tokens $\mathcal{E}$ flow only indirectly through cross-attention. When $\bm{q}_k$ fails to attend to the geometrically corresponding tokens in $\mathcal{E}$, the network has no easy path to correct the mismatch, and cross-attention degenerates into uninformative patterns.

\highlight{Neural Geometry Tokens as Pre-aggregator.}
This asymmetry points to a way out: mesh outputs ultimately require 3D queries, but the burden of cross-view aggregation need not fall on the 3D loss. If geometric clues are aggregated upstream---by training a multi-view Transformer under dense, pixel-aligned geometric supervision---its intermediate tokens will already encode the geometric information a downstream surface model would otherwise have to discover. The 3D queries can then \emph{read out} these geometric tokens rather than aggregate them from raw 2D content. We instantiate this principle through \textbf{Neural Geometry Tokens}, presented next.

\section{Method}
\label{sec:method}

As demonstrated in Sec.~\ref{sec:ffmr}, the structural misalignment between dense 2D observations and sparse 3D surface targets creates a sparsity bottleneck that limits feed-forward mesh regression. To resolve this, we propose {\name}, built on \textbf{Neural Geometry Tokens (NGTs)}---intermediate Transformer representations obtained through dense geometric pretraining that serve as a 2D-aligned bridge to 3D reconstruction. As illustrated in Fig.~\ref{fig:pipeline}, we first construct the NGT representation via geometric pretraining (Sec.~\ref{sec:method:ngt}), extending Eq.~\ref{eq:ff_enc}. Then, we build the \textit{NGTFormer} mesh reconstruction pipeline upon these tokens (Sec.~\ref{sec:method:network}), applying 2D-3D cross attention to obtain latent tokens that implicitly represent shapes (Eq.~\ref{eq:ff_cross}). Pretrained decoders are employed to subsequently extract surfaces, outputting the triangle mesh (Sec.~\ref{sec:method:decode}).
More training details can be found in Appendix~\ref{sec:supp:training}.

\subsection{Neural Geometry Token}
\label{sec:method:ngt}

\highlight{Overview.} Following Sec.~\ref{sec:ffmr}, we construct Neural Geometry Tokens (NGTs) by pretraining a multi-view Transformer under dense, pixel-aligned geometric supervision. Given calibrated images $\mathcal{I} = \{(I_i, C_i)\}_{i=1}^V$ of a target surface $\mathcal{S}$, the model estimates per-view geometric maps $\mathcal{G} = \{\bm{G}_{ij}=(d_{ij}, \bm{n}_{ij}, m_{ij})\}_{i=1}^V$, where $d_{ij}\in\mathbb{R}^+, \bm{n}_{ij}\in\mathbb{R}^{3}_{[-1,1]}, m_{ij}\in\mathbb{R}_{[0,1]},(j=1,2,\dots,HW)$ denote pixel-wise depth, normal, and mask predictions, respectively. These maps serve as strong geometric proxies of $\mathcal{S}$; their 2D pixel-aligned nature affords direct gradient flow to every visual token, sidestepping the structural bottleneck of cross-queried regression. Once pretrained, the Transformer's intermediate activations are exported as NGTs that condition the downstream mesh reconstruction network (Sec.~\ref{sec:method:network}).

\highlight{Geometric Pretraining.} We instantiate the geometric model as a multi-view Transformer, following architectures established for pixel-aligned regression~\cite{jin2024lvsm,hong2023lrm,wang2025vggt}. Camera geometry is incorporated by augmenting each pixel $\{\bm{I}_{ij}\in\mathbb{R}^{3}|i=1,\dots,V,\,j=1,\dots,HW\}$ with \plucker ray embeddings~\cite{plucker1865xvii} $\{\bm{P}_{ij}\in\mathbb{R}^{6}|i=1,\dots,V,\,j=1,\dots,HW\}$:
\begin{equation}
    \bm{P}_{ij} = \textrm{Concatenate}(\bm{o}_{i}\times \bm{d}_{ij},\bm{d}_{ij}) \,.
\end{equation}
Here, $\bm{o}_{i}\in\mathbb{R}^3$ and $\bm{d}_{ij}\in\mathbb{R}^3$ denote pixel-wise camera ray origins and directions in world space, respectively. After encoding the camera-view information, the concatenated 9-channel maps are divided into $N_p=HW/p^2$ patches $\{\bm{I}'_{ij}\in\mathbb{R}^{p\times p\times9}|j=1,\dots,N_p\}$, where $p$ denotes the patch size. A linear tokenizer then projects these patches into latent tokens $\{\bm{T}_{ij}^{(0)}\in\mathbb{R}^D|j=1,\dots,N_p\}$:
\begin{equation}
    \bm{I}'_{ij}=\mathrm{Concatenate}(\bm{I}_{ij},\bm{P}_{ij})\,,\quad
    \bm{T}_{ij}^{(0)} = \textrm{Linear}(\bm{I'}_{ij}) \,.
    \label{eq:attn:0}
\end{equation}
The latent tokens are subsequently flattened into a 1-D sequence and processed by a Transformer consisting of $L$ attention layers. Similar to VGGT~\cite{wang2025vggt}, each layer alternately performs inter-frame and intra-frame attention.
\begin{equation}
    \bm{T}_{11}^{(k+1)},\dots,\bm{T}_{VN_p}^{(k+1)} = \textrm{AttnLayer}^{(k)}\left(\bm{T}_{11}^{(k)},\dots,\bm{T}_{VN_p}^{(k)}\right) \,.
    \label{eq:attn:k}
\end{equation}
Upon reaching the final layer, the tokens $\bm{T}_{ij}^{(L)}$ are detokenized into geometric maps, where each token $\bm{T}_{ij}^{(L)}\in\mathbb{R}^D$ is first mapped into $\bm{G}_{ij}'\in\mathbb{R}^{p\times p\times5}$ and then unflattened into pixel-wise geometric maps, following the RGB prediction mechanism in prior works~\cite{jin2024lvsm}. The output depth, normal, and mask maps are directly supervised against the ground truth ($\hat{d}_{ij},\hat{\bm{n}}_{ij},\hat{{m}}_{ij}$):
\begin{align}
    \mathcal{L}_\textrm{normal}=&\,\lambda_n \|{\bm{n}}_{ij}-\hat{\bm{n}}_{ij}\|_1+\lambda_n'\textrm{Perceptual}({\bm{n}}_{ij},\hat{\bm{n}}_{ij})\,,\\
    \mathcal{L}_\textrm{pretrain}=&\,\lambda_d \|d_{ij}-\hat{d}_{ij}\|_2+\mathcal{L}_\textrm{normal}+\lambda_m \|m_{ij}-\hat{m}_{ij}\|_1\,,
\end{align}
where a perceptual loss~\cite{johnson2016perceptual} is employed to improve normal quality. After pretraining, the model achieves high-precision depth and normal estimation, providing a reliable basis for extracting Neural Geometry Tokens. More details can be found in Appendix~\ref{sec:supp:pretrain}.

\highlight{NGT Representation.} After pretraining, the Transformer serves as a powerful geometric aggregator, enabling the downstream mesh reconstruction network to leverage pre-aggregated features from the posed inputs $\mathcal{I} = \{(I_i, C_i)\}_{i=1}^V$:
\begin{equation}
    \mathcal{T}=\{(\bm{T^*}_{ij}\in\mathbb{R}^D)\}=\textrm{GeoPretrainedTransformer}(\mathcal{I})\,.
    \label{eq:ngt}
\end{equation}
Specifically, we extract intermediate tokens $\bm{T^*}_{ij}=\bm{T}_{ij}^{(k^*)}\in\mathbb{R}^D$ from the final inter-frame attention layer $\textrm{AttnLayer}^{(k^*)}$ to construct our \textit{Neural Geometry Token}. These tokens implicitly encode multi-view geometric cues required to recover the target surface $\mathcal{S}$, acting as a high-dimensional auxiliary input for the mesh reconstruction network. Compared to directly using the geometric predictions $\mathcal{G}$, which compresses the information back into a low-dimensional $\mathbb{R}^5$ space and reintroduces learning difficulties, our NGT representations are highly informative and significantly enhance the final reconstruction performance. Sec.~\ref{sec:analysis:abl} isolates this effect experimentally.

\subsection{\textit{NGTFormer} for Mesh Reconstruction}
\label{sec:method:network}

\highlight{Overview.} With the Neural Geometry Tokens (NGTs) established, we now describe how the mesh reconstruction pipeline leverages these pre-aggregated geometric cues for high-fidelity surface recovery (Eq.~\ref{eq:ff_cross}). Given a set of spatial queries $\bm{q}_k$ and NGTs extracted from input views, NGTFormer is designed to effectively readout the geometry information from NGTs into $\bm{q}_k$, as detailed below.

\highlight{Interleaved Self-cross Attention.}
We keep the query stream $\{\bm{q}_k\in\mathbb{R}^D\}_{k=1}^K$ abstract here: it can be instantiated as either voxel positional embeddings or learnable VecSet queries~\cite{zhang20233dshape}, with the parameterization deferred to Sec.~\ref{sec:method:decode}. The context stream is built from the NGTs by concatenating the frozen $\bm{T}^*_{ij}$ with a trainable patch-wise re-tokenization of the raw 9-channel pose-augmented input $\bm{I}'_{ij}\in\mathbb{R}^{p\times p\times 9}$ (Eq.~\ref{eq:attn:0}):
\begin{equation}
    \bm{c}_{ij} = \textrm{LayerNorm}\left(
    \mathrm{Concatenate}(\textrm{Linear}(\bm{I}'_{ij}),\bm{T}^*_{ij})
    \right)\,.
    \label{eq:ngtformer:context}
\end{equation}
The trainable re-tokenization supplies camera information that the frozen $\bm{T}^*_{ij}$ cannot adapt.
The NGTFormer then conducts interleaved self-cross attention to refine the context stream jointly with the queries at every layer:
\begin{align}
    \bm{c}^{(\ell+1)}_{1\dots V,1\dots N_p} &= \textrm{SelfAttnLayer}\left(\bm{c}^{(\ell)}_{1\dots V,1\dots N_p}\right)\,,\\
    \bm{h}^{(\ell)}_{1\dots K} &= \textrm{SelfAttnLayer}\left(\bm{q}^{(\ell)}_{1\dots K}\right)\,,\\
    \bm{q}_k^{(\ell+1)} &= \textrm{CrossAttnLayer}\left(\bm{h}_k^{(\ell)},\bm{c}^{(\ell+1)}_{1\dots V,1\dots N_p}\right)\,.
    \label{eq:ngtformer:layer}
\end{align}
The interleaving lets the queries read out a context that has been re-aggregated at each step, so the static layout inherited from pretraining will not rigidly determine the downstream 2D-3D assignment.

\highlight{Mask-based Visual Token Pruning.}
A minor benefit of geometric pretraining is the per-view foreground mask $m_{ij}$, which we exploit at no extra cost. Each patch is retained in the context only if any of its pixels has mask probability above a threshold $\tau$ ($\tau{=}0.05$); the surviving tokens are packed into a variable-length conditioning sequence of effective length $N^*\!\ll\!V\!\cdot\!N_p$. This pruning yields two reinforcing benefits. \emph{Computationally}, cross-attention---the dominant FLOPs term of NGTFormer---scales with $N^*$ rather than the full $V\!\cdot\!N_p$, enabling deeper aggregators and finer voxel grids under the same compute budget. \emph{Functionally}, queries no longer compete with empty-background tokens for attention mass, sharpening the geometric readout and improving convergence. We quantify both effects in Sec.~\ref{sec:analysis:abl}.

\subsection{Mesh Decoding and Training}
\label{sec:method:decode}

\highlight{Overview.}
In this section, we finally detail the mesh decoding process (Eqs.~\ref{eq:ff_dec}--\ref{eq:ff_iso}) by explaining (i) how the decoder extracts the unstructured triangle mesh from a set of structured tokens $\bm{z}_k$ and (ii) how to construct spatial queries $\bm{q}_k$, which are dominated by the isosurface $\mathcal{S}'$ parameterization.

\highlight{$\bm{\mathcal{S}}'$ Parameterization.}
A triangle mesh is an \emph{unstructured} collection of vertices and faces, fundamentally incompatible with the structured token tensors produced by NGTFormer. To bridge this gap, the literature has converged on a family of \emph{structured isosurface parameterizations} of $\mathcal{S}$ that admit a standard isosurfacing step at extraction time (Table~\ref{tab:ff_pipeline}). The most basic form is a dense \textbf{Voxel} grid that partitions $[-1,1]^3$ at resolution $R$ and stores a per-vertex SDF (Eq.~\ref{eq:para:vox}), from which meshes are recovered by Marching Cubes~\cite{lorensen1998marching}. To bypass the $\mathcal{O}(R^3)$ memory cost, the \textbf{Sparse Voxel} formulation~\cite{xiang2025trellis2} keeps only $K\!\ll\!R^3$ surface-adjacent voxels and stores dual-grid parameters $\bm{h}_k\in\mathbb{R}^7$ per voxel (Eq.~\ref{eq:para:spvox}), enabling Dual Contouring~\cite{ju2002dual} for mesh extraction. A complementary direction, \textbf{VecSet}~\cite{zhang20233dshape}, drops explicit spatial anchoring altogether and represents the SDF field as an implicit token set (Eq.~\ref{eq:para:vecset}); at extraction time a vertex SDF is queried via cross-attention, $s_k\!=\!\textrm{Network}(\bm{v}_k,\mathcal{S}'_\textrm{VecSet})$, to materialize a regular grid for Marching Cubes.
\begin{align}
    \mathcal{S}'_\textrm{Vox}&=\{\bm{f}_k\in\mathbb{R}^4\}_{k=1}^{(R+1)^3}=\{(\bm{v}_k, s_k)\}_{k=1}^{(R+1)^3}\,,\label{eq:para:vox}\\
    \mathcal{S}'_\textrm{SpVox}&=\{\bm{h}_k\in\mathbb{R}^7\}_{k=1}^{K}\,,&(K\ll R^3)\label{eq:para:spvox}\\
    \mathcal{S}'_\textrm{VecSet}&=\{\bm{h}'_k\in\mathbb{R}^{64}\}_{k=1}^{K'}\,,&(K'\ll R^3)\label{eq:para:vecset}
\end{align}

\highlight{Decoder Instantiation.}
The readout $\bm{z}_k\!\to\!\mathcal{S}'$ is itself a non-trivial 3D-shape regression process, requiring tailored network architecture design. We observe that published 3D generative autoencoders---trained on multi-million-asset 3D collections---already provide pretrained networks. We therefore reuse a pretrained decoder from each backend---TRELLIS2~\cite{xiang2025trellis2} for Sparse Voxel and 3DShape2VecSet~\cite{zhang20233dshape} for VecSet---and freeze it throughout NGTFormer training. This yields two concrete benefits: (i)~pseudo-ground-truth latents $\{\hat{\bm{z}}_k\}$ are obtained for free by running the pretrained \emph{encoder} on the reference mesh, converting end-to-end mesh supervision into a tractable latent-space regression (Eq.~\ref{eq:loss_lat}); and (ii)~NGTFormer can be paired with structurally distinct decoders, making \name{} genuinely decoder-agnostic.

\highlight{Query Construction.}
The query set $\{\bm{q}_k\}_{k=1}^K$ is then dictated by the chosen $\mathcal{S}'$. For the \textbf{Sparse Voxel} backend, $\bm{q}_k$ are voxel positional embeddings anchored on an active sparse-voxel set, so the predicted latents $\bm{z}_k$ inherit an explicit \emph{spatial layout} and are decoded one-to-one into dual-grid parameters. For the \textbf{VecSet} backend, $\bm{q}_k$ are $K$ \emph{learnable embeddings} without spatial anchors, so $\{\bm{z}_k\}$ forms an \emph{unordered} set decoded into the SDF field via cross-attention. Backend-specific implementation details are deferred to Appendix~\ref{sec:supp:decoder}.

\highlight{Training Objectives.}
Since the decoder is pretrained and frozen, we supervise NGTFormer directly in the latent space. Pseudo-ground-truth latents $\{\hat{\bm{z}}_k\}$ are obtained once by encoding each reference mesh with the corresponding pretrained encoder; the only learnable component is the 2D-to-3D readout itself. The latent regression loss adapts to the latent topology:
\begin{equation}
    \mathcal{L}_\text{lat}=\begin{cases}
        \tfrac{1}{K}\sum_{k}\|\bm{z}_k-\hat{\bm{z}}_k\|_2^2 & \text{(SpVox: ordered)}\,,\\[2pt]
        \text{Sinkhorn-OT}\!\left(\{\bm{z}_k\},\{\hat{\bm{z}}_k\}\right) & \text{(VecSet: unordered)}\,.
    \end{cases}
    \label{eq:loss_lat}
\end{equation}
The voxel-aligned SpVox latents are ordered and admit a one-to-one mean-squared-error supervision, whereas the permutation-invariant VecSet latents are unordered, therefore requiring the Sinkhorn optimal-transport~\cite{cuturi2013sinkhorn} loss functions for matching-based supervision. We follow TRELLIS2~\cite{xiang2025trellis2}'s data curation to collect 160K high-quality meshes from Objaverse~\cite{objaverse}. On 8$\times$ NVIDIA RTX Pro6000, the two training stages take about 2 days and 5 days, respectively. During training, the view numbers are set to $V\!=\!8$, while {\name} can generalize to other view numbers at test-time. More details are provided in Appendix~\ref{sec:supp:training}.

\section{Experiment}
\label{sec:exp}

We conduct extensive experiments to validate \name{}.
First, we establish \emph{core results}: Sec.~\ref{sec:exp:perf} benchmarks surface reconstruction quality against state-of-the-art feed-forward, optimization-based, and generative baselines; Sec.~\ref{sec:exp:sparse_vs_dense} contrasts \name{}'s sparse-view feed-forward output with optimization-based methods given dense supervision and probes view-count generalization; Sec.~\ref{sec:exp:recon_vs_gen} provides a focused head-to-head comparison with leading generation methods; Sec.~\ref{sec:exp:fidelity_aware} then narrows that comparison to methods that explicitly target input fidelity from within the generative paradigm, which is the strongest test of whether the gap is intrinsic to the paradigm or merely an objective mismatch; and Sec.~\ref{sec:exp:realworld} evaluates robustness on real-world captured objects.

\subsection{General Comparisons}
\label{sec:exp:perf}

\begin{table*}[t]
    \centering
    \caption{Quantitative comparison on the GSO dataset. We evaluate depth, normal, and mesh quality metrics on both a 6-object subset (compatible with baselines lacking public code) and the full dataset. $\downarrow$ indicates lower is better, $\uparrow$ indicates higher is better.}
    \label{tab:comparison}
    \begin{threeparttable}
    \resizebox{\linewidth}{!}{
        \begin{tabular}{lr|cc|cccccccc|cccccccc}
            \toprule
            \multicolumn{2}{c|}{\multirow{4}{*}{\large{\textbf{Method}}}} & & & \multicolumn{8}{c|}{\textbf{GSO (6 Objects Subset)}} & \multicolumn{8}{c}{\textbf{GSO (Full Benchmark)}} \\
            \multicolumn{2}{c|}{} & \multirow{2}{*}{\#View} & \multirow{2}{*}{item/s$\uparrow$} & \multicolumn{2}{c}{\textbf{Depth Quality}} & \multicolumn{4}{c}{\textbf{Normal Quality}} & \multicolumn{2}{c|}{\textbf{Mesh Quality}} & \multicolumn{2}{c}{\textbf{Depth Quality}} & \multicolumn{4}{c}{\textbf{Normal Quality}} & \multicolumn{2}{c}{\textbf{Mesh Quality}} \\
            \cmidrule(lr){5-6} \cmidrule(lr){7-10} \cmidrule(lr){11-12} \cmidrule(lr){13-14} \cmidrule(lr){15-18} \cmidrule(lr){19-20}
            \multicolumn{2}{c|}{} & & & AbsRel~(\%)$\downarrow$ & Acc~(\%)$\uparrow$ & MAE~($^\circ$)$\downarrow$ & PSNR$\uparrow$ & SSIM$\uparrow$ & LPIPS$\downarrow$ & F1$\uparrow$ & CD$\downarrow$ & AbsRel~(\%)$\downarrow$ & Acc~(\%)$\uparrow$ & MAE~($^\circ$)$\downarrow$ & PSNR$\uparrow$ & SSIM$\uparrow$ & LPIPS$\downarrow$ & F1$\uparrow$ & CD$\downarrow$ \\
            \midrule
            & Ours (SpVox) & 4~ & \textbf{3.417} & 0.338 & 99.86 & 6.763 & 27.53 & {0.9472} & {0.0528} & {0.99858} & {0.160}  & {0.544} & {99.68} & {9.704} & {27.131} & {0.9411} & {0.0595} & {0.99429} & {0.238} \\
            \multirow{2}{*}{Feed-} & Ours (VecSet) & 4~ & 0.154 & 0.292 & 99.91 & \textbf{6.019} & \textbf{28.70} & \textbf{0.9562} & \textbf{0.0500} & \textbf{0.99995} & \textbf{0.066} & \textbf{0.321} & \textbf{99.92} & \textbf{8.123} & \textbf{29.020} & \textbf{0.9529} & \textbf{0.0513} & \textbf{0.99854} & \textbf{0.110} \\
            \multirow{2}{*}{Forward} & PRM & 6\tnote{$\dagger$} & 0.646 & 2.769 & 98.36 & 24.514 & 15.688 & 0.8218 & 0.2413 & 0.93485 & 2.867 & 4.817 & 96.43 & 28.745 & 14.451 & 0.8024 & 0.2815 & 0.78430 &  10.081 \\
            & MeshLRM & 4~ & - & \textbf{0.289} & \textbf{99.93} & 7.241 & 26.332 & 0.9154 & 0.1090 & 0.99387 & 0.233 & - & - & - & - & - & - & - & 2.680 \\
            & MeshFormer & 1\tnote{$\dagger$} & - & 3.372 & 98.95 & 24.802 & 15.270 & 0.8374 & 0.2409 & 0.92174 & 3.594 & 2.732 & 99.24 & 21.512 & 16.768 & 0.8519 & 0.2139 & 0.95082 & 2.559\\
            \midrule
            \multirow{2}{*}{Optim.} & 2DGS & 24\tnote{$\ddagger$}& 0.0018& 0.528 & 99.21 & 16.387 & 23.119 & 0.7601 & 0.1718 & 0.99451 & 1.574 &  - & - & - & - & - & - & - & - \\
            & GOF & 24\tnote{$\ddagger$}& 0.0012& 0.624 & 98.96 & 15.180 & 23.855 & 0.7765 & 0.1693 & 0.99421 & 1.072 & - & - & - & - & - & - & - & -  \\ \midrule
            \multirow{4}{*}{Generation}& Hunyuan 2.5 & 4~ & - & 4.244 & 96.16 & 25.966 & 15.820 & 0.8274 & 0.2536 & 0.85676 & 5.530 & - & - & - & - & - & - & - & - \\
            & Hunyuan 3.1 & 4~& - & 3.528 & 97.54 & 25.660 & 16.096 & 0.8314 & 0.2457 & 0.88221 & 4.382 & - & - & - & - & - & - & - & - \\
            & TRELLIS & 4~ & 0.0111 & 2.166 & 99.45 & 19.101 & 17.922 & 0.8502 & 0.1886 & 0.95949 & 1.971 & - & - & - & - & - & - & - & - \\
            & TRELLIS2 & 1\tnote{$\dagger$} & 0.0095 & 2.570 & 98.61 & 21.531 & 17.440 & 0.8409 & 0.2012 & 0.93844 & 2.552 & - & - & - & - & - & - & - & - \\
            \bottomrule
        \end{tabular}
    }
        \begin{tablenotes}
        \footnotesize
        \item[$\dagger$] They cannot natively support other view numbers like $V\!=\!4$ required by fair comparison.~~~$^\ddagger$ They cannot converge under extremely sparse views like $V\!=\!4$.
        \end{tablenotes}
    \end{threeparttable}
\end{table*}

\begin{figure*}[t]
	\begin{center}
		\includegraphics[width=1.0\textwidth,trim={0cm 1cm 0cm 1cm},clip]{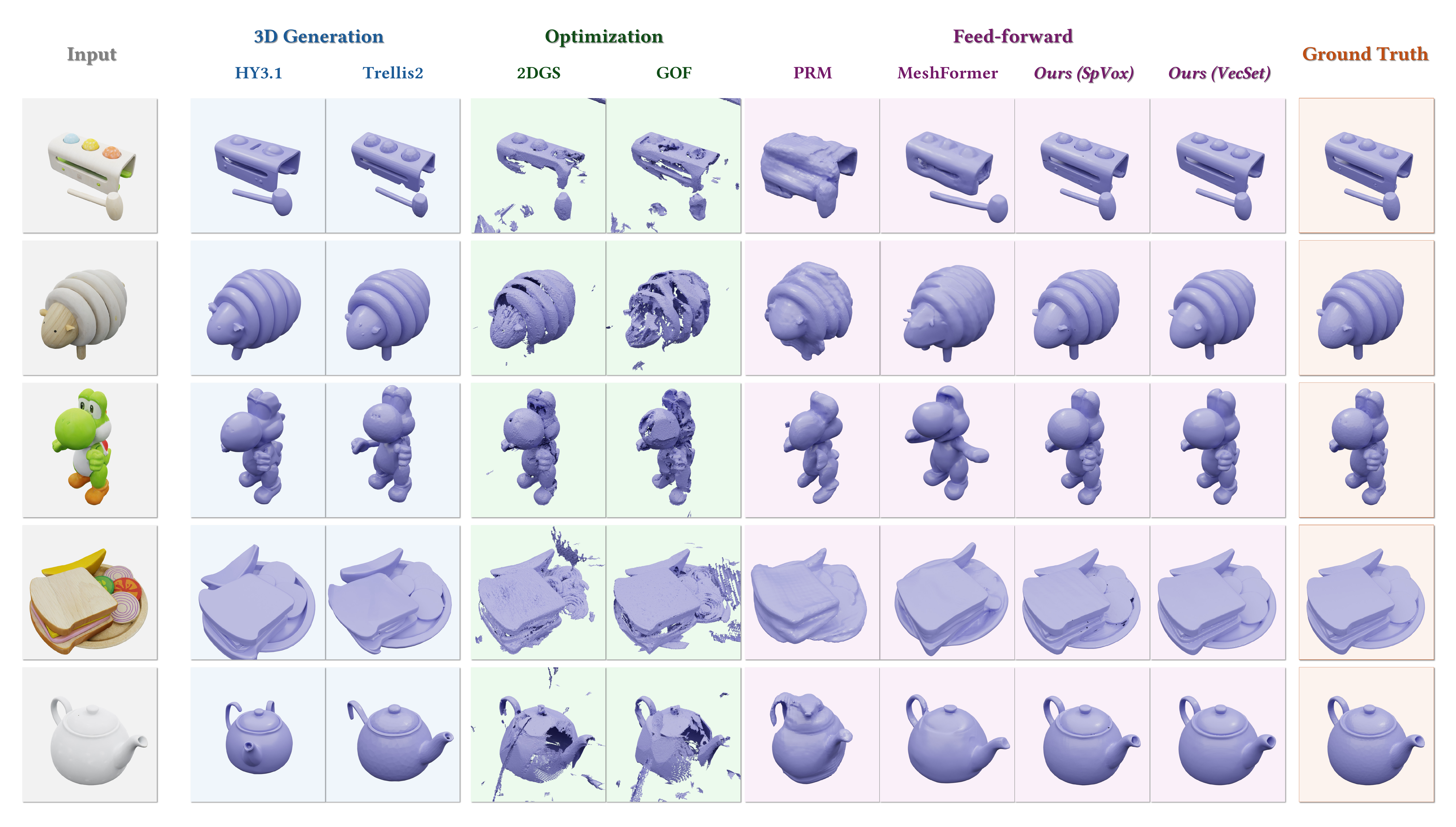}
	\end{center}
	\caption{Compared to a wide range of baselines, our approach delivers significantly more precise reconstruction results.}
    \label{fig:exp:comp:general}
    \Description{A comparison grid with five object rows and ten columns. From left to right, the columns show an input photograph, Hunyuan3D 3.1, TRELLIS2, 2DGS, GOF, PRM, MeshFormer, MeRM with SpVox, MeRM with VecSet, and ground truth. The examples include a toaster, a ridged rounded object, a toy figure, a shoe, and a teapot. Purple mesh renderings reveal differences in holes, handles, surface ridges, and silhouettes. The MeRM columns retain more of these input-specific structures, while several baseline outputs are smoothed, incomplete, or altered.}
\end{figure*}

\highlight{Benchmark.}
To rigorously evaluate the surface reconstruction quality and generalization capability of our method, we follow prior works~\cite{ge2025prm, wei2024meshlrm, liu2024meshformer, jiang2025dimer} to conduct experiments on the Google Scanned Objects (GSO) dataset~\cite{downs2022googlescannedobjectshighquality}, a standard benchmark containing 1,030 high-quality scanned 3D meshes. While our model is trained on Objaverse~\cite{objaverse} with $V\!=\!8$, we directly test it on GSO with $V\!=\!4$ to assess the performance under unseen data distribution.

\highlight{Baselines.}
We compare our {\name} with both decoder backends (\ie, SpVox and VecSet, as detailed in Sec.~\ref{sec:method:decode}) against a comprehensive suite of state-of-the-art baselines categorized into three groups:

\begin{itemize}[leftmargin=*, itemsep=1pt]
    \item \emph{Feed-forward Mesh Regression:} We compare with PRM~\cite{ge2025prm}, MeshLRM~\cite{wei2024meshlrm}, and MeshFormer~\cite{liu2024meshformer}. Note that PRM relies on a Zero123-based backbone sensitive to camera distributions; thus, we follow its optimal setting using 6 fixed canonical views. For MeshLRM and MeshFormer, as their official code or checkpoints are unavailable, we evaluate MeshLRM on the 6-object subset provided on its project page, and evaluate MeshFormer on the single-image-to-3D results, augmented with Zero123++ normals, on its project page.

    \item \emph{Optimization-based Methods:} We include 2DGS~\cite{huang20242d} and GOF~\cite{yu2024gaussian} as representatives. Since per-scene optimization methods typically fail to converge to meaningful geometry under sparse inputs, we run them under 24-view inputs to establish a meaningful comparison.

    \item \emph{Generative 3D Methods:} We compare with Hunyuan3D (v2.5 and v3.1)~\cite{hunyuan3d2025hunyuan3d} and TRELLIS (v1 and v2)~\cite{xiang2025structured, xiang2025trellis2}. These are conditioned on their required inputs (4 views or single view) rendered from the same test objects.

\end{itemize}

\highlight{Metrics.}
We employ a holistic set of metrics covering three aspects: (1) \emph{Depth Quality:} Absolute Relative error (AbsRel) and Accuracy with Threshold (Acc) to measure geometric depth consistency; (2) \emph{Normal Quality:} Mean Angular Error (MAE), PSNR, SSIM, and LPIPS to evaluate surface orientation and fine-grained details; (3) \emph{Mesh Quality:} Chamfer Distance (CD, scaled by \texttt{1e-3}) and F1-score (under threshold of \texttt{5e-3}) to assess the overall fidelity of the reconstructed 3D geometry. We also compare the reconstruction efficiency for both feed-forward and optimization-based methods. Full metric formulas and sampling protocols are detailed in Appendix~\ref{sec:supp:eval}.

\highlight{Results.}
Quantitative comparisons are presented in Table~\ref{tab:comparison}, with qualitative ones provided in Fig.~\ref{fig:exp:comp:general}. To accommodate the availability of baseline results, we report metrics on two test sets: the \emph{GSO (6 Objects Subset in MeshLRM~\cite{wei2024meshlrm})} for comparison with MeshLRM as it does not provide code or models for reproducibility, and the \emph{GSO (Full Dataset)} for a comprehensive evaluation against other feed-forward methods. Also, comparisons with optimization-based and generative methods are only conducted on the subset due to their poor efficiency. As shown in Table~\ref{tab:comparison}, our method consistently outperforms all baselines across depth, normal, and mesh quality metrics on the full dataset, achieving state-of-the-art performance.

\highlight{Discussion.} The results demonstrate that NGTs effectively resolve the structural misalignment between 2D observations and 3D surfaces. Compared to \emph{feed-forward mesh methods} (PRM, MeshLRM), our NGT design recovers significantly finer details. Our data-driven priors also prove more robust than \emph{optimization-based methods} (2DGS, GOF), which overfit photometric cues even under dense supervision. Finally, unlike \emph{generative models} (Hunyuan3D, TRELLIS) that hallucinate unfaithful details, our approach maintains high fidelity to input observations.

\subsection{Sparse View vs.\ Dense View}
\label{sec:exp:sparse_vs_dense}

The general comparison in Sec.~\ref{sec:exp:perf} follows a sparse-view setting $V\!=\!4$. However, baselines like the per-scene optimization approaches \cite{huang20242d,yu2024gaussian} excel at dense photometric supervision. In this section, we evaluate {\name} under varying view numbers to investigate: \emph{(i)} how \name{}'s sparse-view feed-forward output compares to optimization-based methods given dense supervision; \emph{(ii)} how well \name{} will exploit additional views at inference time, despite being trained at a fixed view count.

\highlight{Setup.}
We compare \name{} against 2DGS~\cite{huang20242d} and GOF~\cite{yu2024gaussian} on the GSO subset, sweeping the optimization baselines across $V\!\in\!\{8, 24, 128\}$ views. Both baselines fail to converge under extremely sparse inputs ($V\!=\!8$), so we additionally report $V\!=\!24$ and $V\!=\!128$ to establish a fair dense-supervision upper bound. For \name{}, we evaluate two test-time view counts: $V\!=\!8$ (matching training) and $V\!=\!24$ (a $3\times$ extrapolation beyond the training cardinality, with no retraining or architectural change).

\begin{table}[t]
    \centering
    \caption{Comparison with optimization-based methods across varying view numbers on the GSO subset. \name{} with $V\!=\!8$ already outperforms 2DGS/GOF at $V\!=\!128$ in geometry quality while running much faster.}
    \label{tab:comparison_subset}
    \resizebox{\linewidth}{!}{
        \begin{tabular}{r|cc|ccccc}
            \toprule
            \multirow{2}{*}{\textbf{Method}} & \multirow{2}{*}{\#View}& \multirow{2}{*}{\shortstack{item/s$\uparrow$}}& \textbf{Depth Quality} & \multicolumn{2}{c}{\textbf{Normal Quality}} & \multicolumn{2}{c}{\textbf{Mesh Quality}} \\
            \cmidrule(lr){4-4} \cmidrule(lr){5-6} \cmidrule(lr){7-8} \multicolumn{1}{c|}{} & & & AbsRel~(\%)$\downarrow$ & MAE~($^\circ$)$\downarrow$ & PSNR$\uparrow$ & F1$\uparrow$ & CD$\downarrow$ \\
            \midrule
            \multirow{2}{*}{Ours (SpVox)} & 8 & \textbf{1.982}& 0.224& 6.077& 27.795& 0.99995& 0.069\\
            & 24 & 0.711 & {0.191} & {5.702} & {28.385} & \textbf{0.99999} & \textbf{0.048} \\
            \multirow{2}{*}{Ours (VecSet)} & 8 & {0.151}& 0.286& 5.959& 28.790& 0.99997& 0.063\\
            & 24 & 0.125 & \textbf{0.166} & \textbf{4.702} & \textbf{31.214} & {0.99888} & {0.058} \\
            \midrule
            \multirow{3}{*}{2DGS} & 8 & 0.00263 & 3.742 & 33.154 & 16.210 & 0.89691 & 112.657 \\
             & 24 & 0.00185& 0.528& 16.387& 23.119& 0.99451& 1.574\\
             & 128 & 0.00162& 0.511 & 12.090 & 24.436 & 0.99032 & 0.391\\
            \midrule
            \multirow{3}{*}{GOF} & 8 & 0.00202& 3.977 & 35.245 & 15.400 & 0.83682 & 180.243 \\
             & 24 & 0.00119& 0.624& 15.180& 23.855& 0.99421& 1.072\\
             & 128 & 0.00167& 0.334 & 7.701 & 27.718 & 0.99087 & 0.309 \\
            \bottomrule
        \end{tabular}
    }
\end{table}

\begin{figure*}[t]
	\begin{center}
		\includegraphics[width=1.0\textwidth,trim={0cm 1cm 0cm 1cm},clip]{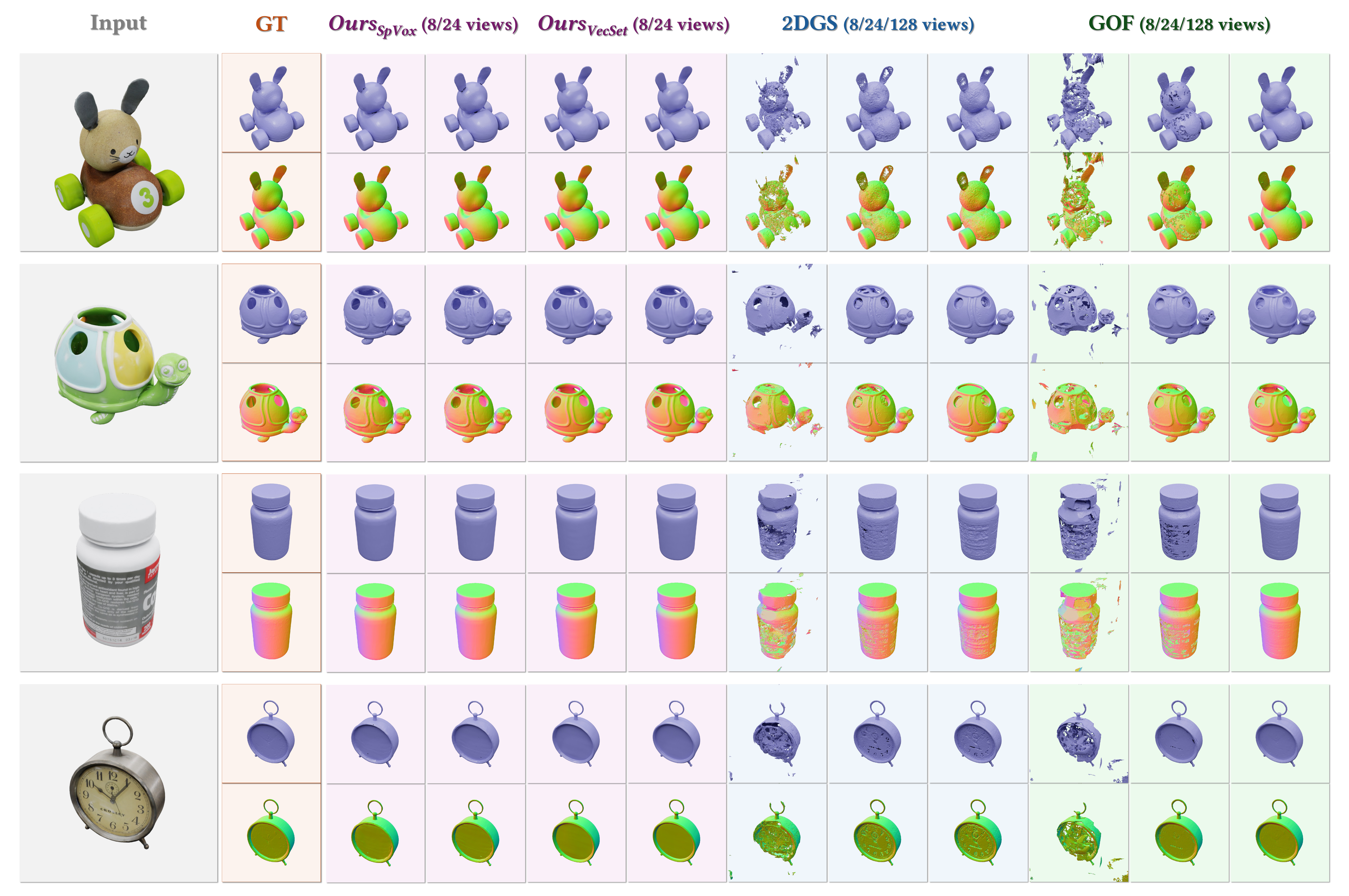}
	\end{center}
	\caption{Compared to optimization-based methods, our approach achieves impressive reconstruction quality, even outperforming state-of-the-art baselines with dense inputs.}
    \label{fig:exp:comp:optim}
    \Description{Four object groups compare a toy rabbit, a rounded toy, a cylindrical container, and a pocket watch. Each group includes a photograph followed by ground truth, MeRM with SpVox and VecSet backends at 8 and 24 views, and 2DGS and GOF at 8, 24, and 128 views. Within each group, an upper row shows shaded meshes and a lower row shows surface-normal colors. MeRM preserves recognizable shapes and fine structures with sparse inputs; optimization-based outputs are fragmented at low view counts and become more complete with denser inputs.}
\end{figure*}

\highlight{Results.}
Quantitative results are reported in Table~\ref{tab:comparison_subset}, with qualitative comparisons in Fig.~\ref{fig:exp:comp:optim}. Optimization methods collapse at $V\!=\!8$ (CD $>\!100$, F1 $<\!0.9$), confirming that photometric supervision alone (\textit{w/o} learning prior) is insufficient under sparse views. With $V\!=\!24$ they recover usable geometry, and at $V\!=\!128$ they approach but do not surpass \name{} at $V\!=\!8$: our model achieves $0.22\%$ AbsRel and $0.069$ CD with eight views, versus $0.33\%$ AbsRel and $0.309$ CD for GOF with $16\times$ as many views. Meanwhile, \name{} runs at $\sim\!2$ items/s, roughly three orders of magnitude faster than the per-scene baselines. Increasing \name{}'s test-time inputs to $V\!=\!24$ further improves every metric (\eg, CD $0.069\!\to\!0.048$) without any retraining.

\subsection{Reconstruction vs.\ Generation}
\label{sec:exp:recon_vs_gen}

\begin{figure*}[t]
    \centering
    \setlength{\tabcolsep}{1.5pt}
    \renewcommand{\arraystretch}{0.6}
    \resizebox{0.83\textwidth}{!}{%
    \begin{tabular}{@{} c ccccc @{}}
        & \textbf{GT}
        & \textbf{\name{} (Ours)}
        & \textbf{TRELLIS2}
        & \textbf{Hunyuan3D v2.5}
        & \textbf{Hunyuan3D v3.1} \\[6pt]
        \rotatebox{90}{\quad\quad\quad\textit{Contour}}
        & \includegraphics[width=3.2cm]{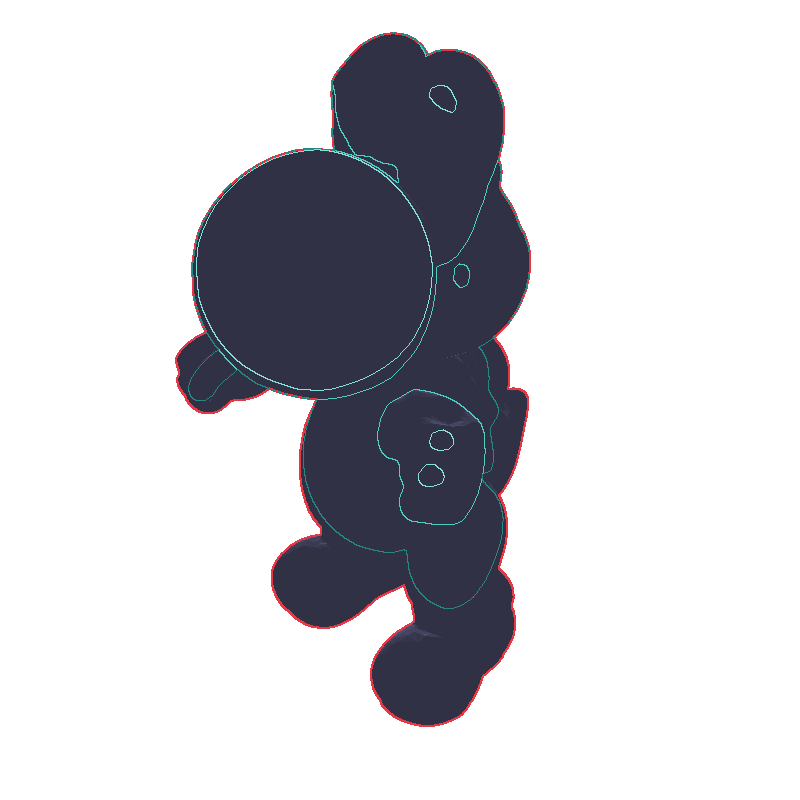}
        & \includegraphics[width=3.2cm]{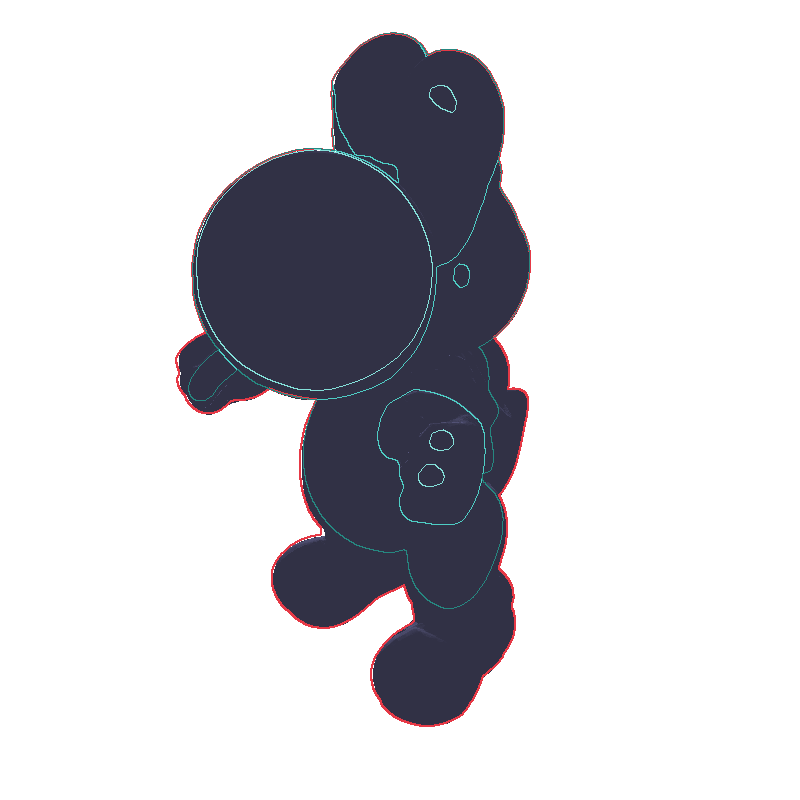}
        & \includegraphics[width=3.2cm]{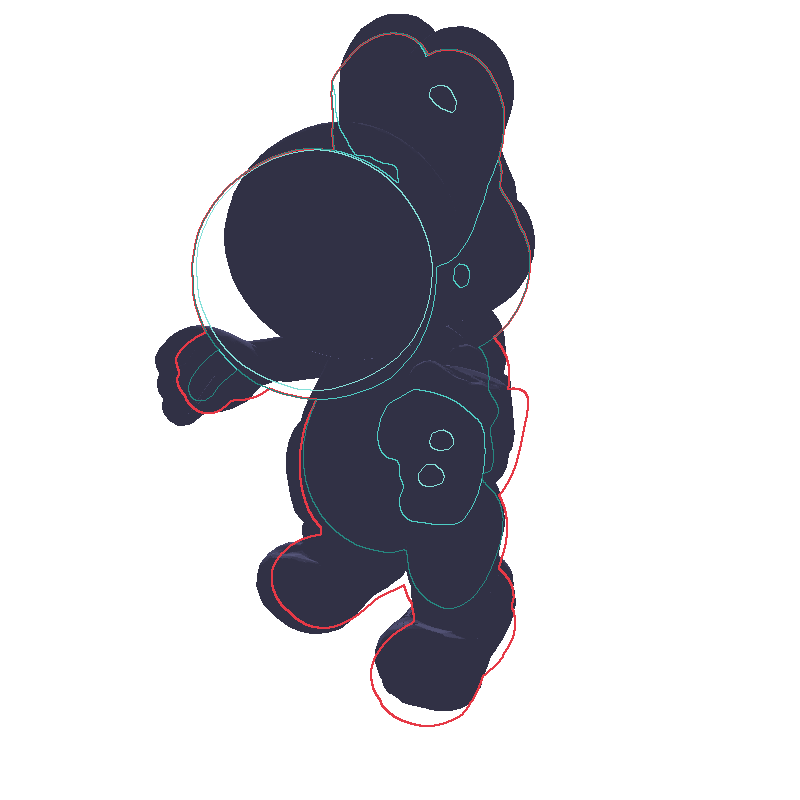}
        & \includegraphics[width=3.2cm]{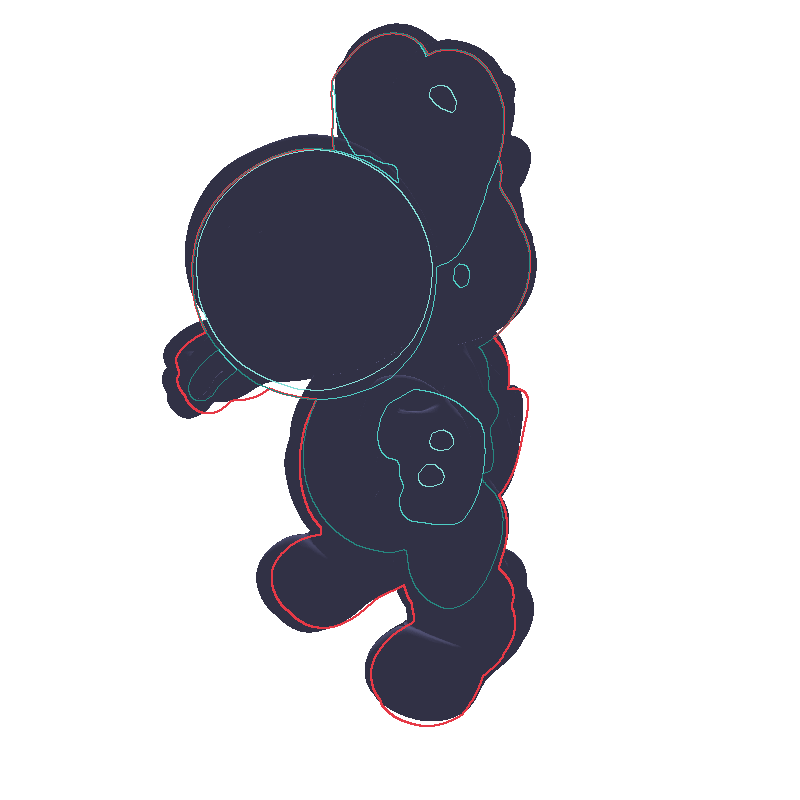}
        & \includegraphics[width=3.2cm]{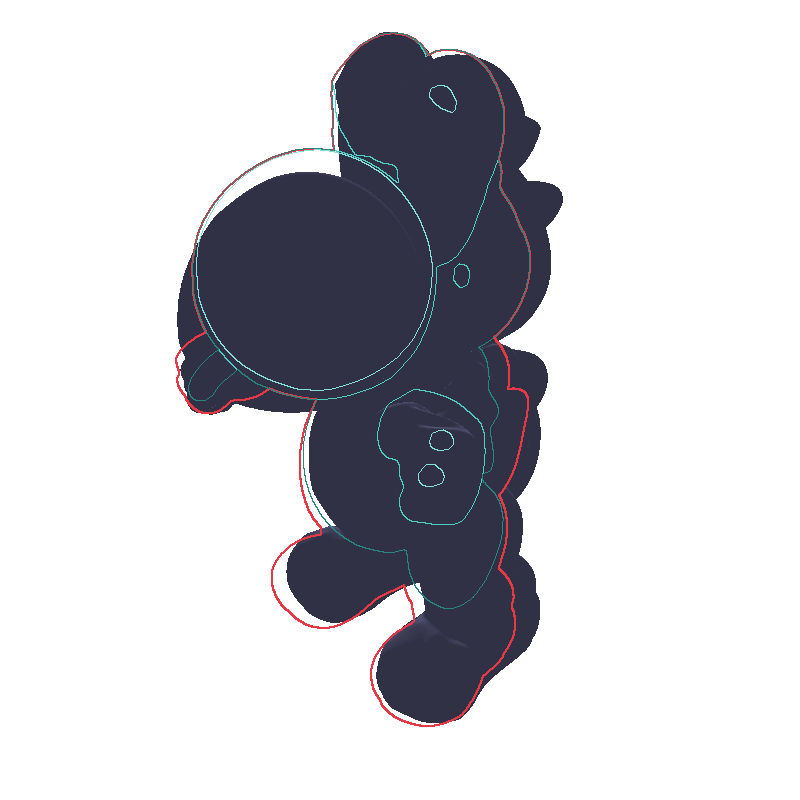} \\[2pt]
        \rotatebox{90}{\quad\quad\quad\textit{Halluc.}}
        &
        & \includegraphics[width=3.2cm]{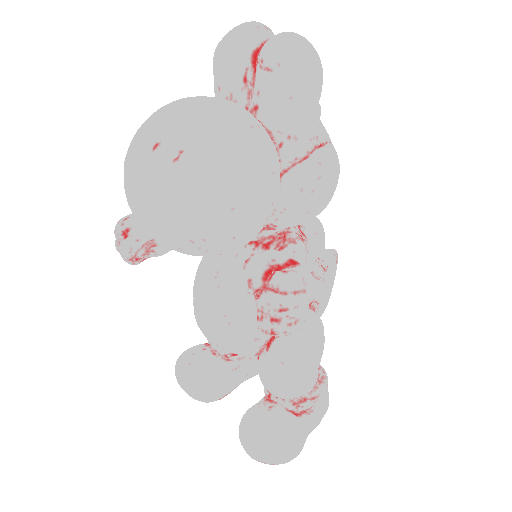}
        & \includegraphics[width=3.2cm]{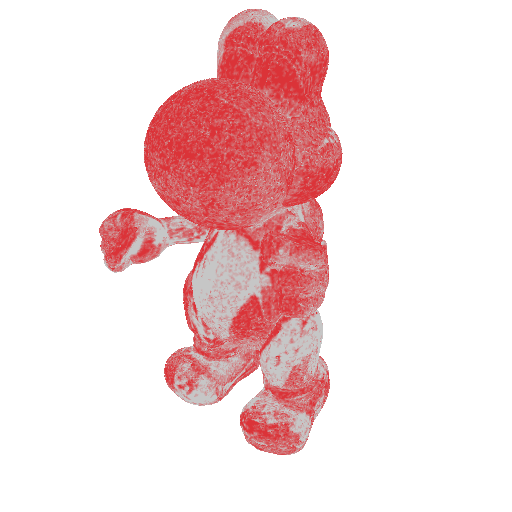}
        & \includegraphics[width=3.2cm]{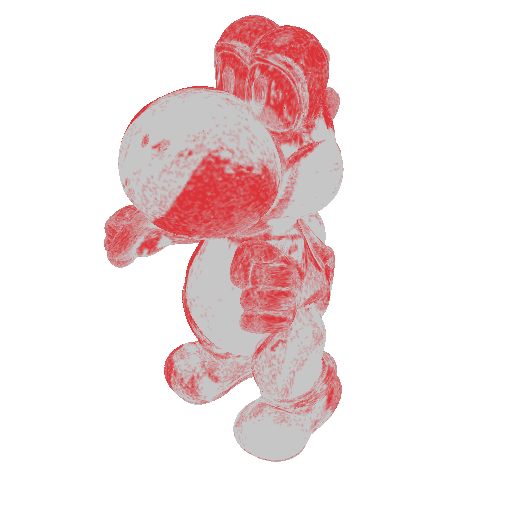}
        & \includegraphics[width=3.2cm]{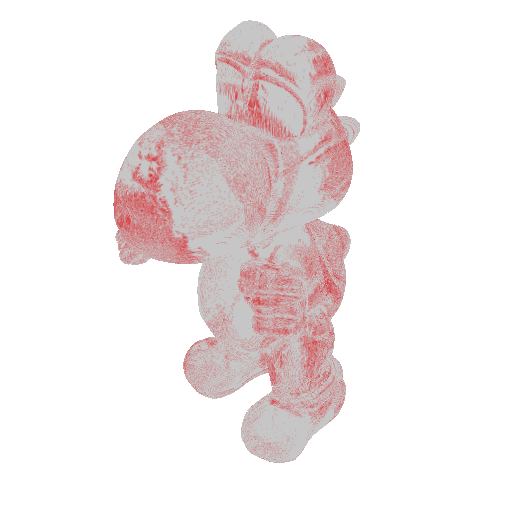} \\[2pt]
        \rotatebox{90}{\quad\quad\quad\textit{Specular}}
        & \includegraphics[width=3.2cm]{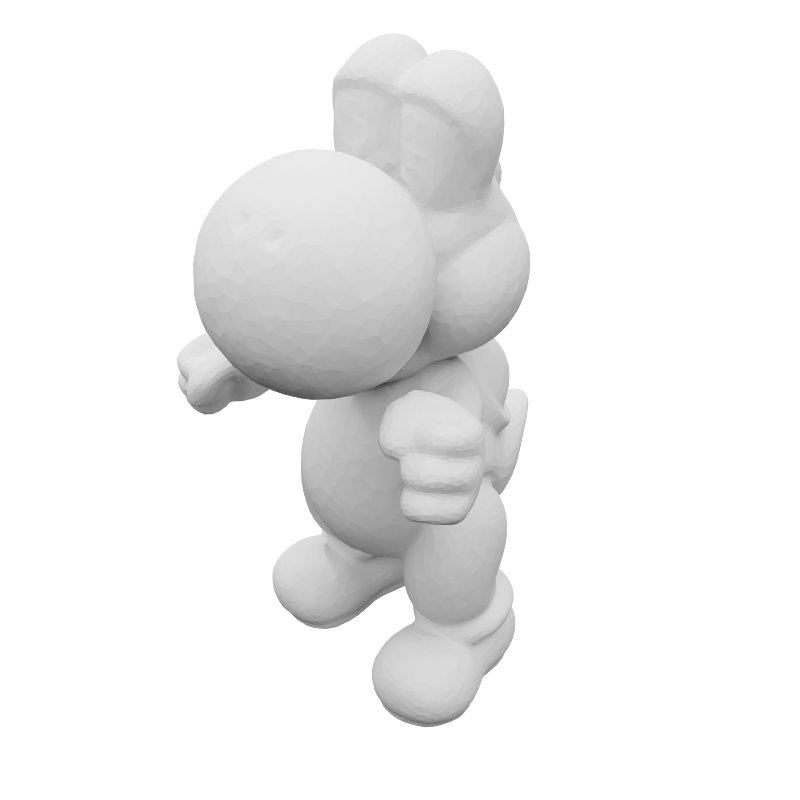}
        & \includegraphics[width=3.2cm]{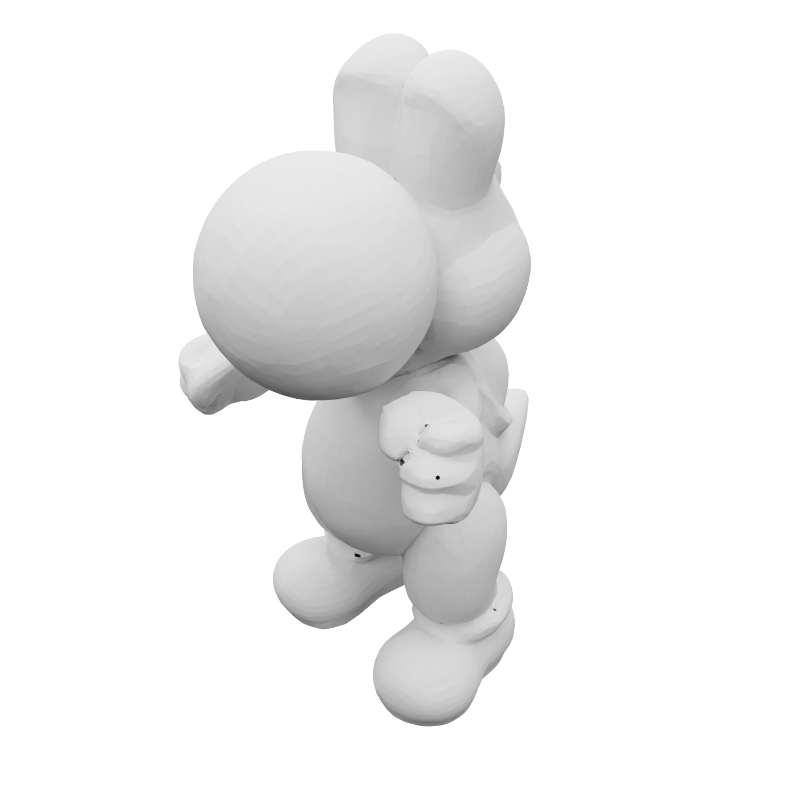}
        & \includegraphics[width=3.2cm]{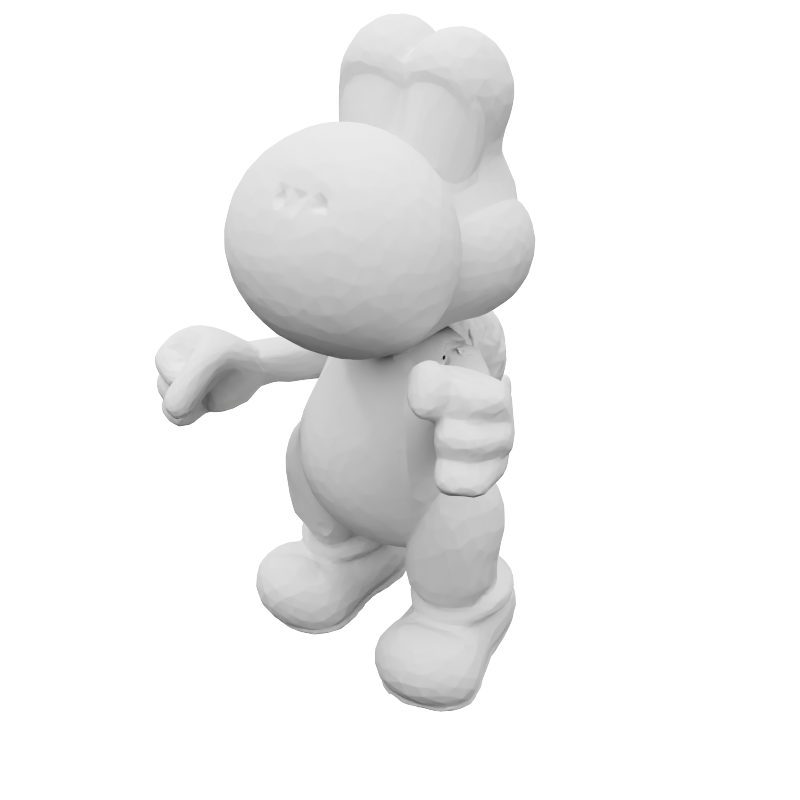}
        & \includegraphics[width=3.2cm]{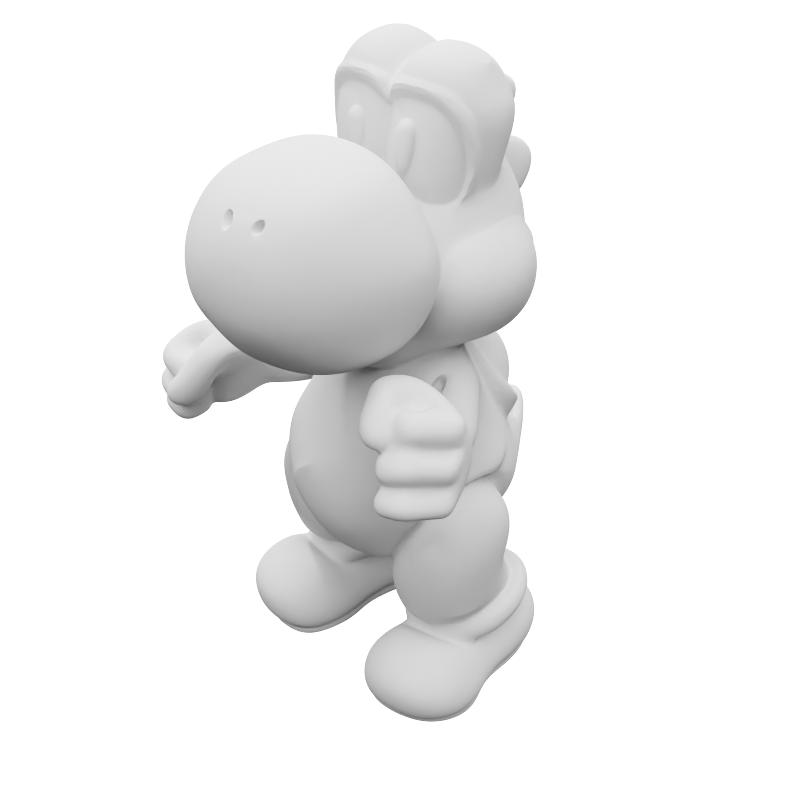}
        & \includegraphics[width=3.2cm]{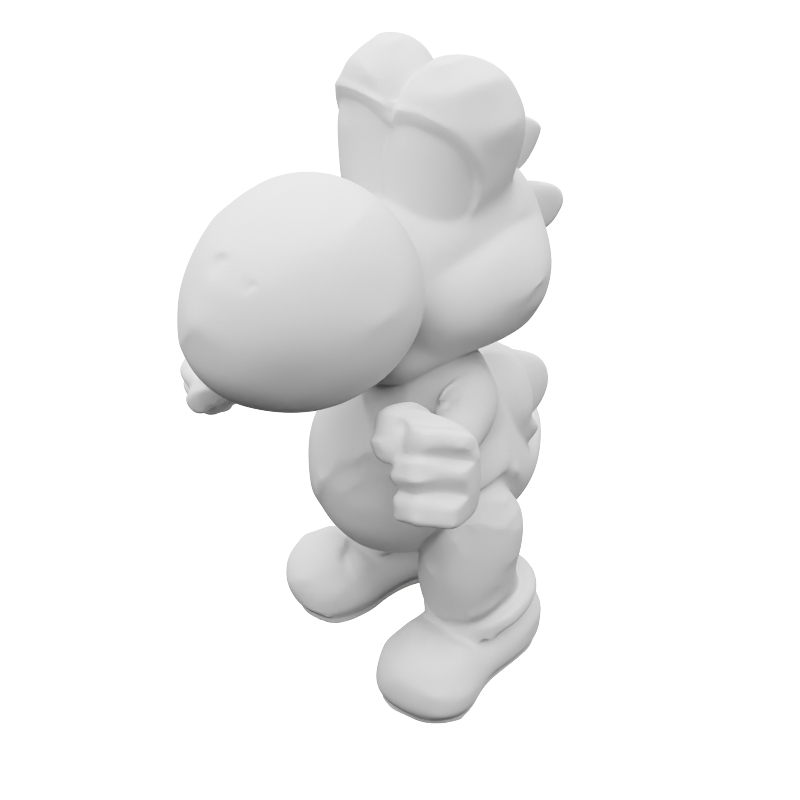} \\
    \end{tabular}%
    }
    \caption{Reconstruction vs.\ generation on a GSO object across \name{} and three state-of-the-art generative baselines, with GT for reference. All renders share the same orbit camera. \emph{Contour overlay} (top): matte render composited with the GT silhouette outline (\textcolor{red}{red}) and GT depth iso-contours (\textcolor{cyan}{cyan}); contours hug \name{} but visibly peel off the generative baselines. \emph{Hallucination mask} (middle): faces visible from some input view yet inconsistent in depth and normal with every view are coloured red; \name{} is near-uniformly grey while baselines show saturated red patches on directly observed surfaces. \emph{Specular render} (bottom): a reflective shader exposes curvature deviations that matte shading averages out.}
    \label{fig:exp:comp:gen}
    \Description{A three-row grid compares a toy figure across ground truth, MeRM, TRELLIS2, Hunyuan3D v2.5, and Hunyuan3D v3.1. The top row overlays red ground-truth silhouettes and cyan depth contours on dark mesh renderings. The middle row marks observation-inconsistent faces in red on otherwise grey meshes, with no ground-truth panel. MeRM has scattered red areas, whereas the generative baselines have broader red patches on the figure's body and limbs. The bottom row uses a reflective material to expose differences in curvature and small surface details.}
\end{figure*}

The general comparison in Sec.~\ref{sec:exp:perf} already places \name{} an order of magnitude ahead of state-of-the-art generative 3D models on Chamfer Distance and F1 (Table~\ref{tab:comparison}). Yet generative methods render visually compelling surfaces with high-resolution details---where does the fidelity gap actually surface, and what does it look like? This section answers visually.

\highlight{Setup \& Results.}
We render the predicted mesh from a held-out orbit camera and apply three diagnostics. \emph{(i)~Contour overlay:} the matte render is composited with the GT silhouette outline (red) and three GT depth iso-contours (cyan) from the same camera; agreement between the rendered surface and these GT-derived contours is a direct visual measure of input-consistency. \emph{(ii)~Hallucination mask:} every face is checked against all input views for joint depth- and normal-agreement, and faces that are visible from some view yet inconsistent with every view's observations are coloured red, while consistent or unobserved faces remain neutral grey. \emph{(iii)~Specular render:} a reflective shader replaces the matte material so that high-frequency highlights expose curvature deviations matte shading averages out. Figure~\ref{fig:exp:comp:gen} applies all three to a single GSO object. With the contour overlay active, GT depth and silhouette contours tightly track \name{}'s rendered surface but visibly peel off the generative baselines, which mis-scale the object and miss surface depressions present in the inputs. The hallucination mask paints \name{}'s mesh near-uniformly grey, whereas TRELLIS2 and Hunyuan3D show large saturated-red patches \emph{on directly observed surfaces}, not merely on unseen backsides. Under specular shading, \name{}'s reflections track the GT highlights cleanly while the baselines exhibit visible curvature ripples. Together these explain the order-of-magnitude CD gap from Sec.~\ref{sec:exp:perf}.

\highlight{Discussion.}
Reconstruction and generation address complementary needs. Generation excels at synthesizing plausible content from minimal input and remains the right tool when the goal is to produce a clean shape coherent with a single observation. Reconstruction is the right tool when downstream tasks---digital twins, simulation, measurement, robotic manipulation---cannot tolerate confidently-rendered but unfaithful surfaces. \name{} does not compete with generative pipelines; it fills the orthogonal need by treating sparse views as observations to honour, not as conditioning to be embellished. Detailed positioning can be found in Appendix~\ref{sec:supp:recon_vs_gen}.

\subsection{Comparison with Fidelity-Aware Generation}
\label{sec:exp:fidelity_aware}

The generative baselines above make no explicit fidelity claim, so a sceptical reading is that the gap merely reflects an objective mismatch. The stronger test is against methods that \emph{do} target input fidelity from within the generative paradigm---the two strategies of Sec.~\ref{sec:related} --- together with a post-hoc remeshing pipeline. Table~\ref{tab:fidelity_aware} reports this comparison on the GSO subset under a matched four-view protocol, with one additional dense-view row for the optimization-based entry.

\begin{table}[t]
    \centering
    \caption{Comparison with fidelity-aware generative methods and a post-hoc remeshing pipeline on the GSO subset. All methods receive four input views, except the second Matcha-Gaussians row, which is given $24$ views to show what per-scene optimization attains with dense supervision. \emph{Gen.} marks methods in which a generative predictor produces the output geometry. Fidelity-oriented designs substantially narrow the gap to reconstruction but do not close it: the remaining margin is the cost of recovering observed surfaces by sampling rather than by construction.}
    \label{tab:fidelity_aware}
    \resizebox{\linewidth}{!}{
        \begin{tabular}{lc|ccccc}
            \toprule
            \textbf{Method} & \textbf{Gen.} & AbsRel~(\%)$\downarrow$ & MAE~($^\circ$)$\downarrow$ & Normal-PSNR$\uparrow$ & F1$\uparrow$ & CD$\downarrow$ \\
            \midrule
            Unique3D~\cite{wu2024unique3d}            & \checkmark & 1.941 & 24.92 & 18.90 & 0.95126 & 1.527 \\
            ReconViaGen~\cite{chang2025reconviagen}         & \checkmark & 0.659 & 14.76 & 21.24 & 0.98220 & 0.295 \\
            Pixal3D~\cite{li2026pixal3d}             & \checkmark & 0.609 & 11.64 & 22.70 & 0.98778 & 0.178 \\
            Matcha-Gaussians(V=4)~\cite{guedon2025matcha}    & --         & 2.364 & 26.77 & 18.66 & 0.94155 & 1.984 \\
            Matcha-Gaussians(V=24)~\cite{guedon2025matcha}    & --         & 0.386 & 10.17 & 24.81 & 0.99612 & 0.105 \\
            \midrule
            \name{} (VecSet)    & --         & \textbf{0.292} & \textbf{6.019} & \textbf{28.70} & \textbf{0.99995} & \textbf{0.066} \\
            \bottomrule
        \end{tabular}
    }
\end{table}

\begin{figure*}[t]
    \centering
    \setlength{\tabcolsep}{1pt}
    \renewcommand{\arraystretch}{0.6}
    \resizebox{\textwidth}{!}{%
    \begin{tabular}{@{} c ccccc c ccccc @{}}
        & \textbf{GT} & \textbf{\name{}} & \textbf{ReconViaGen} & \textbf{Pixal3D} & \textbf{Matcha}
        & & \textbf{GT} & \textbf{\name{}} & \textbf{ReconViaGen} & \textbf{Pixal3D} & \textbf{Matcha} \\[4pt]
        \rotatebox{90}{\quad\quad\textit{Contour}}
        & \includegraphics[width=2.02cm]{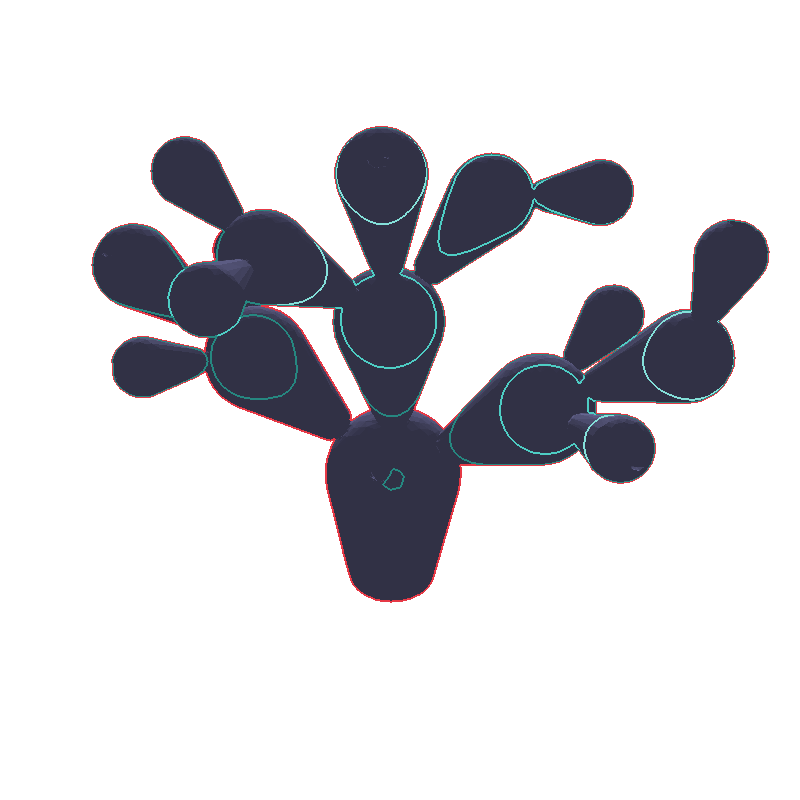} & \includegraphics[width=2.02cm]{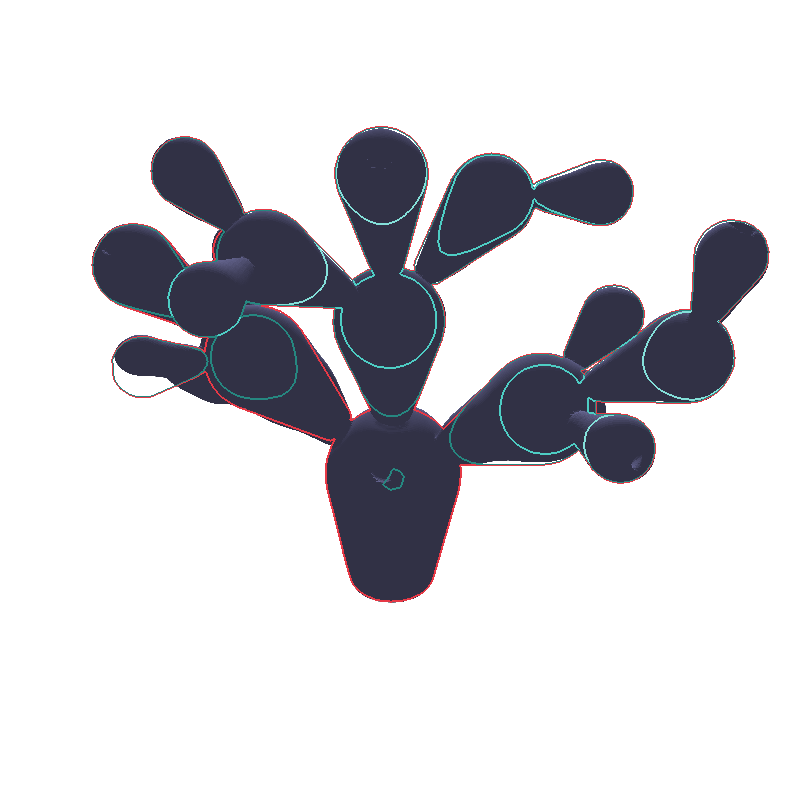} & \includegraphics[width=2.02cm]{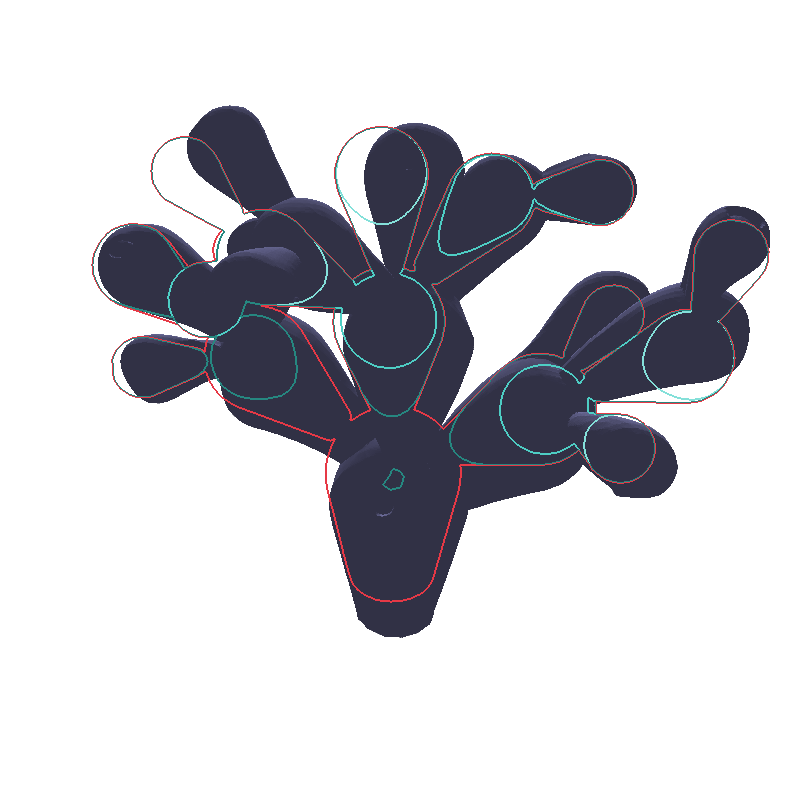}
        & \includegraphics[width=2.02cm]{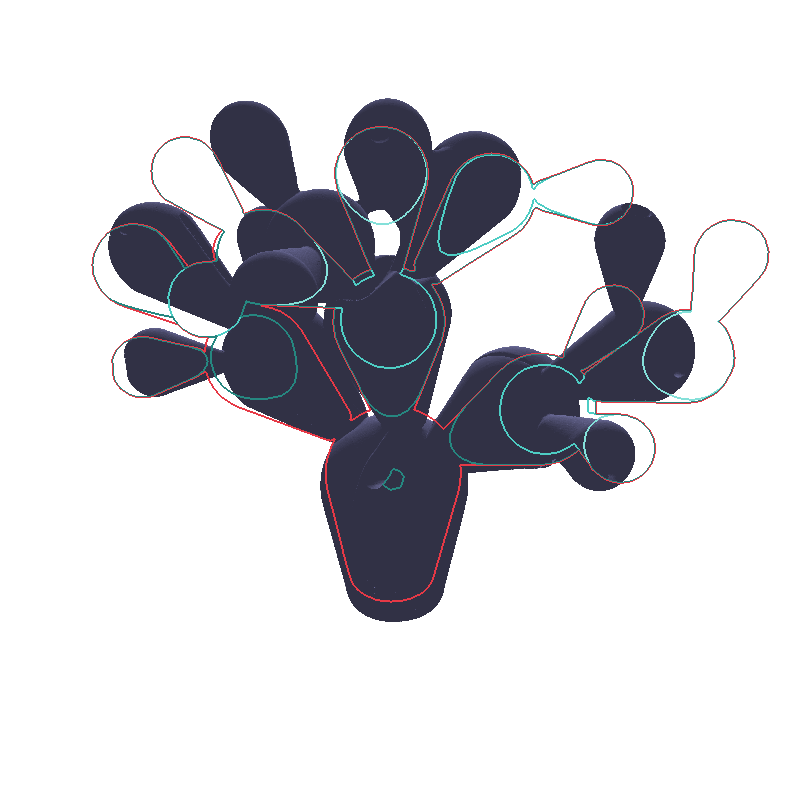} & \includegraphics[width=2.02cm]{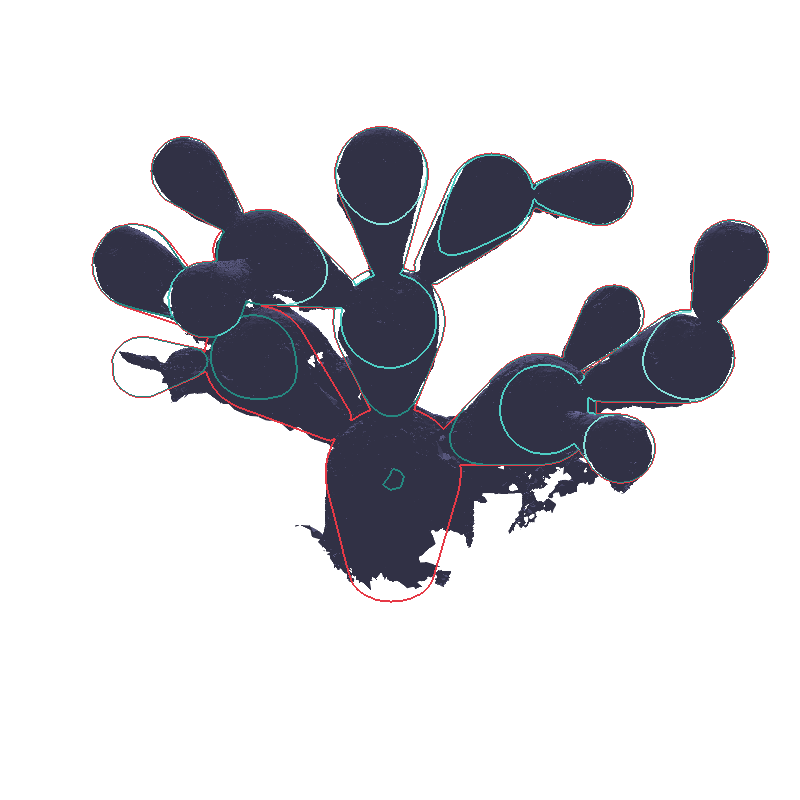}
        & \hspace{1mm}
        & \includegraphics[width=2.02cm]{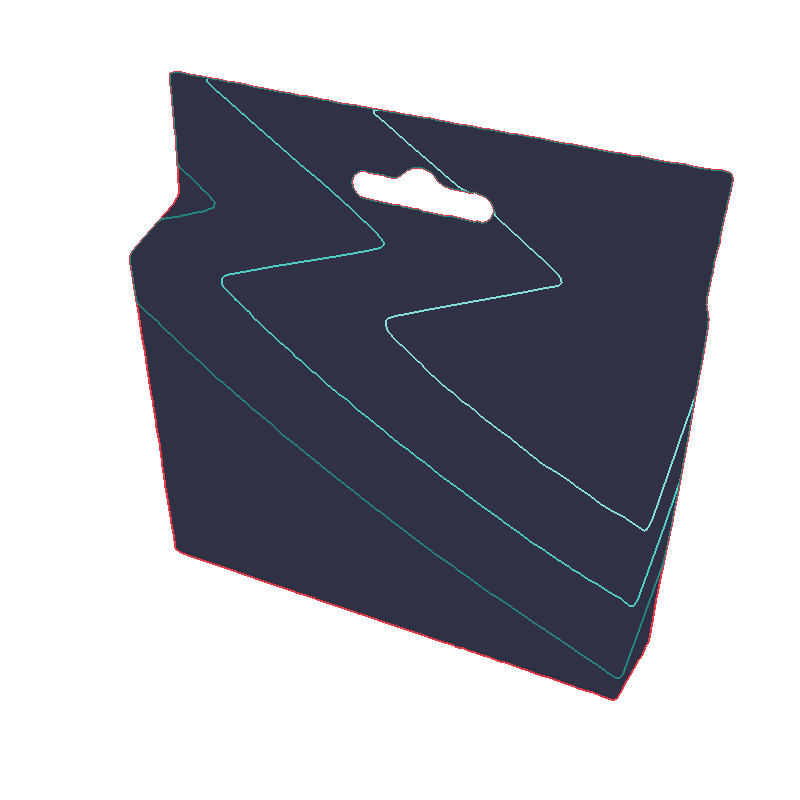} & \includegraphics[width=2.02cm]{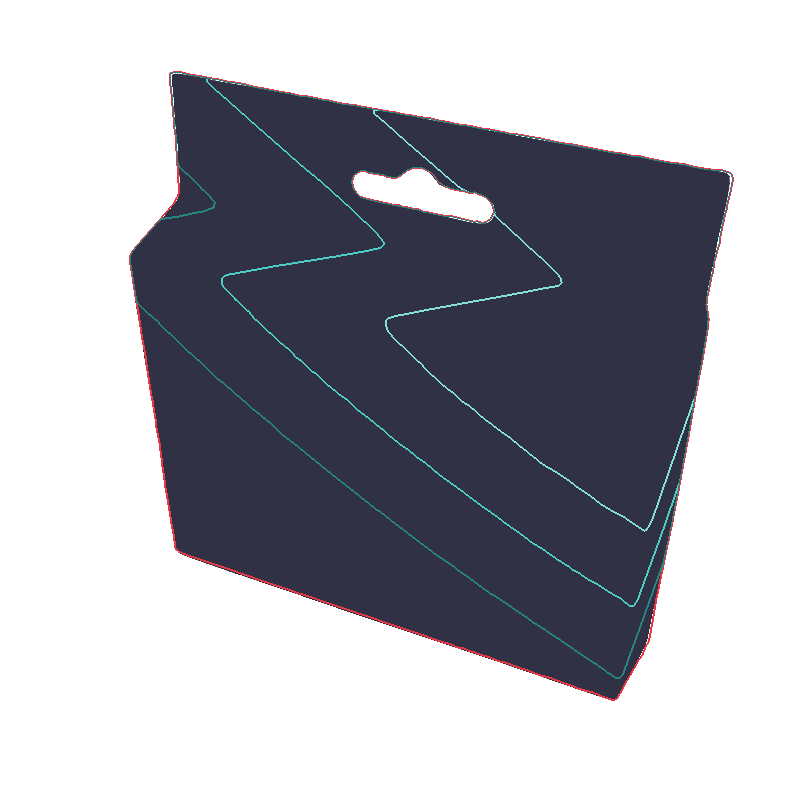} & \includegraphics[width=2.02cm]{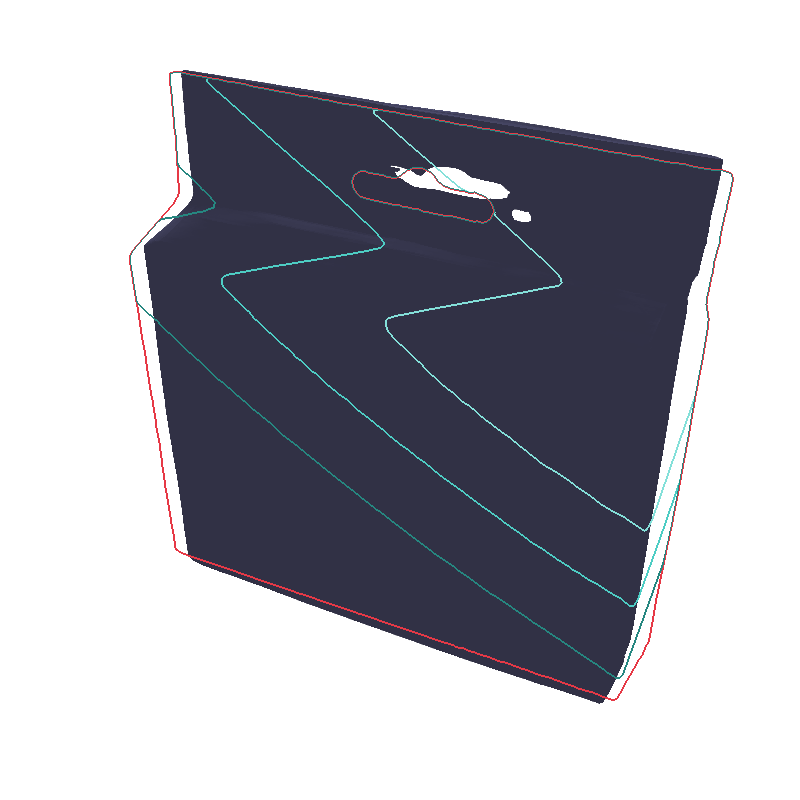}
        & \includegraphics[width=2.02cm]{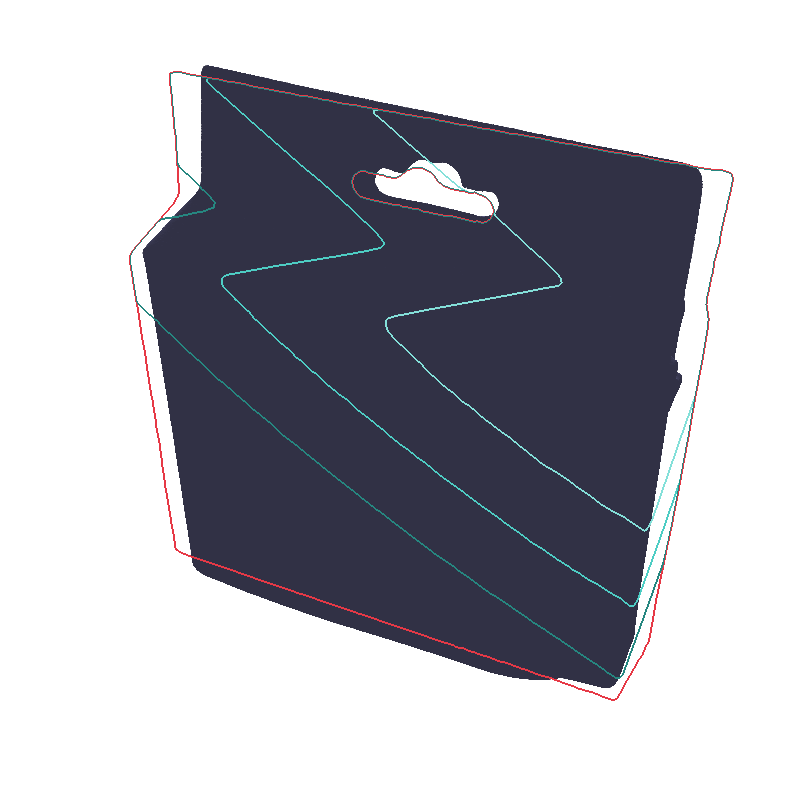} & \includegraphics[width=2.02cm]{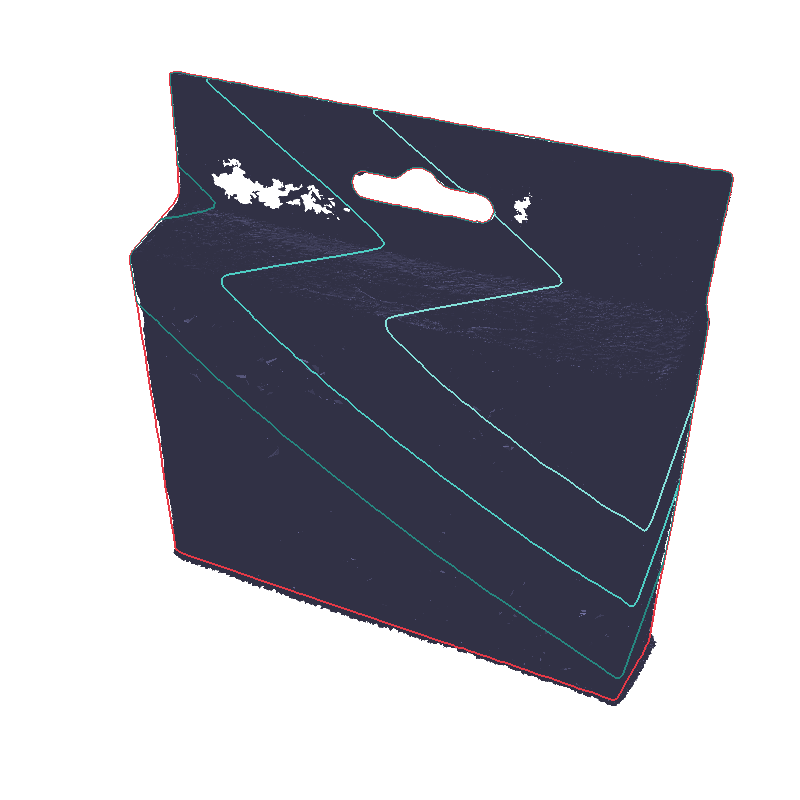} \\[1pt]
        \rotatebox{90}{\quad\quad\textit{Halluc.}}
        & & \includegraphics[width=2.02cm]{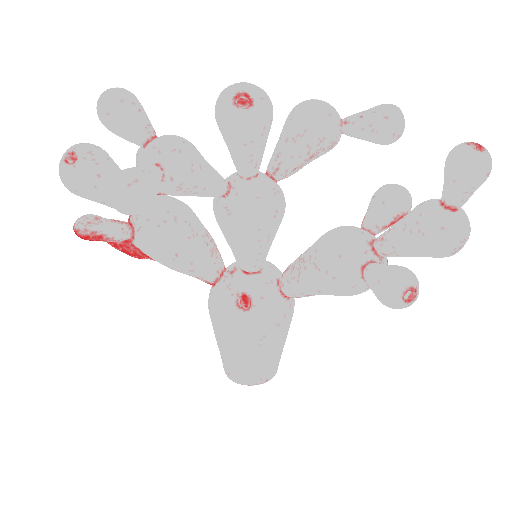} & \includegraphics[width=2.02cm]{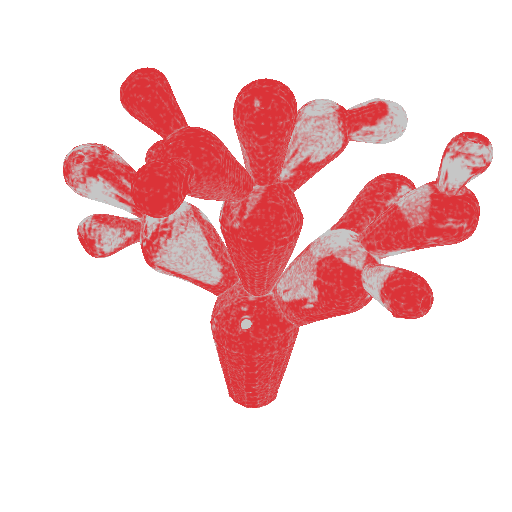}
        & \includegraphics[width=2.02cm]{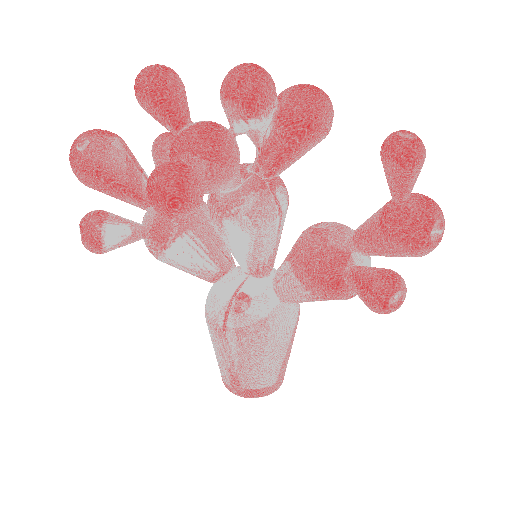} & \includegraphics[width=2.02cm]{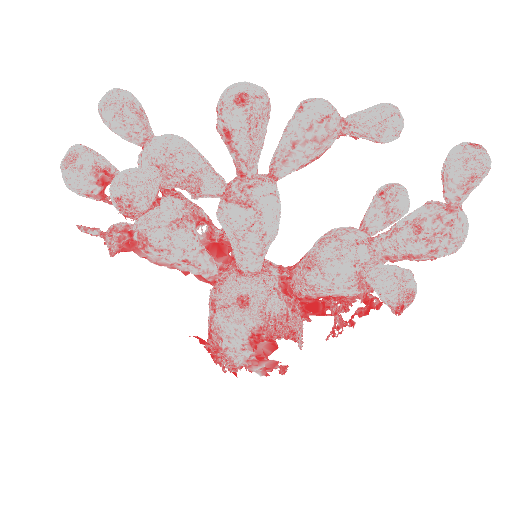}
        & \hspace{1mm}
        & & \includegraphics[width=2.02cm]{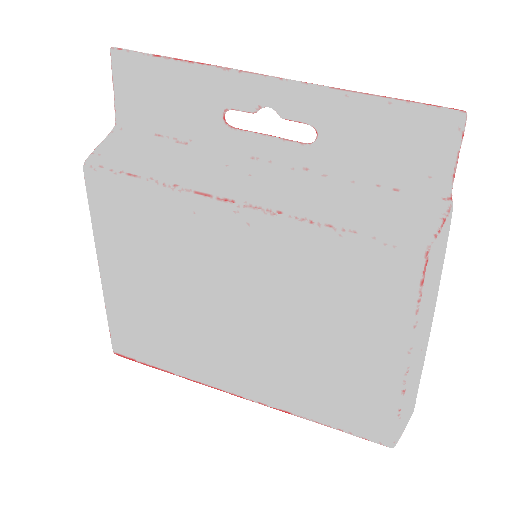} & \includegraphics[width=2.02cm]{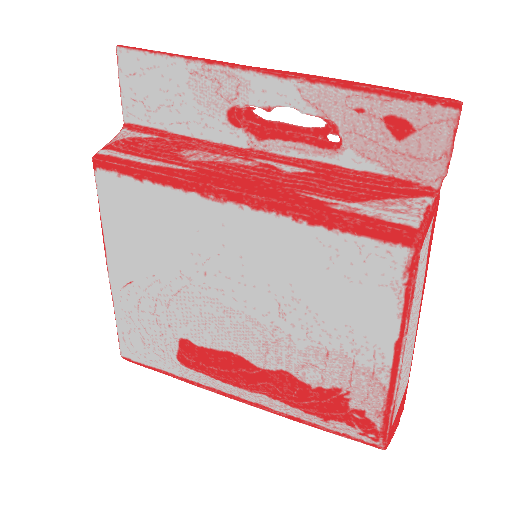}
        & \includegraphics[width=2.02cm]{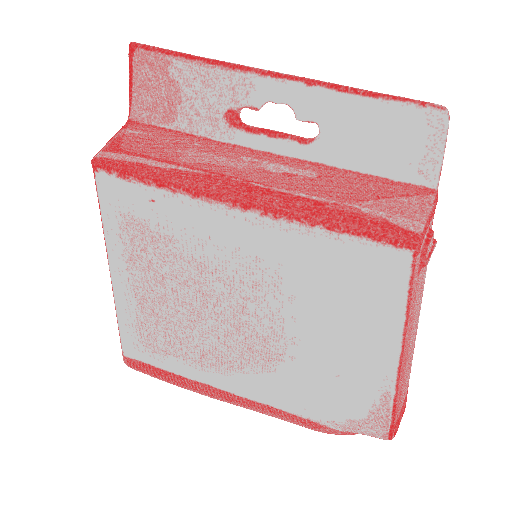} & \includegraphics[width=2.02cm]{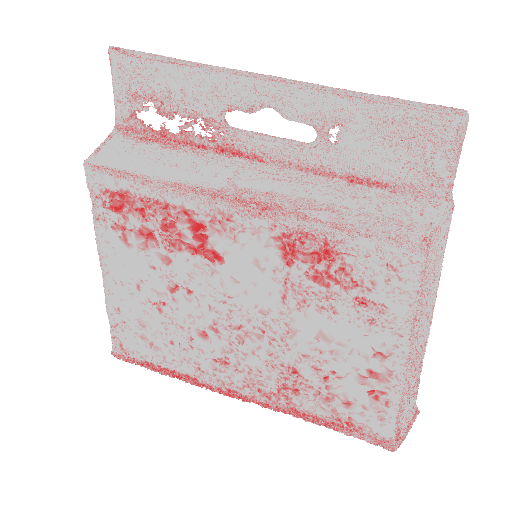} \\[1pt]
        \rotatebox{90}{\quad\quad\textit{Specular}}
        & \includegraphics[width=2.02cm]{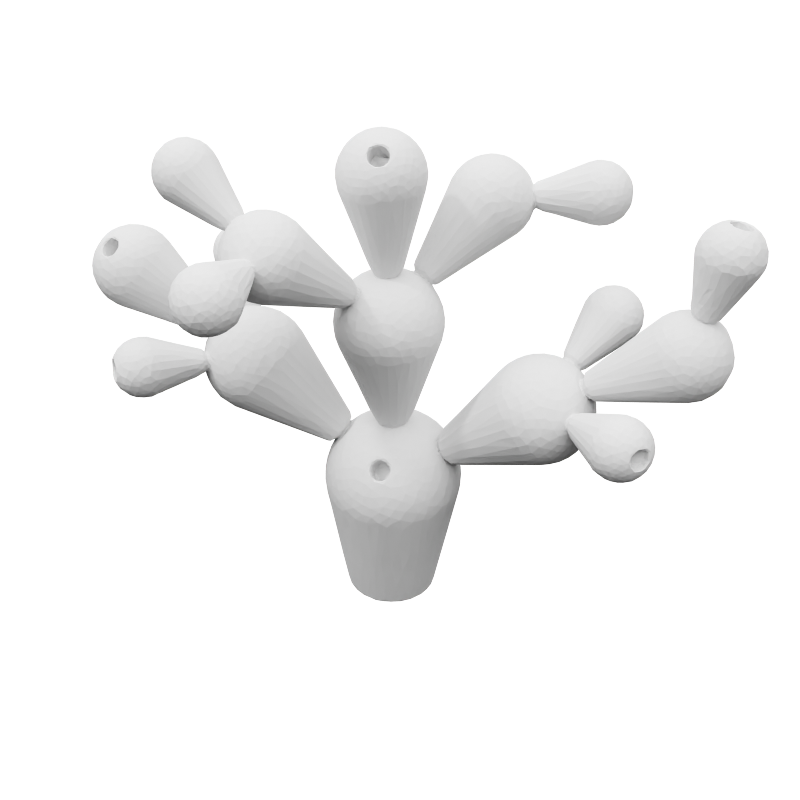} & \includegraphics[width=2.02cm]{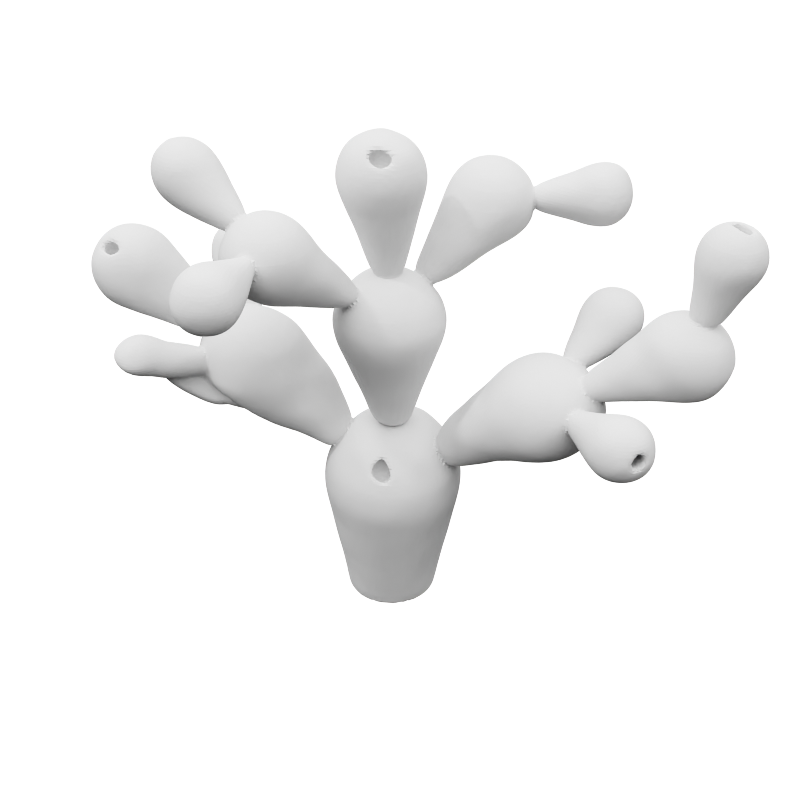} & \includegraphics[width=2.02cm]{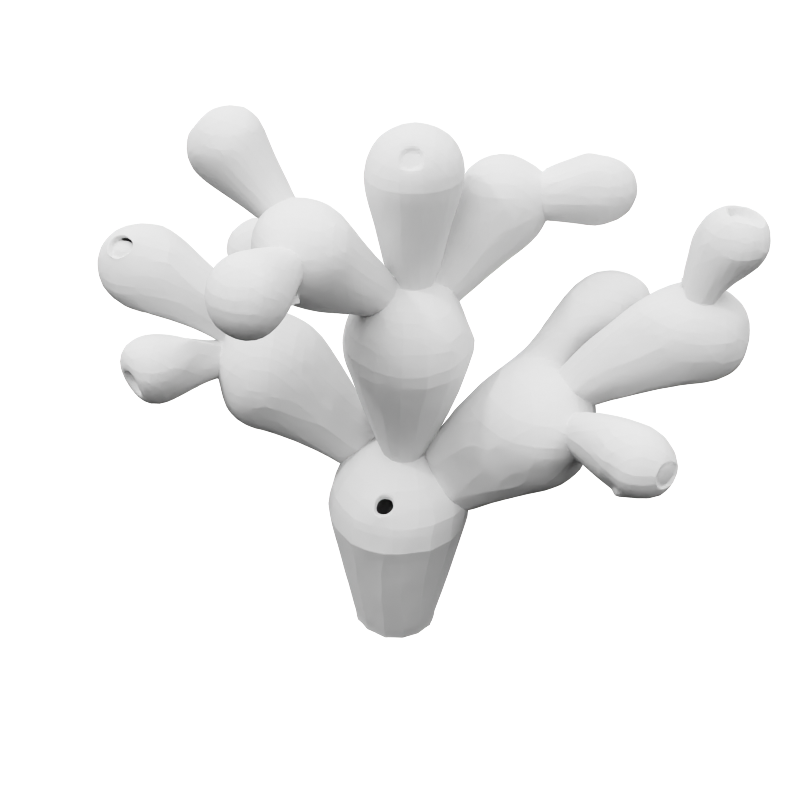}
        & \includegraphics[width=2.02cm]{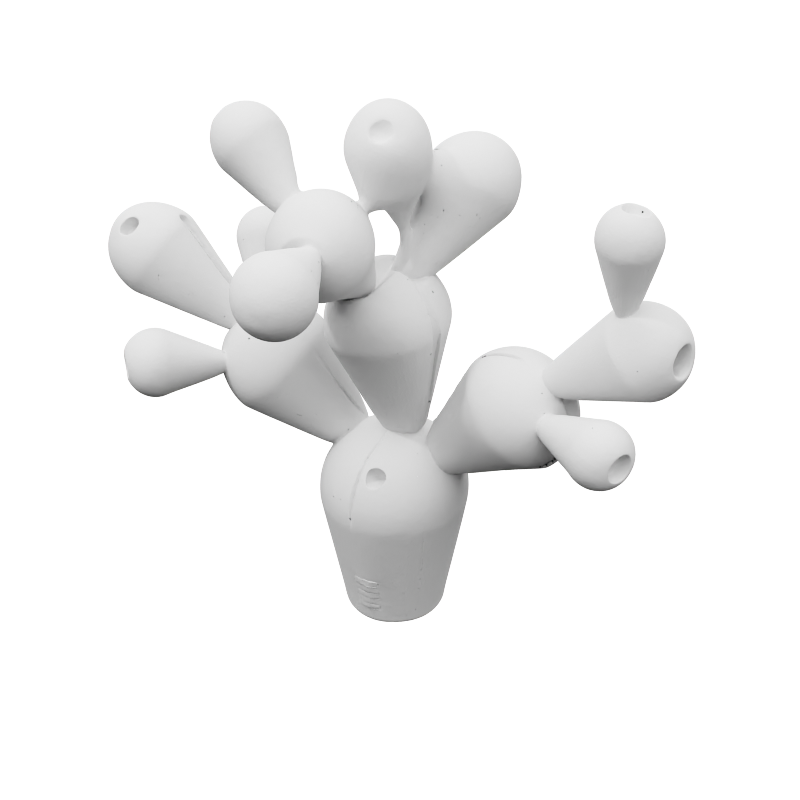} & \includegraphics[width=2.02cm]{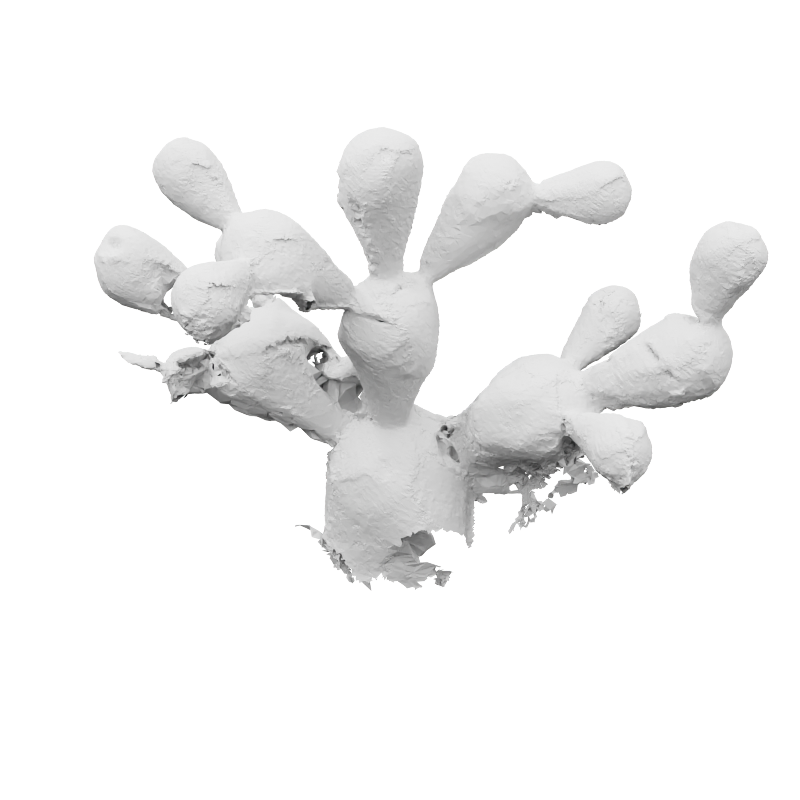}
        & \hspace{1mm}
        & \includegraphics[width=2.02cm]{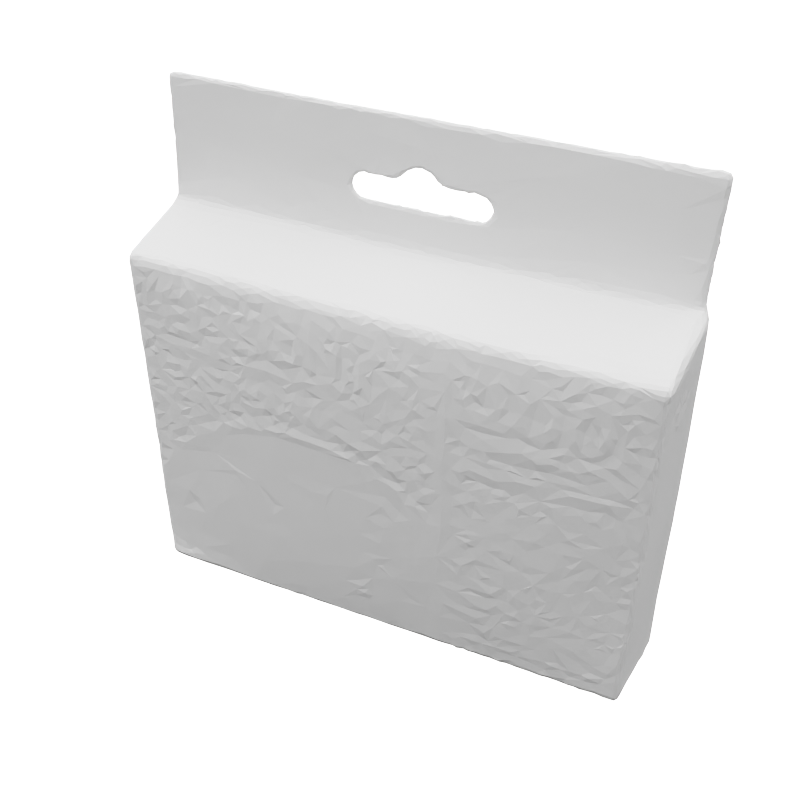} & \includegraphics[width=2.02cm]{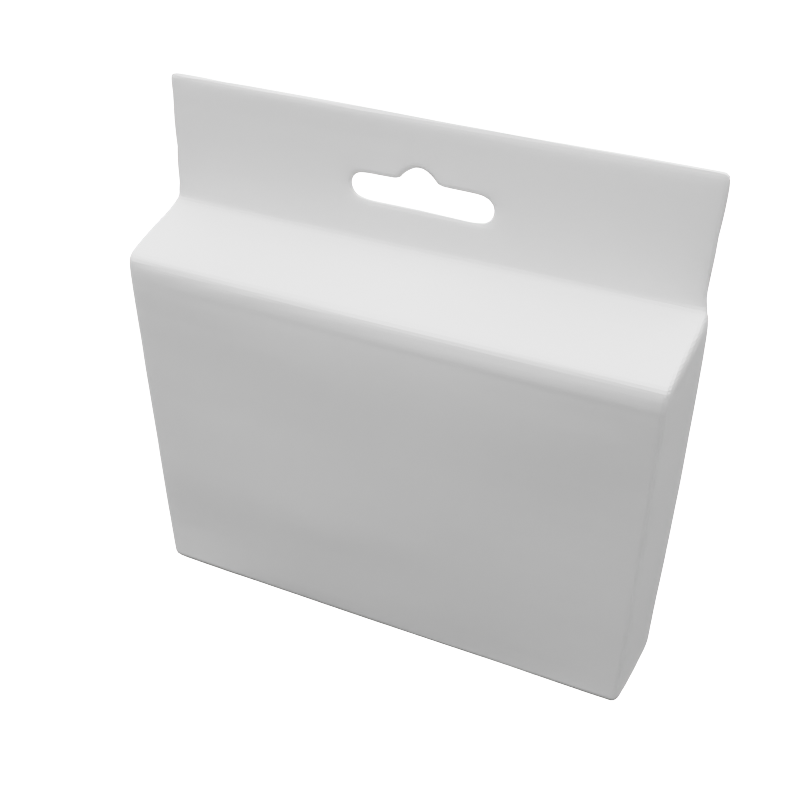} & \includegraphics[width=2.02cm]{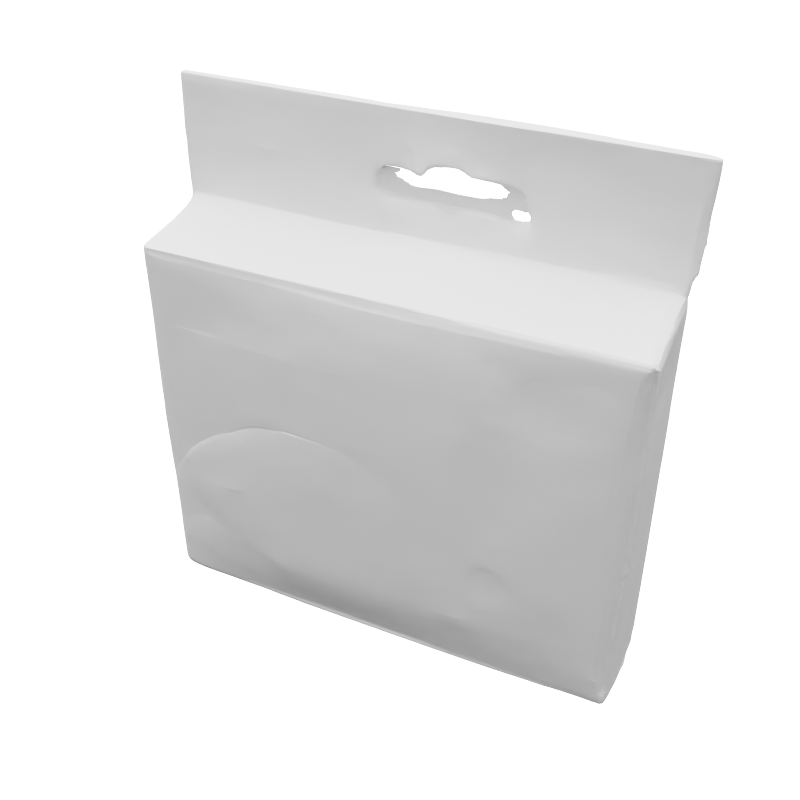}
        & \includegraphics[width=2.02cm]{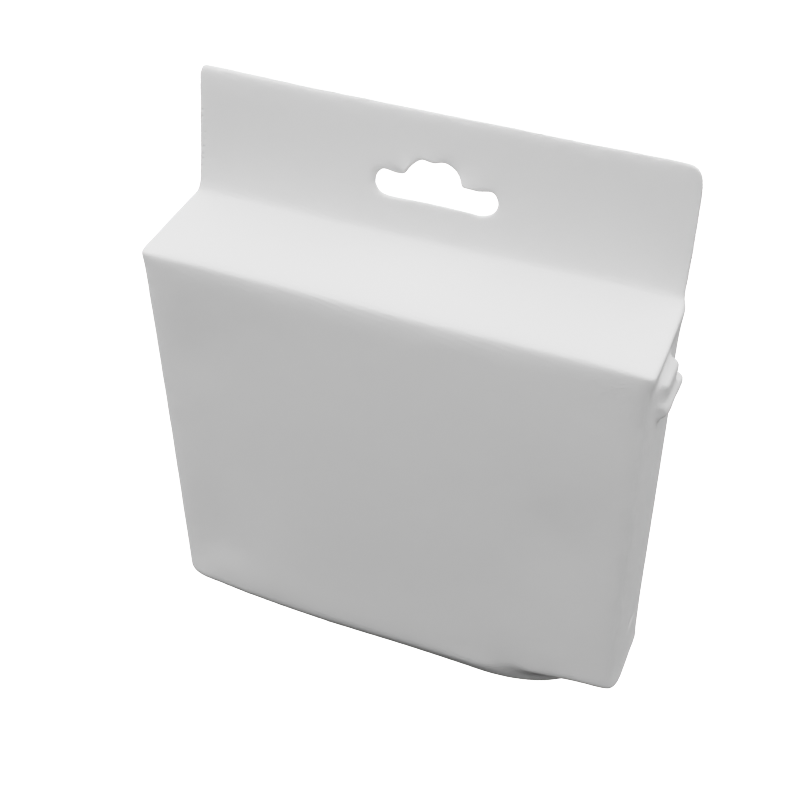} & \includegraphics[width=2.02cm]{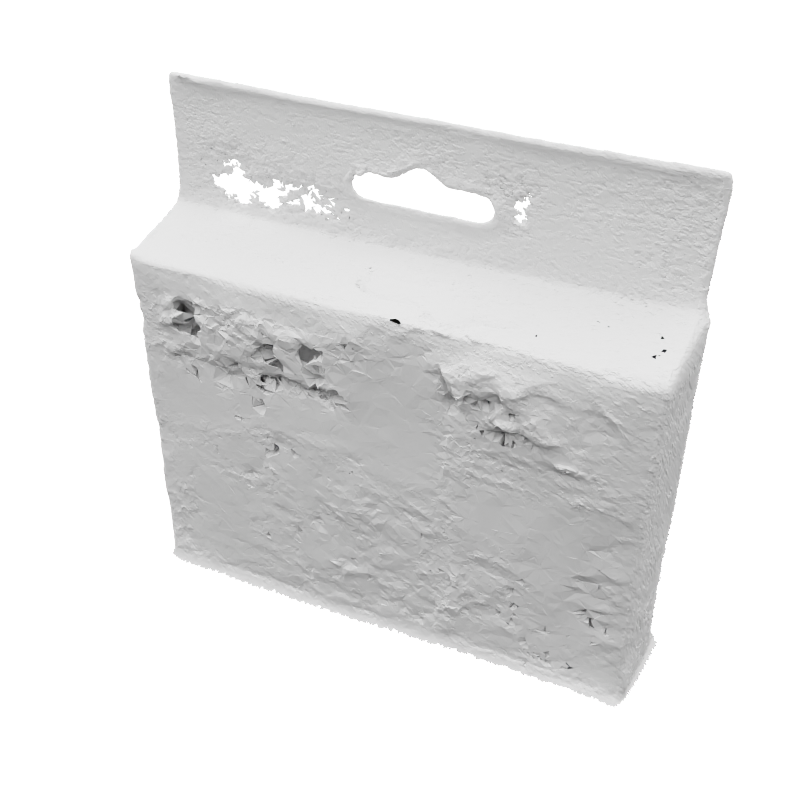} \\[3pt]
        & \multicolumn{5}{c}{\small (a) \textsc{Balancing Cactus}} & &
          \multicolumn{5}{c}{\small (b) \textsc{Epson Ink Cartridge}} \\
    \end{tabular}%
    }
    \caption{Fidelity-aware baselines under the three diagnostics of Fig.~\ref{fig:exp:comp:gen}, on two GSO objects that stress complementary regimes: thin branching structure~(a) and large flat panels~(b). \name{} here is the VecSet backend, Matcha-Gaussians is the $V\!=\!4$ configuration of Tab.~\ref{tab:fidelity_aware}, and all reconstructions use four input views. Within each group the five columns share one held-out orbit camera. \emph{Contour} (top): matte render with the GT silhouette (\textcolor{red}{red}) and GT depth iso-contours (\textcolor{cyan}{cyan}); the contours hug \name{} and visibly peel off the baselines, which thicken thin branches and bevel sharp panel edges. \emph{Hallucination mask} (middle): faces visible from at least one of the eight evaluation views yet inconsistent in depth and normal with \emph{every} one of them are coloured red. This is a deliberately conservative criterion---a face disagreeing with only some views is not flagged---and it still separates the methods by an order of magnitude: $7.3\%$/$1.9\%$ of visible faces for \name{} on (a)/(b), versus $75.3\%$/$28.5\%$ for ReconViaGen and $77.5\%$/$32.2\%$ for Pixal3D. The flagged regions lie on directly observed front surfaces, not on unseen backsides. \emph{Specular} (bottom): a reflective shader exposes curvature deviations that matte shading averages out. GT is omitted from the middle row because the mask is defined relative to the observations, against which GT is consistent by construction.}
    \label{fig:exp:fidelity_aware_qual}
    \Description{Two side-by-side comparison grids show a branching cactus toy and a rectangular ink cartridge. Each grid has columns for ground truth, MeRM, ReconViaGen, Pixal3D, and Matcha-Gaussians. The three rows show silhouette and depth-contour overlays, red observation-inconsistency masks on grey surfaces, and reflective renderings. The ground-truth mask panels are omitted. Differences are visible in the thickness and separation of cactus branches and in the cartridge's flat panels and edges. MeRM has fewer red regions; the other methods show red patches along branches or across the cartridge's front panels.}
\end{figure*}

\highlight{Results.}
The ordering is informative in both directions. Among the generative entries, the two fidelity-oriented designs improve over post-hoc remeshing by roughly an order of magnitude in CD ($0.295$ and $0.178$ versus $1.527$), confirming that their mechanisms work as intended. Yet the ranking is identical on every column: whichever diagnostic is used---depth, normal, or surface---the fidelity-aware generative methods sit between unconstrained generation and \name{}, and the residual margin never closes. The comparison against Matcha-Gaussians separates the two axes that CD alone conflates. Given the same four views, per-scene optimization is the weakest entry in the table ($2.364$ AbsRel, CD $1.984$); given $24$ views it becomes the strongest baseline ($0.386$, CD $0.105$), still short of what \name{} attains from four. The deficit for the generative methods is therefore not a shortage of observations but the architecture: a generative predictor remains the sole producer of all output geometry, so an observed surface is recovered only insofar as the sampler rediscovers it through the conditioning channel. \name{} regresses it directly, and recovers it by construction.

\highlight{Discussion.}

Fig.~\ref{fig:exp:fidelity_aware_qual} applies the same three diagnostics to two GSO objects that stress complementary regimes: a thin-branched cactus, where the failure mode is structural, and a flat-panelled cartridge, where it is a matter of surface fidelity on large smooth faces. The pattern from Tab.~\ref{tab:fidelity_aware} is legible in both. The GT contours track \name{} closely and peel away from the fidelity-aware baselines---on the cactus, ReconViaGen and Pixal3D thicken the branches and round off the tips, so the silhouette outline crosses their interior rather than following it; on the cartridge, both bevel the panel edges that the input views show as sharp. Under the hallucination mask the separation is quantitative: on the cactus, $7.3\%$ of \name{}'s visible faces are flagged, against $75.3\%$ for ReconViaGen and $77.5\%$ for Pixal3D; on the cartridge, $1.9\%$ against $28.5\%$ and $32.2\%$. These are faces visible from some evaluation view yet inconsistent with every one of them---the baselines' red regions sit on the object's directly observed front faces, not on unseen backsides, which is the signature of geometry produced by sampling rather than measurement. Matcha-Gaussians at $V\!=\!4$ fails in a visibly different manner, fragmenting into noise rather than producing a clean but unfaithful surface, consistent with its position at the bottom of the table.

\subsection{Real-World Evaluation}
\label{sec:exp:realworld}

Besides synthetic benchmarks like GSO, we evaluate {\name} on real captured objects with ground-truth 3D scans to fully reflect the challenges encountered in real-world settings, where input images exhibit sensor noise, imperfect segmentation, and inconsistent illumination.

\highlight{Setup.}
We adopt the Stanford-ORB benchmark~\cite{kuang2024stanford}, which provides real-world objects captured under natural conditions along with high-quality ground-truth meshes obtained via structured-light scanning. For each object, we select 8 input views from the provided multi-view captures. We compare \name{} against PRM~\cite{ge2025prm} as a feed-forward baseline, as well as 2DGS \cite{huang20242d} and GOF~\cite{yu2024gaussian} (using 24/64 dense views) as optimization-based references. We report the same holistic suite of depth, normal, and mesh metrics as in Sec.~\ref{sec:exp:perf}.

\begin{table}[t]
    \centering
    \caption{Quantitative evaluation on the Stanford-ORB benchmark with real-world captured objects and ground-truth 3D scans.}
    \label{tab:realworld}
    \resizebox{\linewidth}{!}{
        \begin{tabular}{lr|c|cc|cccc|cc}
            \toprule
            \multicolumn{2}{c|}{\multirow{3}{*}{\large{\textbf{Method}}}} & \multirow{3}{*}{\#View} & \multicolumn{2}{c|}{\textbf{Depth Quality}} & \multicolumn{4}{c|}{\textbf{Normal Quality}} & \multicolumn{2}{c}{\textbf{Mesh Quality}} \\
            \cmidrule(lr){4-5} \cmidrule(lr){6-9} \cmidrule(lr){10-11}
            \multicolumn{2}{c|}{} & & AbsRel~(\%)$\downarrow$ & Acc~(\%)$\uparrow$ & MAE~($^\circ$)$\downarrow$ & PSNR$\uparrow$ & SSIM$\uparrow$ & LPIPS$\downarrow$ & F1$\uparrow$ & CD$\downarrow$ \\
            \midrule
            & Ours (VecSet) & 8 & 0.408 & \textbf{99.98} & \textbf{12.20} & \textbf{27.87} & \textbf{0.962} & \textbf{0.038} & 0.971 & 1.50 \\
            \midrule
            & \multirow{1}{*}{PRM~\cite{ge2025prm}} & 6 & 2.062 & 99.96 & 35.92 & 18.08 & 0.901 & 0.126 & 0.538 & 24.55 \\
            & \multirow{1}{*}{2DGS~\cite{huang20242d}} & 24 & 0.366 & 99.82 & 18.57 & 26.53 & 0.927 & 0.058 & 0.976 & 55.59 \\
            & \multirow{1}{*}{2DGS~\cite{huang20242d}} & 64 & 0.268 & \textbf{99.98} & 15.26 & 27.35 & 0.938 & 0.052 & 0.979 & 0.90 \\
            & \multirow{1}{*}{GOF~\cite{yu2024gaussian}}  & 24 & 0.365 & 99.76 & 16.40 & 26.94 & 0.933 & 0.055 & 0.984 & 70.61 \\
            & \multirow{1}{*}{GOF~\cite{yu2024gaussian}} & 64 & \textbf{0.234} & \textbf{99.98} & 13.50 & 27.70 & 0.944 & 0.048 & \textbf{0.990} & \textbf{0.57} \\
            \bottomrule
        \end{tabular}
    }
\end{table}

\begin{figure*}[t]
    \begin{center}
        \includegraphics[width=1.0\textwidth,trim={0cm 6cm 0cm 0cm},clip]{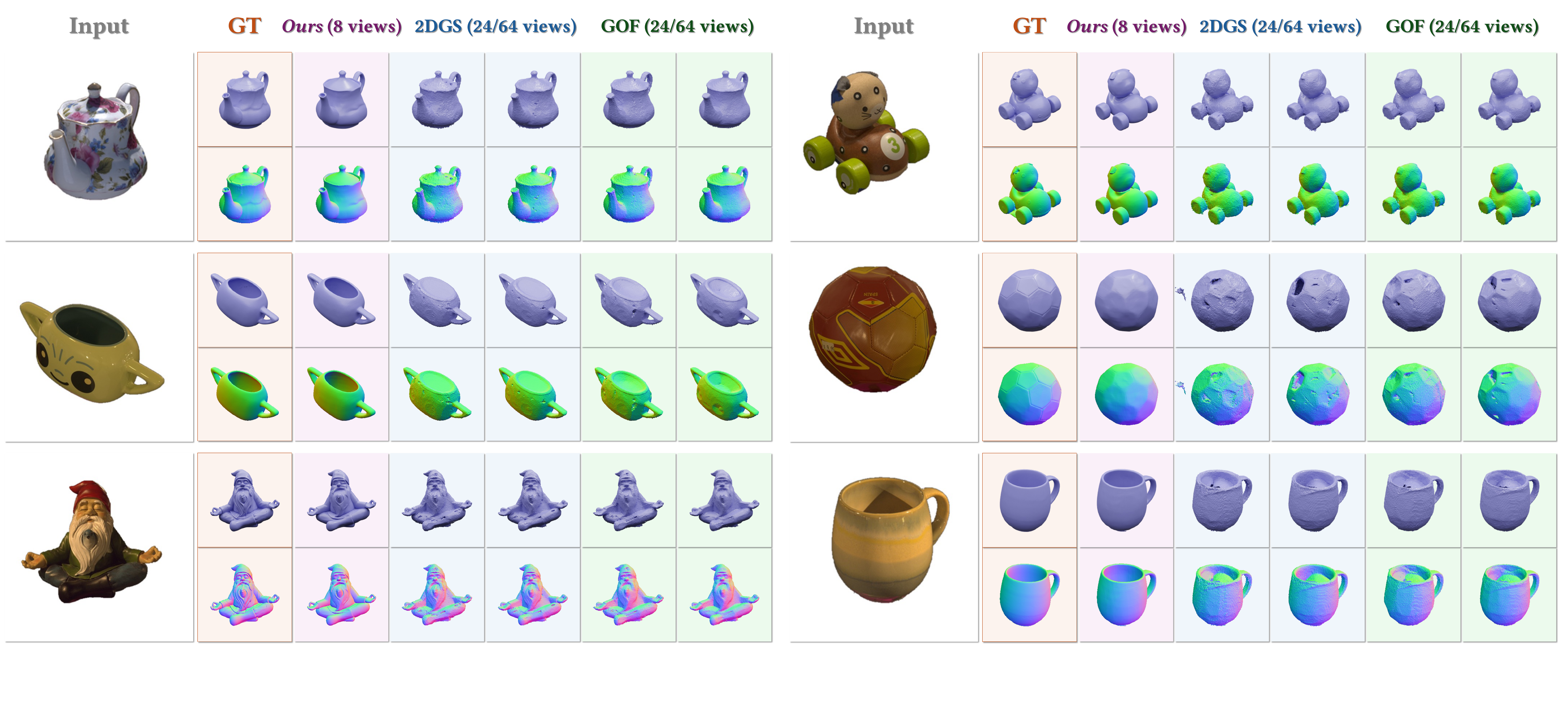}
	\end{center}
    \caption{Real-world reconstruction gallery on Stanford-ORB. \name{} generalizes robustly to in-the-wild captures with complex materials and lighting conditions.}
    \label{fig:exp:comp:real}
    \Description{A gallery of six real-world objects arranged in two groups of three rows, including patterned vessels, a toy animal, a figurine, a ball, and a mug. Each object group begins with an input photograph, followed by ground truth, MeRM using 8 views, and 2DGS and GOF using 2048 views. Each reconstruction is shown with shaded geometry above and surface-normal colors below. The comparisons expose differences in outlines, handles, rims, and fine relief across objects with reflective, textured, or unevenly illuminated surfaces.}
\end{figure*}

\highlight{Results.}
Table~\ref{tab:realworld} reports the quantitative results, with qualitative comparisons in Fig.~\ref{fig:exp:comp:real}. Despite being trained exclusively on synthetic Objaverse data, \name{} achieves competitive or superior performance on real captured objects. Notably, our method with only 8 views outperforms optimization-based methods using 24 views, confirming that NGT-learned geometric priors transfer effectively across the synthetic-to-real domain gap.

\highlight{Discussion.}
The strong real-world performance confirms that the geometric representations learned by {\name{}} are not overfit to synthetic data. The NGT encoder learns view-consistent geometric features rather than appearance-dependent cues, making it naturally invariant to domain-specific photometric variations.

\section{Analysis of Neural Geometry Tokens}
\label{sec:analysis}

The experiments of Sec.~\ref{sec:exp} establish \emph{that} \name{} reconstructs faithfully; this section establishes \emph{why}. We substantiate the structural-bottleneck diagnosis of Sec.~\ref{sec:ffmr} with a controlled comparison of the two supervision pathways (Sec.~\ref{sec:analysis:diag}), isolate what each NGT-related component contributes (Sec.~\ref{sec:analysis:abl}), and probe the encoder layer by layer to see what NGTs encode (Sec.~\ref{sec:analysis:probing}). Backend-specific ablations, which concern the decoder rather than the NGT interface, remain in Appendix~\ref{sec:supp:decoder}.

\subsection{Bottleneck Diagnosis: Cross-Queried vs.\ 2D-Aligned}
\label{sec:analysis:diag}

\highlight{Setup.}
Fig.~\ref{fig:bottleneck} contrasts the cross-queried mesh-regression pipeline with a model regressing 2D-aligned targets. The two endpoints share the same multi-view Transformer backbone and the same Objaverse training data, and differ only in the supervision pathway. The \emph{cross-queried baseline} (Fig.~\ref{fig:bottleneck}a) pairs that backbone with a cross-queried head predicting SpVox latents end-to-end, trained under SpVox latent supervision with no geometric pretraining of any kind; we evaluate two visual-encoder variants, a from-scratch backbone and a frozen DINOv2 backbone. The \emph{2D-aligned target model} (Fig.~\ref{fig:bottleneck}b) is the geometric pretraining model of Appendix~\ref{sec:supp:pretrain}: the same backbone topped with a lightweight per-view header decoding pixel-aligned depth, normal, and mask maps under direct $L_1$/perceptual supervision. Because it instantiates no 3D primitive, it carries far fewer parameters than the cross-queried baseline.

For the cross-queried baselines, the depth and normal columns of Tab.~\ref{tab:fig2_compare} report metrics computed on the geometry \emph{implicit} to the predicted SpVox latents: we decode each predicted latent through the frozen TRELLIS2 decoder to recover a mesh, re-render its depth and normal maps from the input cameras, and evaluate against the ground truth. This equalizes the comparison across endpoints whose native outputs differ.

\begin{table}[t]
    \centering
    \caption{Head-to-head between the cross-queried baseline (Fig.~\ref{fig:bottleneck}a), the 2D-aligned target model (Fig.~\ref{fig:bottleneck}b), and our full pipeline. The 2D-aligned model attains the best geometric accuracy with the smallest model; our full pipeline retains that accuracy at the mesh level.}
    \label{tab:fig2_compare}
    \resizebox{\linewidth}{!}{
        \begin{tabular}{l|cccc}
            \toprule
            \textbf{Configuration} & Depth AbsRel (\%)$\downarrow$ & Normal MAE ($^\circ$)$\downarrow$ & Mesh CD$\downarrow$ & Mesh F1$\uparrow$ \\
            \midrule
            Fig.~\ref{fig:bottleneck}a, cross-queried, from-scratch & 1.352 & 21.435 & 0.235 & 0.96341 \\
            Fig.~\ref{fig:bottleneck}a, cross-queried, DINOv2 frozen & 1.261 & 22.168 & 0.257 & 0.95193 \\
            Fig.~\ref{fig:bottleneck}b, 2D-aligned target (Ours pretrain) & 0.306 & 7.137 & --- & --- \\
            \name{} (full pipeline) & 0.369 & 9.061 & 0.129 & 0.99760 \\
            \bottomrule
        \end{tabular}
    }
\end{table}

\highlight{Results.}

\begin{figure*}[t]
    \centering
    \includegraphics[width=\linewidth]{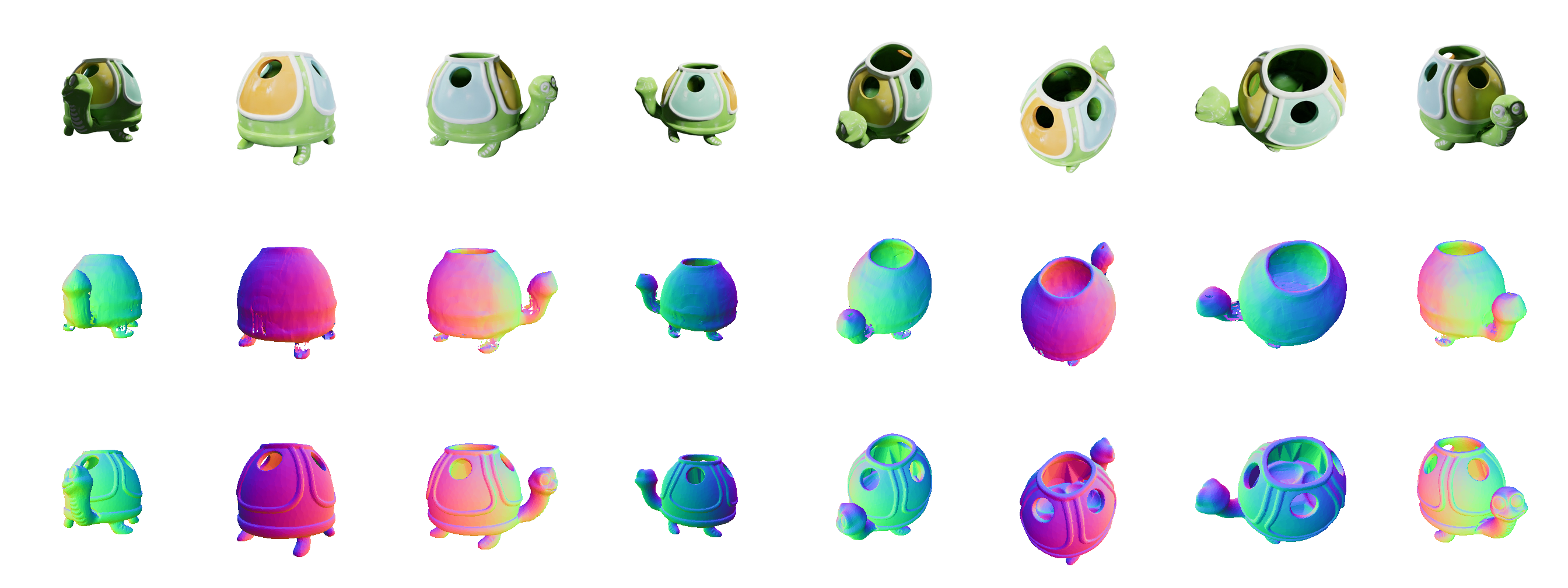}
    \caption{Qualitative mesh outputs of the cross-queried baseline (row (c.1) of Tab.~\ref{tab:ablation}, end-to-end from scratch). \textbf{Top:} input multi-view RGB. \textbf{Middle:} reconstructed mesh; surfaces are oversmoothed and lose fine geometric detail despite matched data and capacity. \textbf{Bottom:} ground-truth mesh, exhibiting sharp edges, holes, and surface ornaments that the cross-queried pipeline fails to recover.}
    \label{fig:supp_scratch_geom}
    \Description{Three rows show eight views of a green mechanical toy. The top row contains input photographs, the middle row contains the cross-queried baseline's reconstructed surfaces colored by normal direction, and the bottom row contains the corresponding ground-truth surfaces. The baseline retains the rounded body but smooths away recessed openings, raised rims, and small mechanical details that remain distinct in the ground truth.}
\end{figure*}

Two observations stand out in Tab.~\ref{tab:fig2_compare}. First, the 2D-aligned model improves depth AbsRel by more than $4\times$ and normal MAE by ${\sim}3\times$ over both cross-queried variants---and does so with a smaller parameter count. That a \emph{smaller} model trained on \emph{identical} data wins so decisively rules out capacity and data volume as explanations and localizes the deficit in the supervision pathway itself. Second, our full pipeline matches the 2D-aligned model's depth and normal quality to within noise while additionally producing high-quality meshes: the NGT-conditioned aggregator transfers the easy-to-learn 2D supervision into the difficult-to-learn 3D output without incurring the structural bottleneck of end-to-end cross-queried regression. Qualitatively, the cross-queried baseline produces oversmoothed surfaces that lose fine geometric detail despite matched data and capacity (Fig.~\ref{fig:supp_scratch_geom}).

\highlight{Implicit Cross-View Correspondence.}

\begin{figure*}[t]
    \centering
    \includegraphics[width=\linewidth]{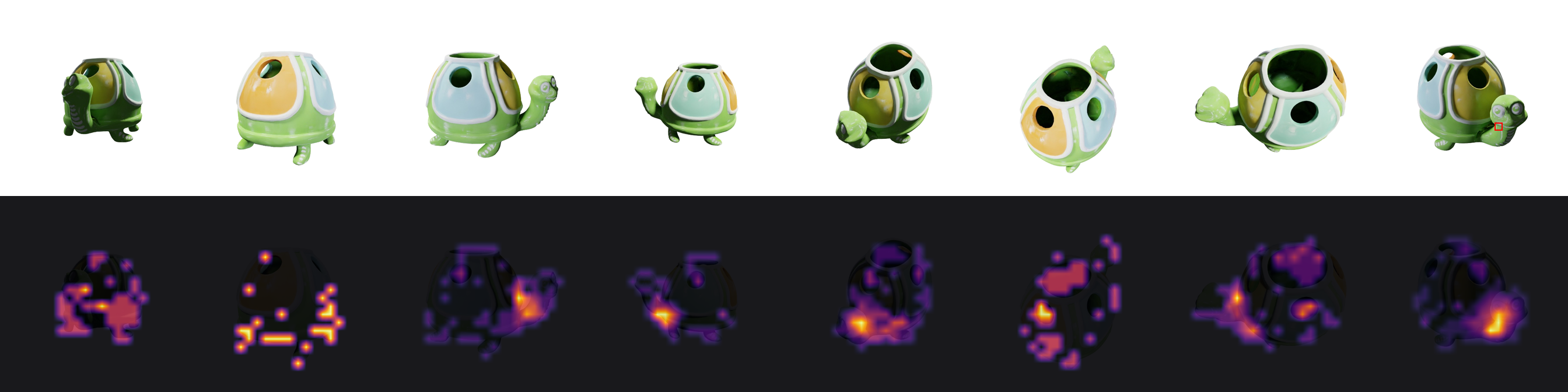}\\[1pt]
    {\small (a) Cross-queried baseline, end-to-end from scratch (no geometric pretraining).}\\[3mm]
    \includegraphics[width=\linewidth]{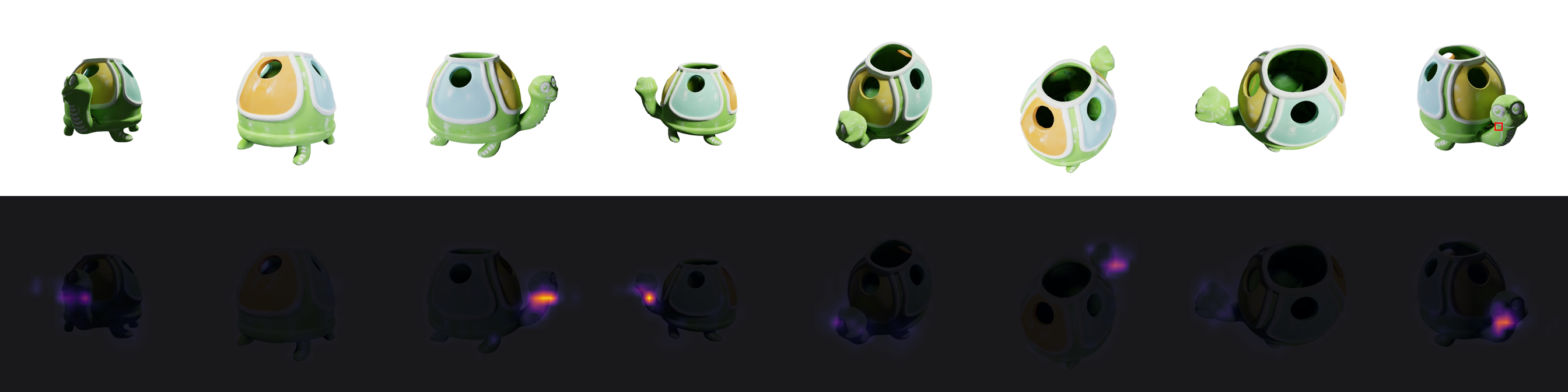}\\[1pt]
    {\small (b) Our 2D-aligned geometric pretraining.}
    \caption{Attention visualization from a fixed query pixel (red marker on the rightmost view) to all input-view patches. \textbf{Top half of each panel:} the input RGB views. \textbf{Bottom half of each panel:} per-patch attention map from the query, overlaid on each view. (a) Without geometric pretraining, attention is distracted across photometrically similar but geometrically unrelated regions. (b) With 2D-aligned supervision, attention concentrates on the corresponding 3D surface point's projection in every view, evidencing implicit cross-view correspondence.}
    \label{fig:supp_attn}
    \Description{Two stacked panels each show eight RGB views of the same toy above eight attention heatmaps. A marked query in the rightmost image is used for all views. Bright yellow and purple areas indicate stronger attention against a dark background. In the upper panel, the model trained from scratch produces dispersed bright patches across unrelated parts of the object. In the lower panel, geometric pretraining yields localized peaks at corresponding surface regions across views.}
\end{figure*}

A deeper read of the 2D-aligned model's attention reveals \emph{why} pixel-aligned supervision works. Fig.~\ref{fig:supp_attn} visualizes, for a fixed query pixel marked on the rightmost input view, the attention weights from that pixel to all patches in every input view. For the 2D-aligned model (Fig.~\ref{fig:supp_attn}b) the attention concentrates tightly on the corresponding 3D surface point's projections in the other views; for the cross-queried baseline (Fig.~\ref{fig:supp_attn}a) the same query attention is diffuse, scattered across photometrically similar but geometrically unrelated patches. The backbone therefore learns multi-view correspondence \emph{implicitly}, under pure 2D supervision---the exact aggregation skill that the cross-queried pipeline would otherwise have to bootstrap from a sparse 3D loss. Sec.~\ref{sec:analysis:probing} confirms the same capability quantitatively: 3D position becomes explicitly decodable from per-patch features by mid-encoder, which is only possible if cross-view correspondence has been established.

\subsection{Ablations}
\label{sec:analysis:abl}

\highlight{Setup.}
We ablate the three NGT-related components shared across both decoder backends. Backend-specific mechanisms (depth-lifting, occupancy filter) are deferred to Appendix~\ref{sec:supp:decoder}. All ablations are trained under the same data, optimization, and view-count protocol as the full model (Appendix~\ref{sec:supp:training}); only the indicated component is altered.
\begin{itemize}[leftmargin=*, itemsep=1pt]
    \item (b) \textit{w/o} \textit{Pruning}: \name{} uses mask-based visual token pruning to drop background patches from the conditioning context (Sec.~\ref{sec:method:network}, threshold $\tau{=}0.05$). Here we disable pruning and feed the full pre-pruning context $\bm{c}$ to both the filter and the aggregator.
    \item (c) \textit{w/o} \textit{Pretraining}: \name{} adopts a two-stage geometric pretraining paradigm (Sec.~\ref{sec:method:ngt}). We compare against end-to-end variants that bypass it: (c.1) trains the entire pipeline end-to-end from scratch; (c.2) replaces the geometric encoder with a frozen DINOv2 backbone.
    \item (d) \textit{w/o} \textit{NGT}: \name{} feeds high-dimensional NGTs into the mesh decoder. Here we replace the NGT context with the explicit depth and normal maps decoded by the pretraining header, eliminating NGTs and splitting the system into two strictly cascaded models.
\end{itemize}

\begin{table}[t]
    \centering
    \caption{NGT-related ablations on shared components. Row (b) probes the mask-based pruning of Sec.~\ref{sec:method:network}; rows (c.1)/(c.2) probe geometric pretraining; row (d) probes the NGT representation itself. Backend-specific ablations on depth-lifting are reported in Appendix~\ref{sec:supp:decoder}.}
    \label{tab:ablation}
    \resizebox{\linewidth}{!}{
        \begin{tabular}{c|l|c|c|cc|cc}
            \toprule
             & & Training & Depth Quality & \multicolumn{2}{c|}{Normal Quality} & \multicolumn{2}{c}{Mesh Quality} \\
            ID & Setting & iter/s$\uparrow$ & AbsRel~(\%)$\downarrow$ & MAE~($^\circ$)$\downarrow$ & PSNR$\uparrow$ & F1$\uparrow$ & CD$\downarrow$ \\
            \midrule
            - & Ours (Full Model) & 2.42 & 0.369 & 9.061 & 26.914 & 0.99760 & 0.129 \\
            \midrule
            (b) & \textit{w/o} Pruning & 4.71 & 0.374 & 10.238 & 26.752 & 0.99651 & 0.130 \\
            (c.1) & \textit{w/o} Pretrain~--~raw & 5.79 & 1.352 & 21.435 & 19.267 & 0.96341 & 0.235 \\
            (c.2) & \textit{w/o} Pretrain~--~DINO & 5.10 & 1.261 & 22.168 & 19.726 & 0.95193 & 0.257 \\
            (d) & \textit{w/o} NGT & 2.21 & 0.692 & 16.232 & 23.495 & 0.98452 & 0.194 \\
            \bottomrule
        \end{tabular}
    }
\end{table}

\highlight{Results.}
Table~\ref{tab:ablation} separates three distinct effects. Removing pruning (b) leaves quality essentially unchanged (AbsRel $0.369\!\to\!0.374$, CD $0.129\!\to\!0.130$) while nearly doubling training throughput ($2.42\!\to\!4.71$ iter/s). Pruning is thus an efficiency lever, not a quality regularizer; the quality contribution of the SpVox sparse path comes instead from depth-lifting, isolated in Appendix~\ref{sec:supp:decoder}.

Both variants that bypass geometric pretraining (c.1, c.2) collapse on every metric---including the one built on the strong frozen DINOv2 encoder. This is the sharpest available test of the structural-bottleneck hypothesis of Sec.~\ref{sec:ffmr}. A high-quality visual representation does not rescue the cross-queried pipeline, because the deficit lies not in what the encoder sees, but in whether the loss can reach it. Absent dense 2D supervision, the backbone cannot ground 3D geometry from sparse surface signals alone, whatever the encoder.

Row (d) isolates the NGT representation itself and answers a question the other rows cannot: is the benefit merely a matter of \emph{having} accurate depth and normal estimates, or does it depend on the high-dimensional token carrier? Here the model retains the full two-stage pretraining and consumes the very depth and normal maps its own header decodes---only the carrier changes, from NGTs to explicit maps. Every metric still degrades by $1.5{-}1.9\times$ (AbsRel $0.369\!\to\!0.692$, MAE $9.061\!\to\!16.232$, CD $0.129\!\to\!0.194$). Accurate explicit geometry is therefore not a substitute for NGTs: compressing the encoder's representation into the low-dimensional $\mathbb{R}^5$ explicit-map space discards geometric information that the downstream aggregator can otherwise exploit, and re-introduces precisely the learning difficulty that the NGT interface was designed to remove.

\subsection{What NGTs Encode: Layer-Wise Probing}
\label{sec:analysis:probing}

The ablations above establish that NGTs are critical, but not \emph{what} information they carry or \emph{how} it evolves through the network. To build an interpretable picture we probe every layer of the pretrained \emph{GeoTransformer} encoder and the downstream \emph{NGTFormer} aggregator, in the spirit of recent multi-view Transformer interpretability work~\cite{stary2025understanding}.

\begin{figure*}[t]
\centering
\includegraphics[width=\linewidth,trim={0cm 2cm 0cm 0cm},clip]{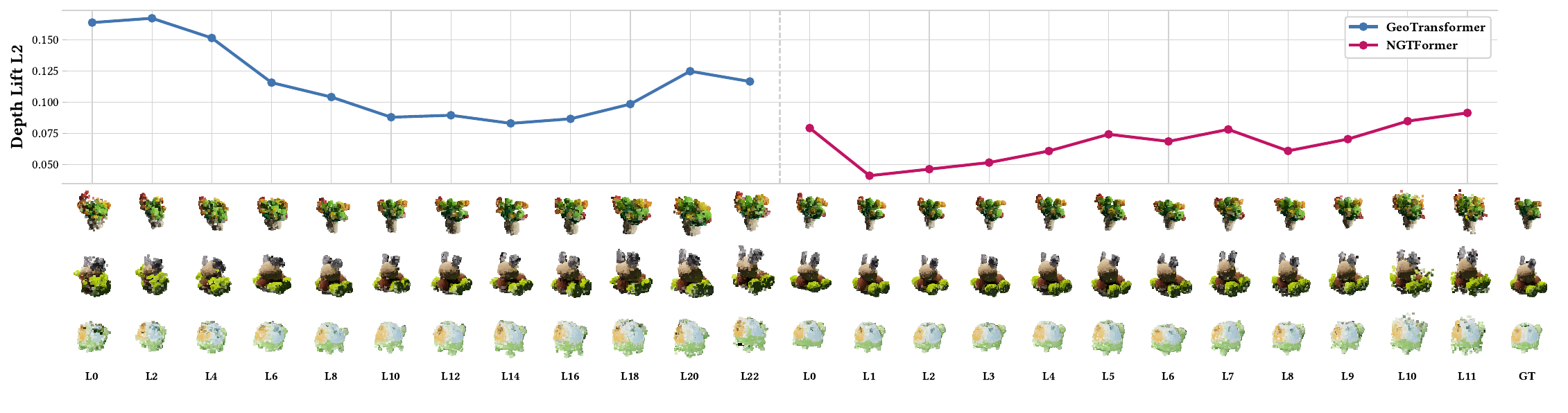}
\caption{Layer-wise probing of GeoTransformer (blue, 24 layers; even indices shown on the curve) and NGTFormer (magenta, 12 layers). \emph{Top:} per-layer depth-lift L2 of frozen-feature probes. The encoder follows a U-shaped profile, dipping from $0.164$ at \texttt{geo\_L0} to ${\sim}0.083$ near the middle of the stack and rebounding to $0.097$ at \texttt{geo\_L23}; the aggregator drops below any encoder layer at its first block and reaches its minimum at \texttt{ngt\_L1} ($0.041$, roughly half the best encoder value). \emph{Bottom:} per-layer pointclouds for three GSO objects, rendered by lifting each patch by its probed depth and coloring it with the source image's mean RGB; the rightmost column is the ground-truth pointcloud. Early-encoder pointclouds float as duplicated patches at inconsistent depths; aggregator layers consolidate them into a coherent textured shell that matches the GT.}
\label{fig:ngt_probing}
\Description{A line chart above three rows of colored point-cloud reconstructions compares features from successive network layers. The horizontal axis lists GeoTransformer layers followed by NGTFormer layers; the vertical axis is depth-lift L2 error. The blue GeoTransformer curve falls through the early layers, is lowest near the middle, and rises again toward the end. The magenta NGTFormer curve starts below the encoder curve and reaches its minimum at its second block before increasing. Below, early-layer point clouds show separated or duplicated patches, while aggregator-layer point clouds form more coherent objects resembling the ground truth in the rightmost column.}
\end{figure*}

\highlight{Setup.}

For every layer of the frozen GeoTransformer (24 layers) and NGTFormer (12 layers) we train a per-patch MLP probe (5 layers, hidden 256, AdamW with $\text{lr}{=}10^{-4}$, $\text{wd}{=}0.05$, 10 epochs) to predict the absolute world-space 3D position of the surface point intersected by each patch's camera ray. We call this prediction error the \emph{depth-lift L2}: it measures how directly a layer's representation encodes view-consistent 3D location. The backbones stay strictly frozen; only the probe heads are trained. Probes are fit on a 48-object subset of GSO---disjoint from the encoder's Objaverse pretraining distribution---and evaluated on held-out GSO objects, so the protocol stress-tests cross-distribution generalization alongside layer-wise decodability.

\highlight{Results.}

Figure~\ref{fig:ngt_probing} reports depth-lift L2 across all extraction depths. Within GeoTransformer, early blocks yield poor position predictions, consistent with encoding low-level visual rather than geometric content. Accuracy improves through the intermediate layers as the encoder integrates multi-view correspondences, reaches a minimum near the middle of the stack ($0.083$ at \texttt{geo\_L14}), and then \emph{rebounds} toward the terminal layers (peaking at $0.143$ at \texttt{geo\_L19} before settling at $0.097$ at \texttt{geo\_L23}). The rebound indicates that the deepest encoder blocks specialize for the pretraining objective in ways not strictly aligned with raw geometric decodability---which is why we extract NGTs from an attention layer near, but not at, the encoder's terminus.

\highlight{Discussion.}
NGTFormer breaks this profile immediately: depth-lift error drops below the encoder's \emph{best} value at its very first block ($0.079$ at \texttt{ngt\_L0}) and tightens further at \texttt{ngt\_L1} ($0.041$, roughly half the best encoder value), before drifting upward in the deeper aggregator blocks as features rotate toward the latent-space training objective. The qualitative pointcloud rows confirm the metric story: early-encoder pointclouds float as duplicated, view-misaligned patches, while aggregator outputs collapse onto a single coherent surface whose patch coloring matches the ground truth. This is direct evidence for the division of labour that motivates the whole design---the encoder \emph{searches} (localizing the surface within each view) and the aggregator \emph{clusters} (fusing per-view evidence into one consistent 3D structure)---and it shows that no per-token encoder layer produces the consolidated positional information on its own.

\section{Conclusion}
\label{sec:conclusion}

We presented \name{}, a feed-forward mesh reconstruction model that addresses the fundamental obstacle limiting prior sparse-view methods: the \emph{sparsity bottleneck} arising from the structural misalignment between dense 2D visual inputs and sparse 3D surface targets. Our key insight is that Neural Geometry Tokens---intermediate Transformer representations pretrained on dense geometric map prediction---serve as a decoder-agnostic bridge that resolves this misalignment by aggregating rich geometric cues in 2D-aligned form. Extensive experiments validate that \name{} achieves state-of-the-art feed-forward mesh reconstruction on GSO, delivers geometric quality surpassing 128-view optimization from only 8 input views, and generalizes robustly to real-world objects on Stanford-ORB.

\highlight{Limitations and Future Work.}
\name{} inherits several limitations from its feed-forward, photometry-driven design. Thin structures, heavily self-occluded geometry, and transparent or reflective surfaces remain challenging, and reconstruction quality degrades when test-time camera distributions fall far outside the training envelope. These limitations suggest several future directions: hybrid recon-plus-generation pipelines that use NGTs for visible regions and generative priors for self-occluded ones; multi-modal cues (\eg, polarization) to handle non-Lambertian surfaces; and pose-agnostic embeddings or wider training-camera distributions to extend the operating envelope.

\begin{acks}
This work was supported by the National Natural Science Foundation of China (Project Number 62595772).
\end{acks}

\bibliographystyle{ACM-Reference-Format}
\bibliography{references}

\appendix
\emergencystretch=1.5em

\section{Data Curation}
\label{sec:supp:data}

This section details the data pipeline that feeds both the Stage~1 geometric pretraining (Sec.~4.1 \& Appendix~\ref{sec:supp:pretrain}) and the Stage~2 NGTFormer training (Sec.~4.2/4.3 \& Appendix~\ref{sec:supp:training}). The same renders supply pretraining targets (depth, normal, mask) and the multi-view inputs at inference; the same object meshes supply the latent ground truth used for Stage~2 supervision.

\highlight{Asset Source and Curation Inheritance.}
We start from Objaverse-XL~\cite{objaverse} and inherit the TRELLIS~\cite{xiang2025structured,xiang2025trellis2} curation protocol that filters for manifoldness, texture quality, and geometric complexity. We select 160K high-quality objects from the Objaverse subset released by TRELLIS, which are partitioned at the object level into disjoint training and held-out validation splits to prevent leakage.

\highlight{Rendering Protocol.}
For each object, we cache a pool of multi-view renders from which $V{=}24$ views are resampled per training instance. Cameras are placed on the upper hemisphere: elevation is uniform in $[0^\circ,90^\circ]$, azimuth uniform in $[0^\circ, 360^\circ]$, and the camera distance is randomized so that the object inscribed in the unit sphere occupies a controlled fraction of the frame. Intrinsics are randomized per view: the field of view is sampled in roughly $[25^\circ, 45^\circ]$ with the principal point at the image center; the image resolution is $518{\times}518$ to match the patch grid used by the Transformer. Lighting is provided by randomly sampled HDRI environment maps that supply realistic indirect illumination and soft shadows; an optional sun-like directional light is added with low probability to broaden the shadow distribution. Backgrounds are alpha-masked and composited onto pure white, so foreground masks are unambiguous. Each view exports four aligned channels: RGB ($3$~ch), metric depth ($1$~ch float), surface normal in camera space ($3$~ch), and foreground mask ($1$~ch).

\highlight{Latent Preparation.}
The NGTFormer is trained against \emph{pseudo}-ground-truth latents pre-computed by the frozen pretrained decoder family of each backend (Sec.~4.3 \& Appendix~\ref{sec:supp:decoder}). Each object's mesh is encoded once and cached for the entire training run.

For the SpVox backend, the ground-truth mesh is passed through the pretrained Trellis2~\cite{xiang2025trellis2} encoder to obtain a Structured-Latent (SLAT) representation. Each active voxel carries a $d_z{=}32$-dimensional latent feature, and the union of active voxels defines the ground-truth occupancy used to supervise the filter transformer of Appendix~\ref{sec:supp:decoder}. Per-channel mean and standard-deviation statistics $\bm{\mu}\in\mathbb{R}^{32}$, $\bm{\sigma}\in\mathbb{R}^{32}$ are pre-computed over the entire training set and frozen during NGTFormer training; the aggregator predicts the normalized latent and the network output is denormalized as $\bm{z}_k = \bm{\hat{z}}_k\odot\bm{\sigma}+\bm{\mu}$.

For the VecSet backend, the same ground-truth mesh is passed through the pretrained 3DShape2VecSet~\cite{zhang20233dshape} encoder, yielding $N_q{=}4096$ latent tokens of dimension $d_v{=}64$ per object. Unlike SpVox latents, VecSet latents have no canonical ordering, which is the reason the Stage~2 loss uses Sinkhorn-OT rather than per-token MSE (Appendix~\ref{sec:supp:training}). Per-channel statistics $\bm{\mu}'\in\mathbb{R}^{64}$, $\bm{\sigma}'\in\mathbb{R}^{64}$ are pre-computed independently for this backend.

\highlight{Storage Layout.}
Each per-object cache contains the multi-view render pool, the pre-encoded SpVox latents (with their topology), and the pre-encoded VecSet latents. Data are served via indexed memory-mapped shards for efficient random access during training.

\section{Training Details}
\label{sec:supp:training}

This section documents the shared training setup used across all stages of \name{}. Per-decoder loss formulations and inference loops are deferred to Appendix~\ref{sec:supp:decoder}; geometric-pretraining specifics are in Appendix~\ref{sec:supp:pretrain}; data and rendering protocols are in Appendix~\ref{sec:supp:data}.

\highlight{Two-Stage Pipeline.}
\name{} is trained in two stages. \emph{Stage~1 (Geometric Pretraining)} trains the multi-view Transformer backbone end-to-end under dense pixel-aligned supervision (depth, normal, and foreground mask), producing the NGT encoder used in Stage~2. \emph{Stage~2 (NGTFormer Training)} freezes the pretrained backbone and the chosen mesh decoder, and trains only the NGTFormer aggregator (and, for the SpVox backend, the occupancy filter) under latent-space supervision. The two stages share the same dataset, renderer, and image resolution; only the loss and the learnable parameters differ.

\highlight{Shared Architectural Constants.}
Across both stages, we use image resolution $H{\times}W{=}518{\times}518$, patch size $p{=}14$ ($N_p{=}(518/14)^2{=}1369$ patches per view), and view count $V{=}8$ randomly sampled per training instance from a larger cached pool per object (Appendix~\ref{sec:supp:data}). The geometry encoder is a $24$-layer Transformer with hidden dimension $D{=}768$ and $12$ attention heads, alternating intra-frame and inter-frame attention following VGGT~\cite{wang2025vggt}. The NGTFormer aggregator uses $L_a{=}12$ interleaved self/cross-attention layers at hidden dimension $D{+}D'{=}1536$; for the SpVox backend, a separate occupancy filter of $L_f{=}6$ sparse Transformer layers operates at dimension $D$.

\highlight{Shared Training Schedule.}
Both stages use AdamW with fused kernels, base learning rate $4{\times}10^{-4}$, betas $(0.9, 0.95)$, and weight decay $0.05$. The schedule is cosine decay preceded by a linear warm-up of $3{,}000$ steps. Mixed-precision training uses \texttt{bfloat16} throughout. Each stage runs for $100{,}000$ optimizer steps. Stage~1 uses a per-GPU batch size of $8$ multi-view samples; Stage~2 uses $16$. Training is conducted on $8{\times}$ NVIDIA RTX Pro6000 GPUs via standard DDP, yielding an effective global batch size of $64$ (Stage~1) or $128$ (Stage~2).

\highlight{Stage~1 Loss Weights.}
The Stage~1 compound loss is
\begin{equation}
\begin{aligned}
\mathcal{L}_\textrm{pretrain}&= \lambda_d\,\|d-\hat{d}\|_2 + \lambda_n\,\|\bm{n}-\hat{\bm{n}}\|_1 \\& + \lambda_n'\,\textrm{Perceptual}(\bm{n},\hat{\bm{n}}) + \lambda_m\,\|m-\hat{m}\|_2 \, ,
\end{aligned}
\end{equation}
with $\lambda_d{=}1.0$ (depth $L_2$), $\lambda_n{=}0.5$ (angular $L_1$ on normal), $\lambda_n'{=}0.5$ (perceptual loss on the normal map~\cite{johnson2016perceptual}), and $\lambda_m{=}1.0$ (mask MSE). All losses are computed over valid pixels (foreground mask $\geq 0.5$ for depth/normal) and averaged across the eight views.

\highlight{Stage~2 Loss Weights.}
The Stage~2 latent-regression loss adapts to the chosen $\mathcal{S}'$ parameterization (Eq.~18). For the SpVox backend, the total loss combines a per-voxel latent MSE term and an occupancy BCE term,
\begin{equation}
\mathcal{L}_\textrm{SpVox} = \lambda_\textrm{lat}\,\tfrac{1}{K}\sum_{k}\|\bm{z}_k-\hat{\bm{z}}_k\|_2^2 + \lambda_\textrm{flag}\,\textrm{BCE}(o, \hat{o}) \, ,
\end{equation}
with $\lambda_\textrm{lat}{=}0.04$ and $\lambda_\textrm{flag}{=}0.4$. The latent term is computed against pseudo-ground-truth latents $\{\hat{\bm{z}}_k\}$ pre-encoded once per object by the frozen Trellis2 encoder. For the VecSet backend, the loss is the Sinkhorn optimal-transport distance~\cite{cuturi2013sinkhorn} between predicted and target latent sets, with entropic regularization $\varepsilon{=}1.0$ and $10$ Sinkhorn iterations in log-space; gradients flow through the cost matrix but not through the transport plan.

\highlight{Reproducibility Notes.}
All experiments use a fixed seed for the dataset shuffling order and the data-augmentation RNGs. Checkpoints are saved every $10{,}000$ steps; the reported numbers use the final ($100{,}000$-step) checkpoint of each run. Evaluation protocols (sampling, alignment, visibility filter, baseline methodology) are detailed in Appendix~\ref{sec:supp:eval}.

\section{Evaluation Details}
\label{sec:supp:eval}

This section pins down the metric definitions, sampling and alignment protocols, per-benchmark camera configurations, and baseline-evaluation methodology used throughout Sec.~5 and these appendices. Critically, it documents the visibility filter used to fairly evaluate models trained with hemisphere-only inputs.

\subsection{Metric Definitions}

\highlight{Depth Quality.}
Let $d_{ij}$ denote the predicted depth at pixel $j$ of view $i$ and $\hat{d}_{ij}$ the ground truth. Over the set $\mathcal{P}$ of valid (foreground) pixels:
\begin{align}
    \textrm{AbsRel} \;&=\; \tfrac{1}{|\mathcal{P}|}\!\sum_{(i,j)\in\mathcal{P}}\frac{|d_{ij}-\hat{d}_{ij}|}{\hat{d}_{ij}}\,,\\
    \textrm{Acc}_\delta \;&=\; \tfrac{1}{|\mathcal{P}|}\!\sum_{(i,j)\in\mathcal{P}}\mathbf{1}\!\left[\max\!\left(\tfrac{d_{ij}}{\hat{d}_{ij}},\tfrac{\hat{d}_{ij}}{d_{ij}}\right) < \delta\right]\,.
\end{align}
We report AbsRel in percent and Acc at threshold $\delta{=}1.25$.

\highlight{Normal Quality.}
With $\bm{n}_{ij}, \hat{\bm{n}}_{ij}\in\mathbb{R}^3$ unit-norm surface normals,
\begin{equation}
    \textrm{MAE}_n \;=\; \tfrac{1}{|\mathcal{P}|}\!\sum_{(i,j)\in\mathcal{P}}\arccos\!\big(\,\langle\bm{n}_{ij},\hat{\bm{n}}_{ij}\rangle\,\big)\;[\textrm{deg}].
\end{equation}
We additionally report PSNR/SSIM/LPIPS on the normal map rendered to RGB via the standard $(\bm{n}{+}1)/2$ encoding into linear sRGB, computed only over foreground pixels.

\highlight{Mesh Quality.}
We sample $N{=}10^5$ surface points from each mesh via Poisson-disk sampling. Let $P$ and $\hat{P}$ be the sampled point clouds from the predicted and reference meshes, $d(p, Q){=}\min_{q\in Q}\|p-q\|$. Chamfer Distance (CD) and F1-score at threshold $\tau$ (with the meshes pre-scaled to fit a unit sphere; we use $\tau{=}0.005$) are
\begin{align}
    \textrm{CD} \;&=\; \tfrac{1}{|P|}\!\sum_{p\in P}d(p,\hat{P}) + \tfrac{1}{|\hat{P}|}\!\sum_{q\in \hat{P}}d(q, P)\,,\\
    \textrm{F1}_\tau \;&=\; \tfrac{2\cdot\textrm{Pre}_\tau\cdot\textrm{Rec}_\tau}{\textrm{Pre}_\tau+\textrm{Rec}_\tau}\,,\\
    \textrm{Pre}_\tau \;&=\; \tfrac{|\{p\in P\,:\,d(p,\hat{P}){<}\tau\}|}{|P|}\,,\\
    \textrm{Rec}_\tau \;&=\; \tfrac{|\{q\in\hat{P}\,:\,d(q,P){<}\tau\}|}{|\hat{P}|}\,.
\end{align}

\subsection{Sampling and Alignment Protocols}

\highlight{Coordinate Normalization.}
All predicted and reference meshes are uniformly scaled to fit a unit bounding box centered at the origin before any point sampling. Methods that produce meshes in a canonical pose distinct from the inputs (\eg, single-image generation methods) additionally receive a manual alignment step.

\highlight{Object Alignment (Recon vs.\ Gen Only).}
For the Reconstruction-vs.-Generation comparison (Sec.~5.3 and Appendix~\ref{sec:supp:recon_vs_gen}), generation baselines output meshes in their own canonical frame, which is not necessarily aligned with the input cameras. To measure geometric fidelity rather than pose error, we align each predicted mesh to the reference manually.

\highlight{Visibility Filter for Hemisphere Inputs.}
Both our training renderer and our sparse-view test protocol place cameras on the upper hemisphere of the unit sphere, leaving the bottom of every object unobserved. Without correction, all methods would be penalized for surface error in regions they could not possibly reconstruct from the inputs. For honest comparison, we therefore evaluate metrics only over mesh regions visible from at least one of the $V$ input cameras. Visibility is computed by ray-mesh intersection: a surface sample $p$ is marked \emph{visible} from view $i$ if the ray from $p$ to the camera center $\bm{o}_i$ is unoccluded by the reference mesh. We discard samples that are visible from zero input views before computing CD/F1. The same filter is applied uniformly to all methods on a benchmark to preserve fairness. For benchmarks dominated by hemisphere inputs (GSO main table, Stanford-ORB), all reported numbers use the visibility-filtered version.

\subsection{Baseline Evaluation Methodology}

For transparency, we document how each baseline number in Tab.~2 was obtained.

\begin{itemize}[leftmargin=*,itemsep=2pt,topsep=2pt]
    \item PRM~\cite{ge2025prm}: official checkpoints, run on our rendered inputs. PRM uses its canonical 6-view input.
    \item MeshLRM~\cite{wei2024meshlrm}, MeshFormer~\cite{liu2024meshformer}: no public code or checkpoint at the time of submission; we therefore report numbers on the 6-object subset directly from the project page.
    \item 2DGS~\cite{huang20242d}, GOF~\cite{yu2024gaussian}: official code, per-scene optimization with the published hyperparameters; we report the 8-view, 24-view, and 128-view configurations.
    \item TRELLIS~\cite{xiang2025structured}, TRELLIS2~\cite{xiang2025trellis2}: official checkpoints, conditioned per the publishers' documentation.
    \item Hunyuan3D~\cite{hunyuan3d2025hunyuan3d}: no official checkpoint. We assess it via website services.
\end{itemize}
Any minor deviations from the original baselines' inputs (\eg, view-count misalignment compared to our protocol) are noted alongside the relevant table cell.

\highlight{Fidelity-Aware Baselines.}
The four baselines added for the comparison of Sec.~\ref{sec:exp:fidelity_aware} were obtained as follows. All four emit meshes in their own canonical frame rather than in the input-camera frame, so each requires an explicit alignment step before metrics are computed; because a residual pose offset would be indistinguishable from a geometric error, we document that step per method.
\begin{itemize}[leftmargin=*,itemsep=2pt,topsep=2pt]
    \item ReconViaGen~\cite{chang2025reconviagen}: we take the refined output of its final stage. Alignment to the evaluation frame is a fixed $180^\circ$ rotation about the vertical axis followed by normalization by the mesh's own maximum bounding-box extent.
    \item Pixal3D~\cite{li2026pixal3d}: its raw output lives in the original GSO model frame, so alignment additionally recenters on the reference bounding box and divides by the reference maximum extent.
    \item Matcha-Gaussians~\cite{guedon2025matcha}: per-scene optimization with the published hyperparameters. Being optimization-based rather than feed-forward, it is reported twice---at $V\!=\!4$ for a protocol matched to all other rows, and at $V\!=\!24$ to show what dense supervision attains.
    \item Unique3D~\cite{wu2024unique3d}: unlike the three methods above it is not a fidelity-aware design but a diffusion front-end followed by ISOMER post-hoc remeshing, and it is included as a reference point for that post-hoc strategy. Its front-end is conditioned on a single image, so it is run under its native protocol rather than on our multi-view input. Its mesh is aligned to the reference in the same manner as ReconViaGen.
\end{itemize}
Both the visibility filter and the coordinate normalization above apply unchanged to all four, and \name{} is evaluated with the identical filter, so no method is credited for geometry it could not observe.

\section{Geometric Pretraining Details}
\label{sec:supp:pretrain}

This section details the Stage~1 geometric pretraining that produces the NGT encoder. The shared optimizer, schedule, and hardware footprint are summarized in Appendix~\ref{sec:supp:training}; the data and rendering protocol are in Appendix~\ref{sec:supp:data}. Here we focus on the pretrain-specific architecture, the per-view ``header'' decoder that converts NGTs into pixel-aligned geometric maps, the quantitative and qualitative evaluation of the pretrained model, and two targeted ablations.

\highlight{Backbone Architecture.}
The geometric encoder is a multi-view Transformer of $L{=}24$ attention layers with hidden dimension $D{=}768$ and $12$ heads, alternating intra-frame and inter-frame self-attention following VGGT~\cite{wang2025vggt}. The input is the 9-channel pose-augmented map $\bm{I}'_{ij}\in\mathbb{R}^{p\times p\times 9}$ defined in Eq.~6 (RGB concatenated with the 6-channel \plucker ray embedding), tokenized by a trainable patch-wise linear projection at patch size $p{=}14$ to produce $N_p{=}(518/14)^2{=}1369$ tokens per view. Tokens from all $V{=}8$ views are flattened into a single sequence and processed jointly by the alternating attention stack.

\highlight{Header Transformer.}
A separate \emph{header} module decodes the final-layer tokens into per-pixel geometric maps. The header is a residual Transformer of $L_h{=}4$ layers at hidden dimension $D$, operating per-view (only intra-frame attention). Its output tokens are linearly mapped to $\mathbb{R}^{p\times p\times 5}$ and unflattened to per-patch $(p,p)$ image grids, jointly carrying depth ($1$ channel, decoded in log-space and exponentiated to enforce positivity), normal ($3$ channels, post-decoded by unit-norm projection), and foreground mask ($1$ channel, post-decoded by sigmoid). Decoupling the per-view header from the bulk backbone is deliberate: it isolates pixel-output specialization from the multi-view aggregation pathway, so the bulk features can be re-used as NGTs at Stage~2 without overfitting to the pretraining decoder.

\highlight{Training.}
Stage~1 optimization (AdamW, $4{\times}10^{-4}$, cosine with $3000$-step warm-up, \texttt{bfloat16}, $100$K total steps on $8{\times}$~NVIDIA RTX Pro6000) and the compound loss $\mathcal{L}_\textrm{pretrain}$ with weights $(\lambda_d, \lambda_n, \lambda_n', \lambda_m){=}(1.0, 0.5, 0.5, 1.0)$ are summarized in Appendix~\ref{sec:supp:training}. The perceptual term~\cite{johnson2016perceptual} is applied to the normal channel, where pure $L_1$ loss tends to oversmooth high-frequency surface detail.

\highlight{Quantitative Evaluation.}
Table~\ref{tab:pretrain_perf} reports depth and normal estimation quality on a held-out 256-object Objaverse~\cite{objaverse} test split (matched in distribution to training) and on the GSO~\cite{downs2022googlescannedobjectshighquality} zero-shot benchmark. The model attains sub-percent AbsRel depth error and single-digit angular normal error across both distributions, confirming that the pretrained backbone produces a reliable geometric representation, which provides the basis for the NGT extraction used downstream.

\begin{table}[t]
    \centering
    \caption{Quantitative evaluation of geometric pretraining. We report depth and normal estimation quality on a separate 256-object test set on Objaverse, and on zero-shot test set GSO.}
    \label{tab:pretrain_perf}
    \resizebox{\linewidth}{!}{
        \begin{tabular}{l|cc|cccc}
            \toprule
             & \multicolumn{2}{c|}{\textbf{Depth Quality}} & \multicolumn{4}{c}{\textbf{Normal Quality}} \\
            \textbf{Dataset} & AbsRel (\%)$\downarrow$ & Acc (\%)$\uparrow$ & MAE ($^\circ$)$\downarrow$ & PSNR$\uparrow$ & SSIM$\uparrow$ & LPIPS$\downarrow$ \\
            \midrule
            Objaverse-256 & 0.411 & 99.95 & 10.236 & 28.584 & 0.8724 & 0.0830 \\
            GSO & 0.306 & 99.99 & 7.137 & 31.46 & 0.8565 & 0.0889 \\
            \bottomrule
        \end{tabular}
    }
\end{table}

\begin{figure}[t]
    \centering
    \setlength{\tabcolsep}{1pt}

    \begin{tabular}{ccccccc}
        \raisebox{1.5mm}{\rotatebox{90}{\scriptsize Input RGB}} &
        \includegraphics[width=0.155\linewidth, height=0.155\linewidth, keepaspectratio, trim={0 0 0 0}, clip]{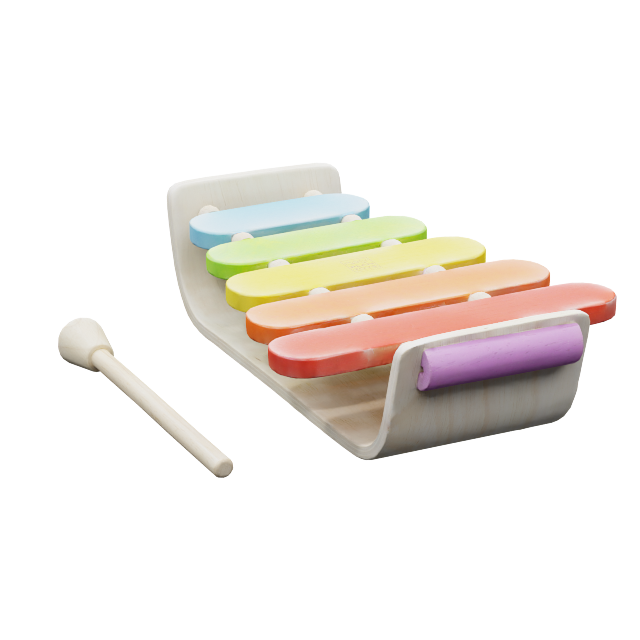} &
        \includegraphics[width=0.155\linewidth, height=0.155\linewidth, keepaspectratio, trim={0 0 0 0}, clip]{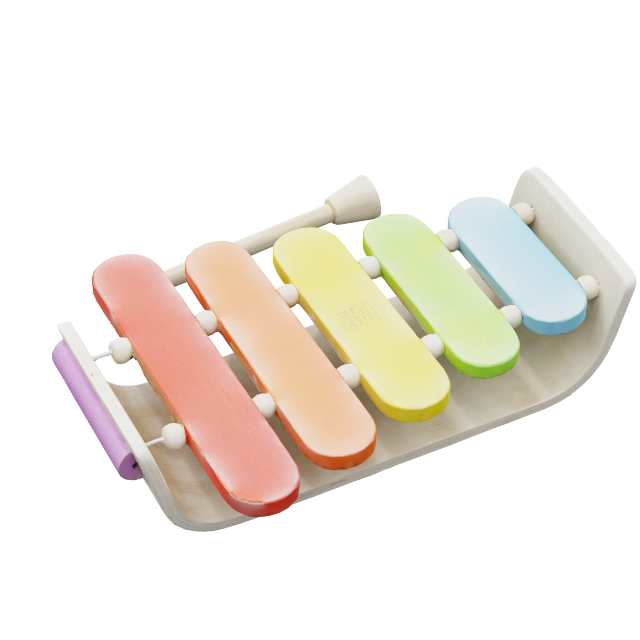} &
        \includegraphics[width=0.155\linewidth, height=0.155\linewidth, keepaspectratio, trim={0 0 0 0}, clip]{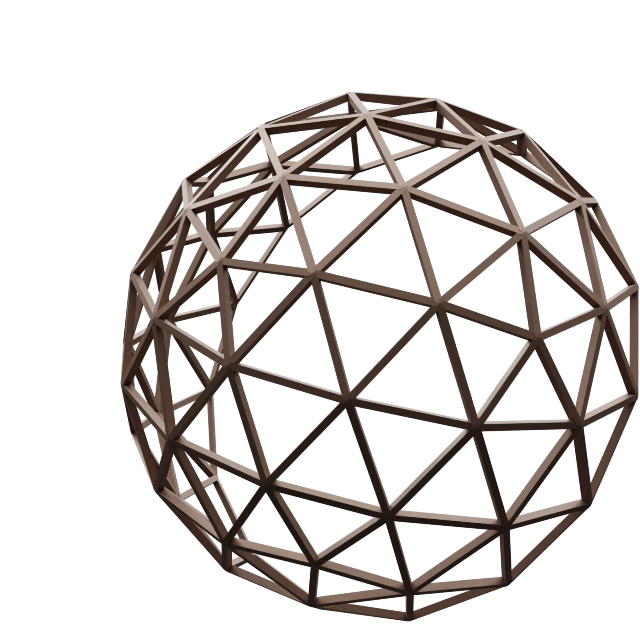} &
        \includegraphics[width=0.155\linewidth, height=0.155\linewidth, keepaspectratio, trim={0 0 0 0}, clip]{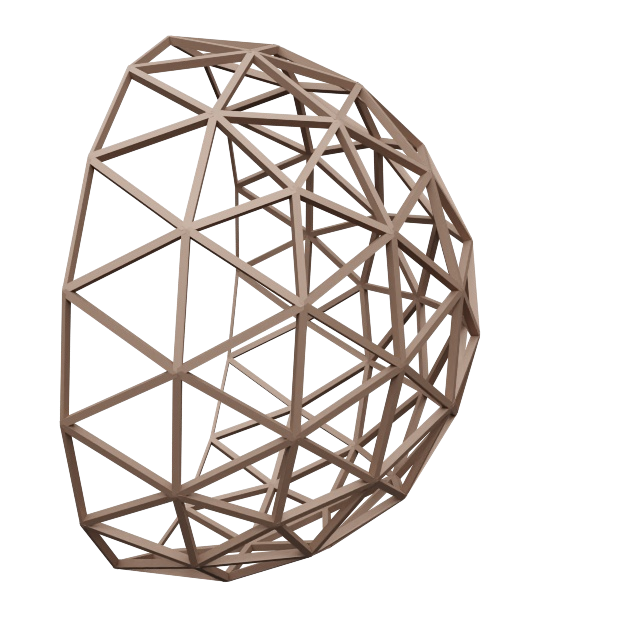} &
        \includegraphics[width=0.155\linewidth, height=0.155\linewidth, keepaspectratio, trim={0 0 0 0}, clip]{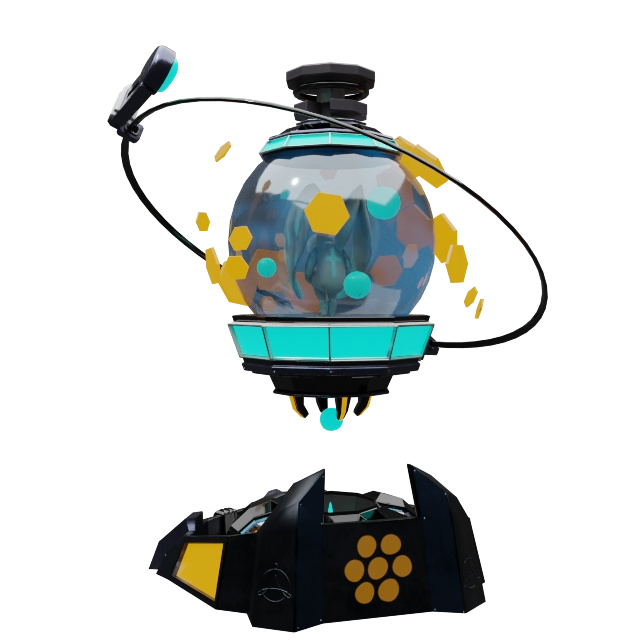} &
        \includegraphics[width=0.155\linewidth, height=0.155\linewidth, keepaspectratio, trim={0 0 0 0}, clip]{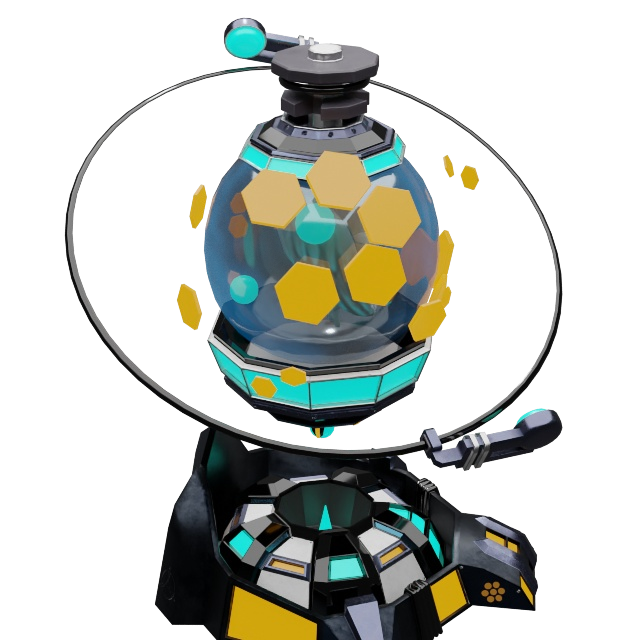} \\

        \raisebox{2mm}{\rotatebox{90}{\scriptsize Pred Normal}} &
        \includegraphics[width=0.155\linewidth, height=0.155\linewidth, keepaspectratio, trim={0 0 0 0}, clip]{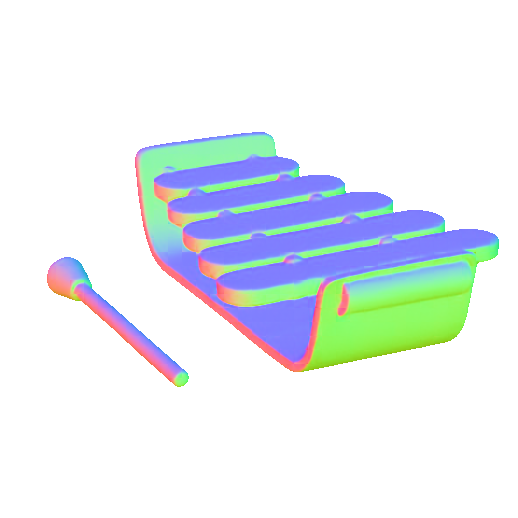} &
        \includegraphics[width=0.155\linewidth, height=0.155\linewidth, keepaspectratio, trim={0 0 0 0}, clip]{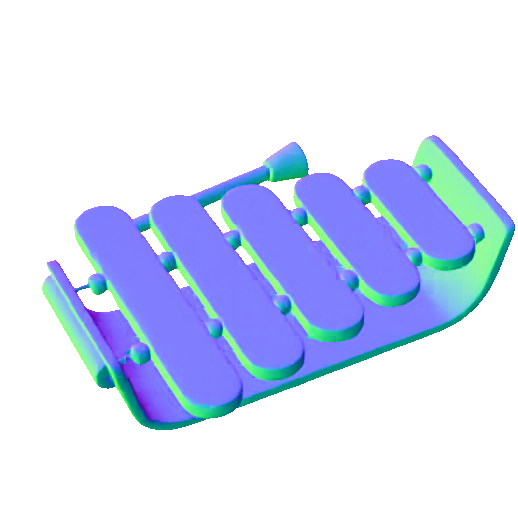} &
        \includegraphics[width=0.155\linewidth, height=0.155\linewidth, keepaspectratio, trim={0 0 0 0}, clip]{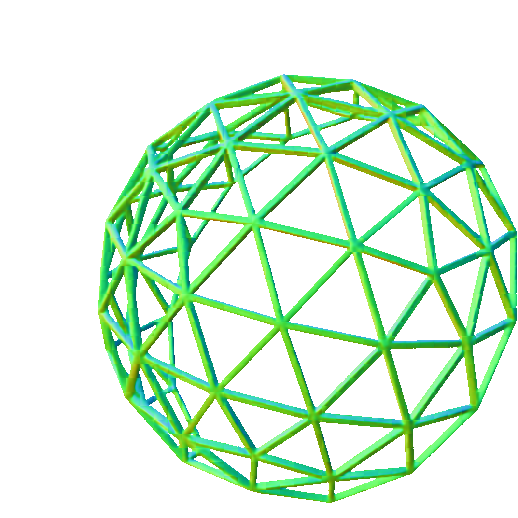} &
        \includegraphics[width=0.155\linewidth, height=0.155\linewidth, keepaspectratio, trim={0 0 0 0}, clip]{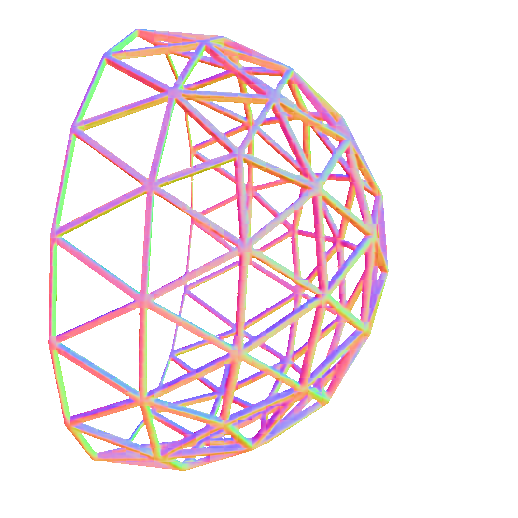} &
        \includegraphics[width=0.155\linewidth, height=0.155\linewidth, keepaspectratio, trim={0 0 0 0}, clip]{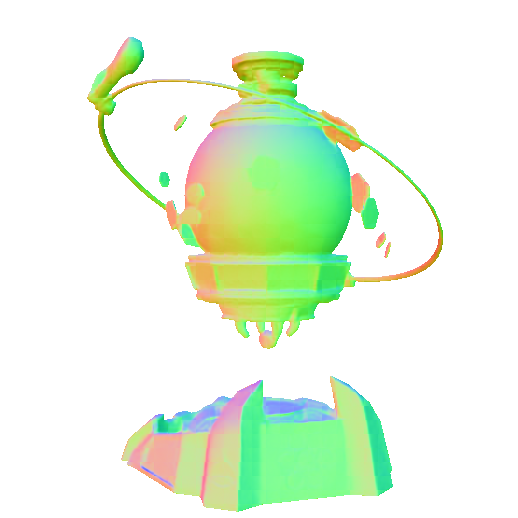} &
        \includegraphics[width=0.155\linewidth, height=0.155\linewidth, keepaspectratio, trim={0 0 0 0}, clip]{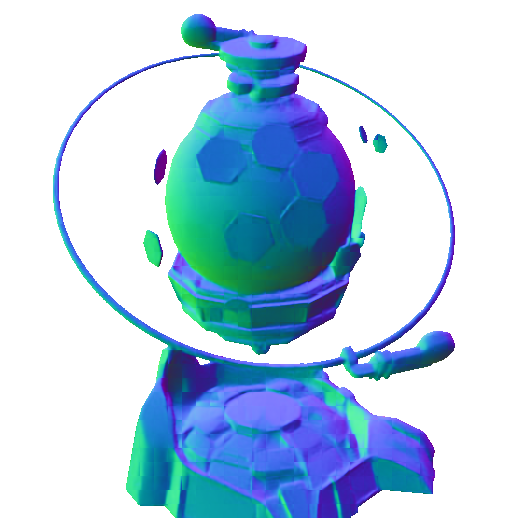} \\

        \raisebox{2mm}{\rotatebox{90}{\scriptsize Pred Depth}} &
        \includegraphics[width=0.155\linewidth, height=0.155\linewidth, keepaspectratio, trim={0 0 0 0}, clip]{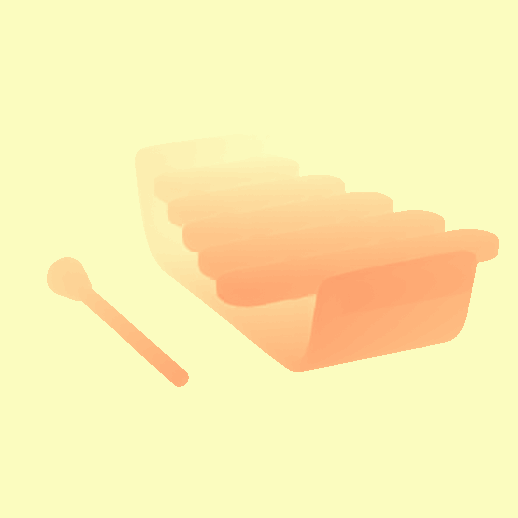} &
        \includegraphics[width=0.155\linewidth, height=0.155\linewidth, keepaspectratio, trim={0 0 0 0}, clip]{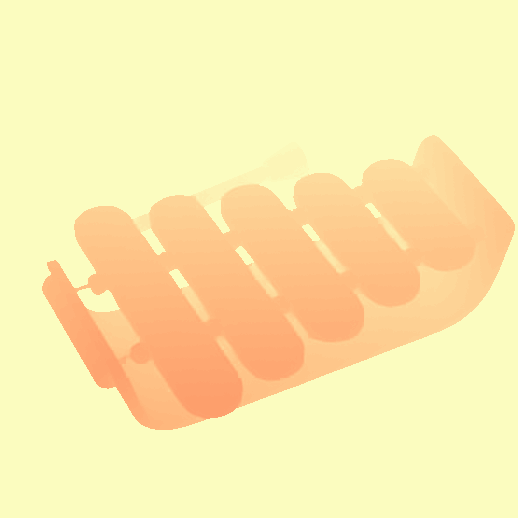} &
        \includegraphics[width=0.155\linewidth, height=0.155\linewidth, keepaspectratio, trim={0 0 0 0}, clip]{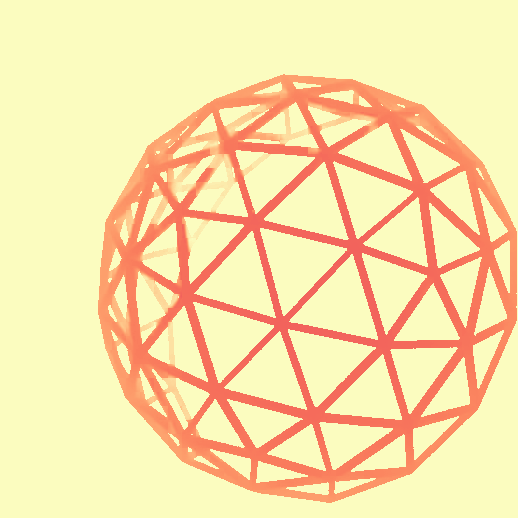} &
        \includegraphics[width=0.155\linewidth, height=0.155\linewidth, keepaspectratio, trim={0 0 0 0}, clip]{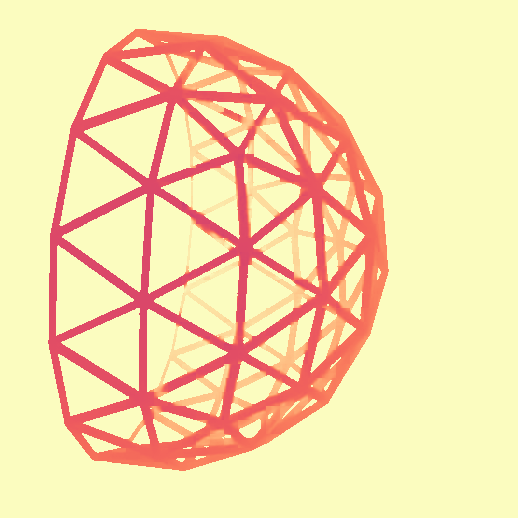} &
        \includegraphics[width=0.155\linewidth, height=0.155\linewidth, keepaspectratio, trim={0 0 0 0}, clip]{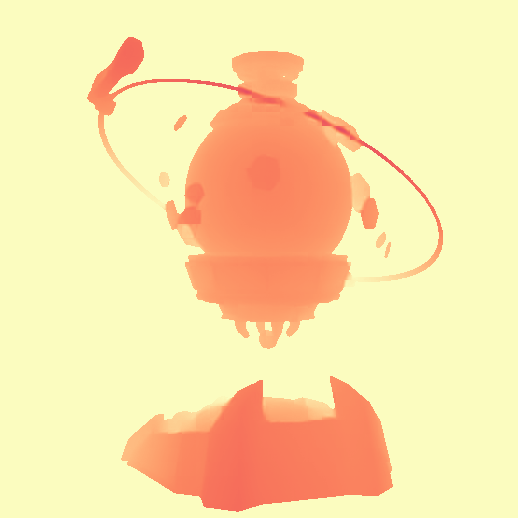} &
        \includegraphics[width=0.155\linewidth, height=0.155\linewidth, keepaspectratio, trim={0 0 0 0}, clip]{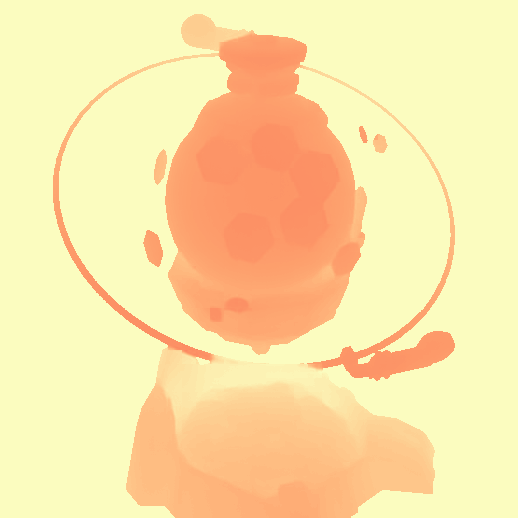} \\
    \end{tabular}
    \caption{Qualitative results of geometric pretraining. We visualize 2 (out of 8) views of input RGB images (top), predicted normal maps (middle), and predicted depth maps (bottom).}
    \label{fig:pretrain_qual}
    \Description{A three-row, six-column grid shows two views each of a compartmented tray, an open spherical framework, and a mechanical toy. The top row contains RGB input images, the middle row contains predicted normal maps colored by surface orientation, and the bottom row contains predicted depth maps shown in warm colors. Changes in the predicted colors follow compartment boundaries, thin framework struts, and the toy's curved and articulated parts.}
\end{figure}

\highlight{Qualitative Evaluation.}
Figure~\ref{fig:pretrain_qual} visualizes representative predictions from the pretrained encoder. For each object, we show two of the eight input views together with the predicted normal map and depth map. The predictions preserve fine surface detail and recover crisp object boundaries across diverse object categories, consistent with the quantitative numbers above.

\section{Decoder-specific Implementation}
\label{sec:supp:decoder}

Section~4.3 introduces two complementary mesh-decoding backends supported by NGTFormer---\emph{Sparse Voxel} (SpVox) and \emph{VecSet}---and develops a unified latent-regression formulation. This section provides the per-backend details that were deferred for space: how the queries $\{\bm{q}_k\}$ are constructed, what auxiliary modules run alongside the aggregator, how the pretrained decoder is invoked at inference, and how the training loss is concretely realized. Shared training hyperparameters and loss weights are summarized in Appendix~\ref{sec:supp:training}; data and latent caches in Appendix~\ref{sec:supp:data}.

Both backends share the same NGT encoder and the same context construction (Eq.~11), and use mask-based visual token pruning with threshold $\tau{=}0.05$ (Sec.~4.2). The differences reside in (i)~the query construction, (ii)~the optional auxiliary modules (an occupancy filter for SpVox), (iii)~the form of the latent loss, and (iv)~the choice of frozen pretrained decoder. Table~\ref{tab:decoder_diff} summarizes these contrasts.

\begin{table}[t]
    \centering
    \caption{Backend-by-backend summary of the two NGTFormer instantiations. Both inherit the same context construction (Eq.~11) and mask-based token pruning.}
    \label{tab:decoder_diff}
    \resizebox{\linewidth}{!}{
        \begin{tabular}{l|cc}
            \toprule
             & \textbf{SpVox} & \textbf{VecSet} \\
            \midrule
            Query type             & Voxel positional embed.\ & Learnable embeddings \\
            Spatial anchoring      & Yes (active voxels)      & No (unordered set) \\
            Aux.\ module           & Occupancy filter         & --- \\
            Latent dim.            & $d_z{=}32$               & $d_v{=}64$ \\
            Latent loss            & MSE (Eq.~18, top) & Sinkhorn-OT (Eq.~18, bot.) \\
            Pretrained decoder     & Trellis2~\cite{xiang2025trellis2} & 3DShape2VecSet~\cite{zhang20233dshape} \\
            Mesh extraction        & Dual Contouring~\cite{ju2002dual} & Marching Cubes~\cite{lorensen1998marching} \\
            \bottomrule
        \end{tabular}
    }
\end{table}

\subsection{Sparse Voxel (SpVox) Backend}

\highlight{Depth-lifted Candidate Voxel Set.}
A direct benefit of geometric pretraining is that the same encoder simultaneously produces per-view depth and mask predictions. We exploit these by unprojecting each predicted depth value along its camera ray, $\bm{x}_{ij} = \bm{o}_i + d_{ij}\bm{d}_{ij}$, and quantizing the multi-view point cloud onto a coarse $R{=}16^3$ voxel grid to obtain a candidate voxel set $\mathcal{V}$. Mask-thresholded pixels are excluded before quantization. This depth-lifting step provides the sparse aggregator with an explicit 3D structural prior and bounds the candidate set far below the dense $R^3{=}4096$ enumeration.

\highlight{Occupancy Filter.}
The lifted set $\mathcal{V}$ inevitably contains noise from imperfect depth predictions. A lightweight filter Transformer of $L_f{=}6$ sparse Transformer layers at hidden dimension $D{=}768$ refines $\mathcal{V}$ by predicting per-voxel occupancy. Voxels are sinusoidally positionally encoded with $M$ log-linearly spaced frequencies, projected to dimension $D$, and cross-attended against the same pruned NGT context used by the aggregator. The output is a per-voxel logit $o_k$; at inference, $\mathcal{V}^*{=}\{v_k\,|\,\sigma(o_k){>}0.5\}$ is forwarded to the aggregator. At training time, we substitute the ground-truth voxel set for $\mathcal{V}^*$ to decouple the filter and aggregator gradients, while the filter itself is supervised with a per-voxel binary cross-entropy term $\mathcal{L}_\textrm{flag}{=}\textrm{BCE}(o, \hat{o})$.

\highlight{Aggregator.}
The aggregator is a stack of $L_a{=}12$ interleaved self/cross-attention layers at hidden dimension $D{+}D'{=}1536$ (Eq.~12). Queries are voxel positional embeddings anchored on $\mathcal{V}^*$, so the predicted latents inherit an explicit one-to-one mapping to the active sparse-voxel set. The output is normalized per-channel and denormalized using frozen statistics derived from the pretrained Trellis2 encoder, $\bm{z}_k = \bm{\hat{z}}_k\odot\bm{\sigma}+\bm{\mu}$, with $\bm{\mu},\bm{\sigma}\in\mathbb{R}^{32}$.

\highlight{Pretrained Decoder and Mesh Extraction.}
The denormalized latents are passed through the frozen pretrained Trellis2 decoder, which performs hierarchical coarse-to-fine subdivision via sparse 3D convolutions and ultimately produces per-voxel dual-grid parameters (vertex offsets and edge flags). The triangle mesh is then extracted by Dual Contouring~\cite{ju2002dual}.

\highlight{Training Loss.}
The total SpVox loss combines the latent regression and the filter cross-entropy as introduced in Appendix~\ref{sec:supp:training}:
\begin{equation}
\mathcal{L}_\textrm{SpVox}{=}\lambda_\textrm{lat}\,\tfrac{1}{K}\sum_{k}\|\bm{z}_k-\hat{\bm{z}}_k\|_2^2 + \lambda_\textrm{flag}\,\textrm{BCE}(o, \hat{o})\,,
\end{equation}
with $\lambda_\textrm{lat}{=}0.04$, $\lambda_\textrm{flag}{=}0.4$. Mild scale-shift augmentation is applied jointly to cameras and latents as described in Appendix~\ref{sec:supp:data}.

\begin{algorithm}[t]
\caption{SpVox Inference Loop}
\label{alg:spvox}
\KwIn{Posed input views $\mathcal{I}=\{(I_i, C_i)\}_{i=1}^{V}$}
\KwOut{Triangle mesh $\mathcal{S}$}
$\mathcal{T}^*\leftarrow\textrm{GeoPretrainedTransformer}(\mathcal{I})$\tcp*{NGTs (Eq.~10)}
$(d_{ij}, m_{ij})\leftarrow\textrm{HeaderTransformer}(\mathcal{T}^*)$\tcp*{depth, mask}
$\mathcal{V}\leftarrow\textrm{QuantizeMultiViewPoints}(\{d_{ij}\}, R)$\tcp*{depth-lifted set}
$\bm{c}\leftarrow\textrm{LN}(\textrm{Concat}(\textrm{Linear}(\bm{I}'), \mathcal{T}^*))$\tcp*{context, Eq.~11}
$\bm{c}\leftarrow\textrm{MaskPrune}(\bm{c}, \{m_{ij}\}, \tau{=}0.05)$\;
$o \leftarrow\textrm{FilterTransformer}(\mathcal{V}, \bm{c})$\;
$\mathcal{V}^*\leftarrow\{v_k:\sigma(o_k)>0.5\}$\;
$\bm{z}\leftarrow\textrm{Aggregator}(\mathcal{V}^*, \bm{c})$\;
$\bm{z}\leftarrow \bm{z}\odot\bm{\sigma}+\bm{\mu}$\tcp*{denormalize}
$\mathcal{S}'\leftarrow\textrm{Trellis2Decoder}(\bm{z}, \mathcal{V}^*)$\;
$\mathcal{S}\leftarrow\textrm{DualContouring}(\mathcal{S}')$\;
\Return $\mathcal{S}$\;
\end{algorithm}

\subsection{VecSet Backend}

\highlight{Learnable Queries.}
The VecSet backend uses $N_q{=}4096$ learnable embeddings $\{\bm{q}_k\in\mathbb{R}^{D+D'}\}$ initialized from $\mathcal{N}(0, 0.02^2)$ as queries. Unlike SpVox queries, these embeddings carry no spatial anchor; the aggregator must learn what each slot encodes purely from the data. No depth-lifting, no occupancy filter, and no scale-shift augmentation are used on this path---the unordered set has no fixed spatial layout to perturb.

\highlight{Aggregator.}
The aggregator architecture is identical to the SpVox case: $L_a{=}12$ interleaved self/cross-attention layers at hidden dimension $D{+}D'{=}1536$. The cross-attention is dense over the pruned NGT context (no ray-voxel sparsity is available since queries have no positions). The output is normalized and denormalized using frozen statistics derived from the pretrained 3DShape2VecSet encoder, with $\bm{\mu}',\bm{\sigma}'\in\mathbb{R}^{64}$.

\highlight{Pretrained Decoder and Mesh Extraction.}
The denormalized latent set is consumed by the frozen pretrained 3DShape2VecSet~\cite{zhang20233dshape} decoder, which represents the SDF as a continuous field queried via cross-attention: for any 3D point $\bm{v}$, the decoder returns $s(\bm{v}){=}\textrm{Network}(\bm{v}, \{\bm{z}_k\})$. We materialize a regular SDF grid over $[-1,1]^3$ and extract the mesh via Marching Cubes~\cite{lorensen1998marching}.

\highlight{Training Loss.}
Because $\{\bm{z}_k\}$ and the pseudo-ground-truth $\{\hat{\bm{z}}_k\}$ are unordered, per-token MSE is not appropriate. We instead use the Sinkhorn optimal-transport loss~\cite{cuturi2013sinkhorn} with squared-$L_2$ ground cost. Let $\bm{C}\in\mathbb{R}^{N_q\times N_q}$ be the pairwise cost matrix $C_{kl}{=}\|\bm{z}_k-\hat{\bm{z}}_l\|_2^2$ and let $\bm{P}$ be the doubly-stochastic transport plan obtained by log-domain Sinkhorn iterations with entropic regularization $\varepsilon{=}1.0$ and $10$ iterations. The loss is
\begin{equation}
\mathcal{L}_\textrm{VecSet} \;=\; \tfrac{1}{N_q\cdot d_v}\,\langle \bm{P}, \bm{C}\rangle\,,
\end{equation}
with the transport plan $\bm{P}$ detached from the gradient graph (gradients flow only through $\bm{C}$, \ie, through the predicted latents).

\begin{algorithm}[t]
\caption{VecSet Inference Loop}
\label{alg:vecset}
\KwIn{Posed input views $\mathcal{I}=\{(I_i, C_i)\}_{i=1}^{V}$}
\KwOut{Triangle mesh $\mathcal{S}$}
$\mathcal{T}^*\leftarrow\textrm{GeoPretrainedTransformer}(\mathcal{I})$\;
$(\_,m_{ij})\leftarrow\textrm{HeaderTransformer}(\mathcal{T}^*)$\tcp*{mask only}
$\bm{c}\leftarrow\textrm{LN}(\textrm{Concat}(\textrm{Linear}(\bm{I}'), \mathcal{T}^*))$\;
$\bm{c}\leftarrow\textrm{MaskPrune}(\bm{c}, \{m_{ij}\}, \tau{=}0.05)$\;
$\bm{q}\leftarrow\textrm{LearnableQueries}(N_q)$\tcp*{$\mathcal{N}(0,0.02^2)$}
$\bm{z}\leftarrow\textrm{Aggregator}(\bm{q}, \bm{c})$\;
$\bm{z}\leftarrow \bm{z}\odot\bm{\sigma}'+\bm{\mu}'$\;
\For{each vertex $\bm{v}$ in regular grid}{
    $s(\bm{v})\leftarrow\textrm{3DShape2VecSetDecoder}(\bm{v}, \{\bm{z}_k\})$\;
}
$\mathcal{S}\leftarrow\textrm{MarchingCubes}(\{s(\bm{v})\})$\;
\Return $\mathcal{S}$\;
\end{algorithm}

\subsection{Backend Ablations: Isolating Depth-Lifting}
\label{sec:supp:decoder:ablation}

\highlight{Setup.}
\name{}'s SpVox backend combines two sparse mechanisms with different scopes. \emph{Pruning} (mask-based visual token filtering, $\tau{=}0.05$) is defined at the NGTFormer level (Sec.~4.2) and is shared with the VecSet backend; its ablation belongs with the NGT-related components and is reported in Sec.~\ref{sec:analysis:abl}. \emph{Depth-lifting} (the candidate voxel set $\mathcal{V}$ and the occupancy filter $\mathcal{V}^{*}$ of Appendix~\ref{sec:supp:decoder}) is SpVox-specific: VecSet uses unanchored learnable queries and has no analog. This subsection isolates Lifting's contribution to SpVox by stepping through three configurations: the full model, ``\textit{w/o} Pruning'' which keeps Lifting, and ``\textit{w/o} Pruning \& Lifting'' which removes both. In the last variant, the SpVox aggregator queries the full dense $R{=}16^3$ voxel grid (4096 voxels) and no occupancy filter is used.

\begin{table}[t]
    \centering
    \caption{Backend-specific ablation on SpVox. The full model uses both Pruning and Lifting. Row (b) is reproduced from Tab.~\ref{tab:ablation} for the Pruning$\to$Lifting decomposition; the (b)$\to$(a) gap isolates Lifting's effect.}
    \label{tab:ablation_backend}
    \resizebox{\linewidth}{!}{
        \begin{tabular}{c|l|c|c|cc|cc}
            \toprule
             & & Training & Depth Quality & \multicolumn{2}{c|}{Normal Quality} & \multicolumn{2}{c}{Mesh Quality} \\
            ID & Setting & iter/s$\uparrow$ & AbsRel~(\%)$\downarrow$ & MAE~($^\circ$)$\downarrow$ & PSNR$\uparrow$ & F1$\uparrow$ & CD$\downarrow$ \\
            \midrule
            - & Ours (Full Model, SpVox) & 2.42 & 0.369 & 9.061 & 26.914 & 0.99760 & 0.129 \\
            (b) & \textit{w/o} Pruning~~~~\textit{(cf.\ Tab.~\ref{tab:ablation})} & 4.71 & 0.374 & 10.238 & 26.752 & 0.99651 & 0.130 \\
            (a) & \textit{w/o} Pruning \& Lifting & 5.26 & 0.483 & 10.238 & 26.752 & 0.99025 & 0.167 \\
            \bottomrule
        \end{tabular}
    }
\end{table}

\highlight{Discussion.}
Reading Tab.~\ref{tab:ablation_backend} top-down decomposes the two mechanisms cleanly. The full$\to$(b) step (only Pruning removed) leaves all quality metrics essentially unchanged while almost doubling training throughput---confirming Pruning's role as an efficiency lever, consistent with its NGTFormer-level scope (Sec.~\ref{sec:analysis:abl}). The (b)$\to$(a) step (Lifting additionally removed) is qualitatively different: AbsRel jumps $0.374\!\to\!0.483$, CD $0.130\!\to\!0.167$, and F1 drops by $0.6$ pp, while iter/s only ticks from $4.71$ to $5.26$. Lifting is therefore a quality lever, not an efficiency one: anchoring SpVox queries on a depth-projected candidate set $\mathcal{V}^{*}$ supplies the structural prior the aggregator needs to localize geometry, and removing it forces a dense $4096$-voxel enumeration that the aggregator cannot disambiguate from the same NGT context. This also explains why the VecSet backend does not require an analog: its $N_q{=}4096$ learnable queries are unordered set elements rather than spatial slots that must be filled---each query learns to specialize across training, and any failure mode would manifest as set-coverage rather than per-voxel localization error.

\section{Cross-Decoder Evaluation}
\label{sec:supp:cross_decoder}

A key design principle of \name{} is that NGTs function as a \emph{general} geometric interface, decoupled from any specific decoder architecture. Sec.~5.1 reports both \name{}~(SpVox) and \name{}~(VecSet) variants, but a clean head-to-head with their no-NGT counterparts is missing. This section closes that gap with a $2{\times}2$ design that directly validates the decoder-agnostic claim.

\highlight{Setup.}
We evaluate four configurations on the full GSO benchmark: (i)~\name{}~+~SpVox, our primary structured decoder built on Trellis2~\cite{xiang2025trellis2}; (ii)~\name{}~+~VecSet, an unstructured set-based decoder built on 3DShape2VecSet~\cite{zhang20233dshape}; (iii)~w/o~NGT~+~SpVox, the same decoder trained end-to-end from raw images without geometric pretraining; and (iv)~w/o~NGT~+~VecSet, the analogous baseline for the VecSet family. Configurations~(i) and~(ii) reuse the numbers from Tab.~2. Configurations~(iii) and~(iv) are independent training runs that share the same architecture and training schedule as their NGT-conditioned counterparts. All variants are evaluated with the identical four-view input protocol described in Appendix~\ref{sec:supp:eval}.

\highlight{Results.}
Table~\ref{tab:cross_decoder} reports the $2{\times}2$ design. Both decoders benefit substantially from NGT conditioning, with consistent improvements across all metrics, indicating that the geometric information carried by NGTs is not decoder-specific. Measured by the absolute degradation incurred when NGT conditioning is removed, the gap is larger for the SpVox family (AbsRel $+1.02$, Normal MAE $+9.57^\circ$, CD $+0.187$) than for VecSet (AbsRel $+0.37$, Normal MAE $+3.90^\circ$, CD $+0.144$). We attribute this to SpVox's hardcoded spatial layout: its queries are anchored to a fixed voxel grid, so an end-to-end model must learn to place geometry into predetermined spatial slots directly from raw images, which is harder than regressing an unanchored latent set. The same structural-bottleneck mechanism analysed in Sec.~3 explains both no-NGT failures.

\begin{table}[t]
    \centering
    \caption{Cross-decoder $2{\times}2$ evaluation on the full GSO benchmark. Both decoder families degrade when NGT conditioning is removed, confirming that NGTs act as a decoder-agnostic geometric interface rather than a Trellis2-specific adaptor. The two GeoPretraining=Yes rows are the \name{}~(SpVox) and \name{}~(VecSet) entries of Tab.~\ref{tab:comparison} (GSO Full columns).}
    \label{tab:cross_decoder}
    \resizebox{\linewidth}{!}{
        \begin{tabular}{ll|ccc}
            \toprule
            \textbf{Decoder} & \textbf{GeoPretraining} & Depth AbsRel~(\%)$\downarrow$ & Normal MAE~($^\circ$)$\downarrow$ & Mesh CD$\downarrow$ \\
            \midrule
            \multirow{2}{*}{SpVox}  & Yes & 0.544 & 9.704  & 0.238 \\
                                    & No  & 1.567 & 19.278 & 0.425 \\
            \midrule
            \multirow{2}{*}{VecSet} & Yes & 0.321 & 8.123  & 0.110 \\
                                    & No  & 0.693 & 12.018 & 0.254 \\
            \bottomrule
        \end{tabular}
    }
\end{table}

\highlight{Qualitative Trade-offs.}
Before contrasting the two backends, we state the scope that governs both. \name{} is a \emph{strict-reconstruction} method: it recovers the surface that the input views actually observe, and it does not attempt to complete geometry that no view sees. Watertightness is therefore not a target, and not a criterion by which either backend should be judged. Sparse views necessarily leave parts of an object unobserved, and a faithful reconstruction of an incompletely observed object is an open surface; closing it would require inventing geometry that the observations do not support. We accordingly evaluate both backends on the \emph{faithfulness of the surfaces they reconstruct from observed regions}, and treat the boundaries left by unobserved regions as the intended behaviour of the method rather than as a defect. Appendix~\ref{sec:supp:failures} characterises how \name{} behaves in those unobserved regions.

Within that scope, the two NGT-conditioned decoders both produce high-fidelity reconstructions, but their residual artifacts are complementary. The SpVox backend, owing to its explicit voxel-anchored layout, excels at recovering fine geometric details---sharp edges, thin handles, and high-frequency surface variations are faithfully preserved. Its principal artifacts are topological: small holes on smooth surfaces, together with the properties inherited from dual contouring on the Trellis2~\cite{xiang2025trellis2} sparse grid---an unclosed triangulation whose per-voxel face orientations are not globally consistent. These are properties of the frozen pretrained decoder and its mesh-extraction rule, not of NGT conditioning, and they leave geometric fidelity essentially untouched; Appendix~\ref{sec:supp:cross_decoder:meshquality} inspects them directly. The VecSet backend, in contrast, attains better aggregate metrics and yields watertight, winding-consistent surfaces, but tends to over-smooth fine details and pays a non-trivial inference-time cost: its unordered latent set must be queried densely at decoding time, making it noticeably slower than the structured SpVox pipeline. In practice, the two backends thus offer a detail-versus-topology trade-off that downstream applications can choose between without changing the upstream NGT encoder.

This complementarity is itself an argument for the decoder-agnostic design: neither backend dominates, and which of their properties matters is a function of the downstream consumer rather than of the reconstruction problem. Appendix~\ref{sec:supp:cross_decoder:meshquality} inspects the SpVox output directly and closes with a concrete recommendation.

\highlight{Discussion.}
The decoder-agnostic property directly counters the concern that NGTs are merely a Trellis2-flavored variant tied to a single decoder. NGTs capture geometric information that is readily consumed by diverse downstream architectures, ranging from spatially structured sparse voxels to unordered latent token sets. This validates NGTs as a \emph{general} geometric interface for feed-forward mesh reconstruction.

\subsection{Mesh Quality of the SpVox Backend}
\label{sec:supp:cross_decoder:meshquality}

The SpVox backend decodes into a voxel-anchored latent whose queries occupy a predetermined spatial grid. A reasonable concern about any such layout is whether the decoded surface is geometrically \emph{sound}---whether the grid induces overlapping shells, doubled walls, or other layout artifacts that shaded surface renders would hide. Aggregate metrics cannot settle this: Chamfer Distance is computed on sampled surface points and is largely insensitive to a duplicated shell lying within sampling tolerance of the true surface. We therefore inspect the meshes directly. Throughout this subsection, ``sound'' refers to the internal consistency of the triangulation and says nothing about completeness: as established above, an open boundary where no view observed the surface is the intended output of a strict-reconstruction method, not a defect.

\highlight{Diagnostics.}
We use three diagnostics, chosen so that the artifacts in question would be plainly visible or countable if present. \emph{(i)~Topological statistics.} We report per-mesh face counts, self-intersecting faces, connected components, boundary edges, watertightness, and winding consistency, aggregated over a six-object GSO subset. \emph{(ii)~Cross-sectional slices.} We cut each reconstructed mesh with axis-aligned planes and render the intersection curves, and we overlay the two backends on the same cut to compare them quantitatively. A single-shell surface produces one contour stroke per wall; an overlapping shell appears as an additional stroke offset from it at the same location. Because the VecSet backend extracts an iso-surface from a continuous field, its cross-section is by construction free of interior strokes, which makes it a useful reference against which to read the SpVox contour. \emph{(iii)~Wireframe renders.} We render the raw triangulation without shading. Dual contouring on a regular grid yields a locally uniform tessellation; a grid-induced layout artifact would break that regularity into scrambled or degenerate neighbourhoods, and would recur at grid-aligned positions across objects.

\highlight{Topological Statistics.}
Table~\ref{tab:mesh_topology} reports the statistics for both backends, measured on the \emph{raw} decoder output before any mesh post-processing, so that what is being characterised is the extraction rule itself. The two backends behave as their mesh-extraction rules predict. VecSet is watertight, winding-consistent, free of self-intersections and almost always a single connected component---the expected consequence of extracting an iso-surface by Marching Cubes from a continuous SDF, which cannot produce a non-manifold or inconsistently oriented result by construction. SpVox is none of these in its raw form. Dual contouring assembles the surface from per-voxel dual-cell primitives whose vertex offsets and edge flags are predicted independently per voxel, so neighbouring cells are not constrained to agree on a globally consistent orientation, and a cell whose predicted occupancy disagrees with its neighbours leaves a boundary rather than a closed seam. The many small components are the direct signature of this: they are fragments near such disagreements, not spurious duplicate objects. Notably, the underlying sparse-voxel decoder of Trellis2~\cite{xiang2025trellis2} does not guarantee watertight output in the first place, so these are properties of the pretrained decoder we adopt rather than of our reconstruction.

Two of these statistics are also inexpensively repairable, and our reported results are post-processed accordingly. Following the same practice as Trellis2, we unify face orientation as a post-process; this is a purely combinatorial relabelling that moves no vertex, and it resolves the winding inconsistency together with the back-facing majority discussed below, leaving the reconstructed geometry bit-identical. The statistics in Tab.~\ref{tab:mesh_topology} deliberately precede that step. We report the raw numbers because they characterise the extraction rule, and because the one property that post-processing does \emph{not} supply is watertightness---closing the surface would require re-meshing that invents geometry the observations do not support, which a strict-reconstruction method should not do.

\begin{table}[t]
    \centering
    \caption{Mesh topology statistics on a six-object GSO subset, measured on the \emph{raw} decoder output before the face-orientation post-process described in the text. VecSet is watertight, winding-consistent and free of self-intersections by construction; raw SpVox is neither watertight nor globally winding-consistent, and fragments into many small components. Self-intersections are reported as both mean and median because their distribution is strongly skewed---one smooth cylindrical object accounts for most of the subset total, while the remaining five stay at or below $0.3\%$ of faces. Neither backend is penalised here for open boundaries \emph{per se}---see the text on why watertightness is not a target for a strict-reconstruction method.}
    \label{tab:mesh_topology}
    \resizebox{0.7\linewidth}{!}{
        \begin{tabular}{l|cc}
            \toprule
            \textbf{Statistic} & \textbf{SpVox} & \textbf{VecSet} \\
            \midrule
            Faces (mean)                      & 340{,}138 & 305{,}153 \\
            Self-intersecting faces (mean)    & 5020~($1.48\%$) & 0~($0.00\%$) \\
            Self-intersecting faces (median)  & 532~($0.15\%$)  & 0~($0.00\%$) \\
            Connected components (mean)       & 196.3     & 2.8 \\
            Boundary edges (mean)             & 3638      & 0 \\
            Watertight                        & 0/6       & 6/6 \\
            Winding-consistent (raw)          & 0/6       & 6/6 \\
            \bottomrule
        \end{tabular}
    }
\end{table}

\begin{figure*}[t]
\centering
\begin{tabular}{@{}c@{~~}c@{~~}c@{}}
\includegraphics[height=3.0cm]{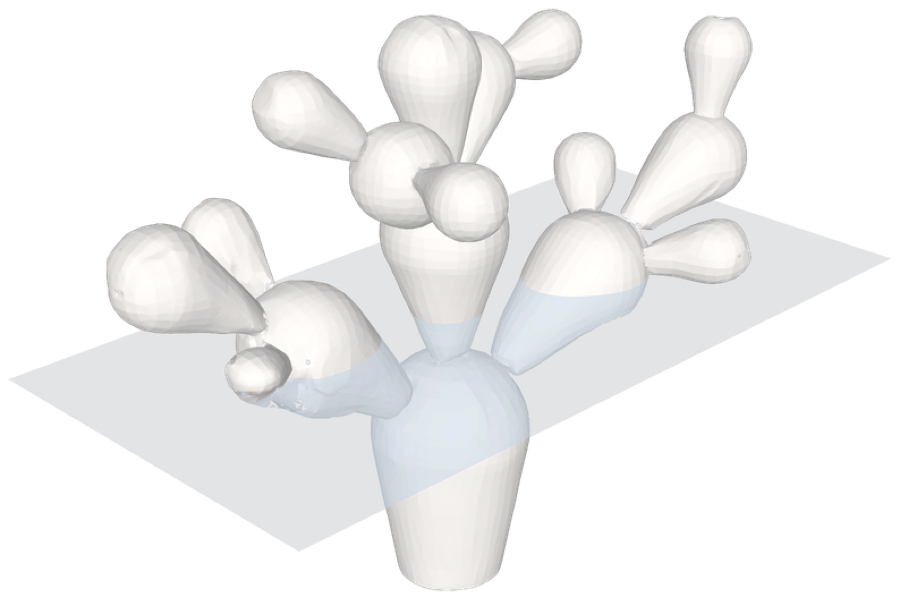} &
\includegraphics[height=3.0cm]{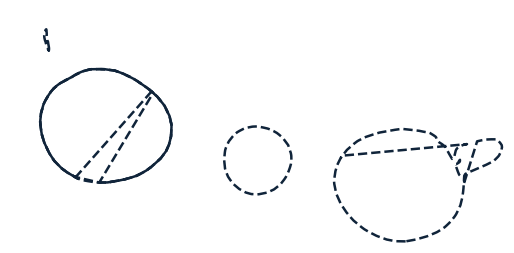} &
\includegraphics[height=3.0cm]{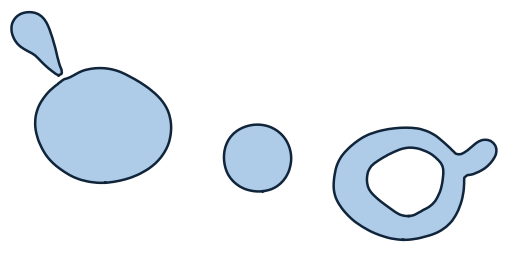} \\
{\scriptsize (a) cut plane} & {\scriptsize (b) SpVox cross-section} & {\scriptsize (c) VecSet cross-section} \\[2mm]
\includegraphics[height=4.0cm]{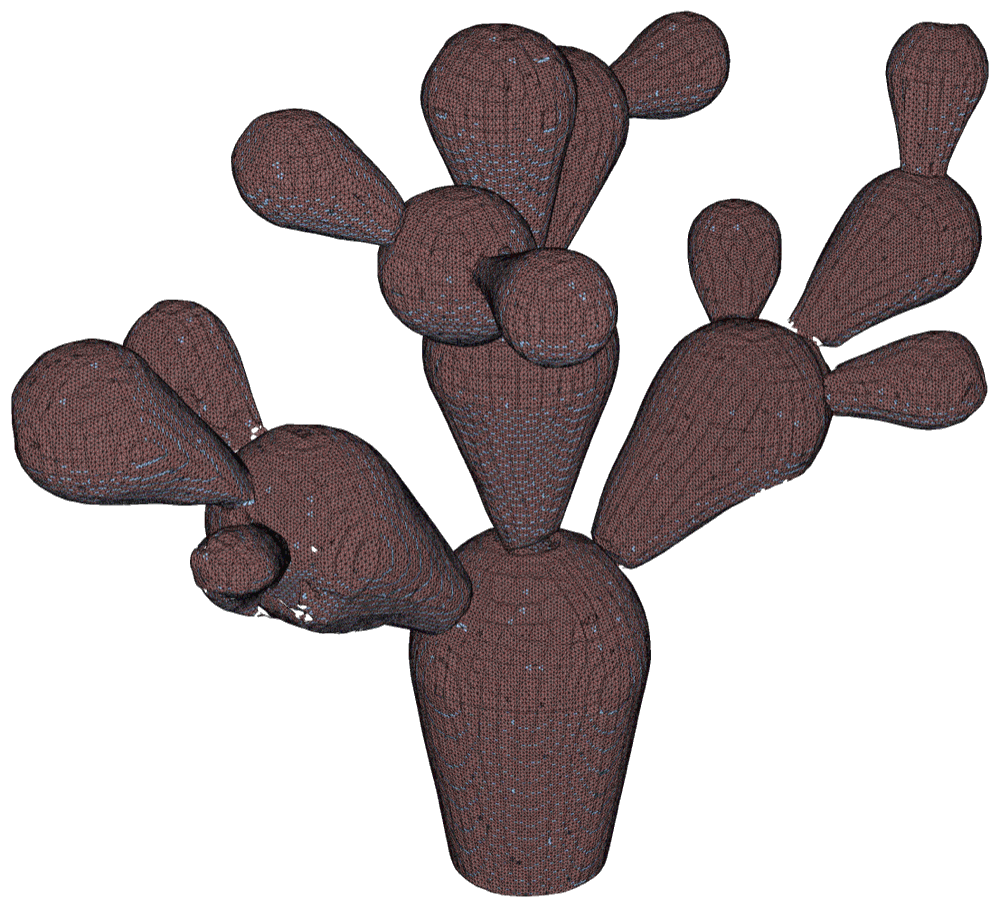} &
\includegraphics[height=4.0cm]{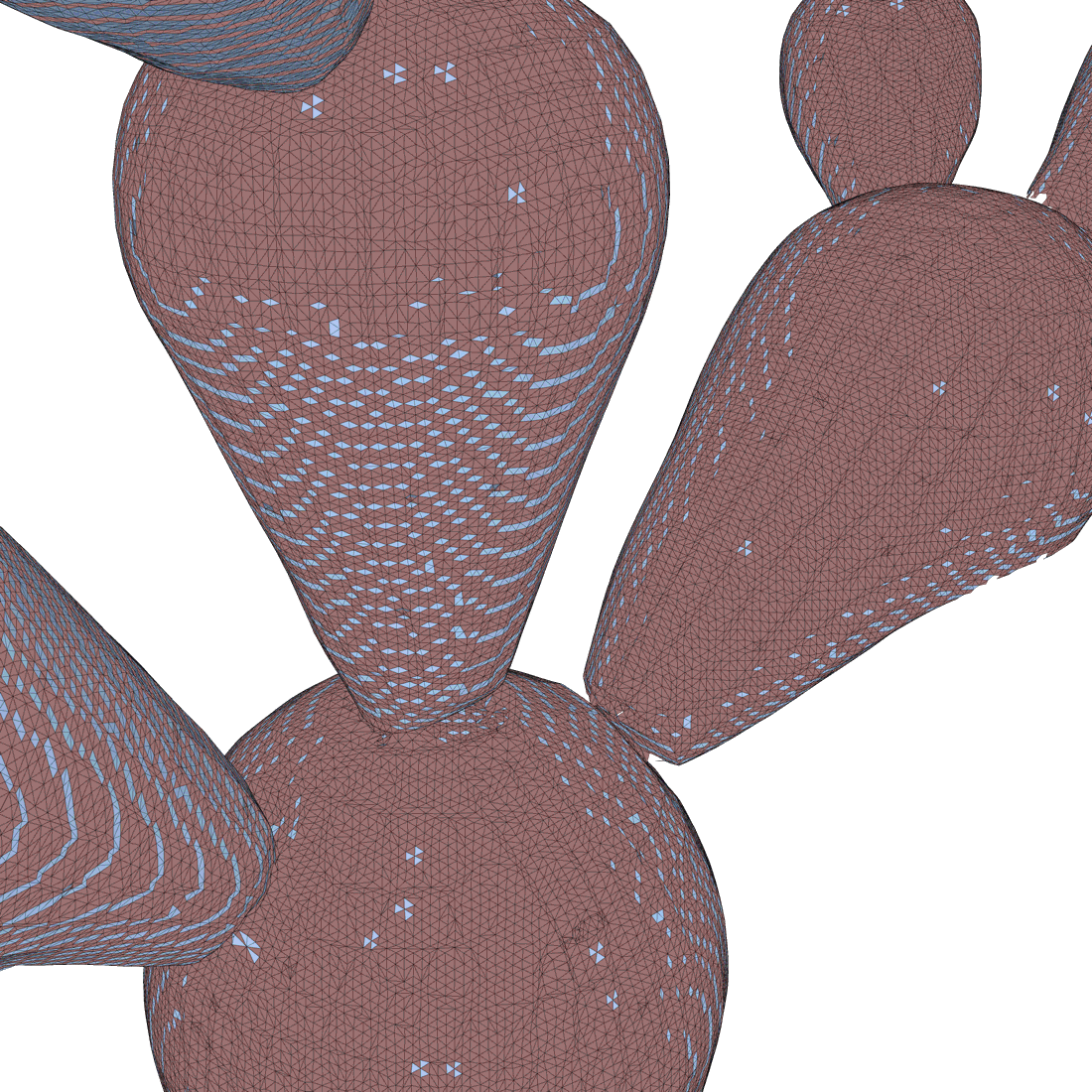} &
\includegraphics[height=4.0cm]{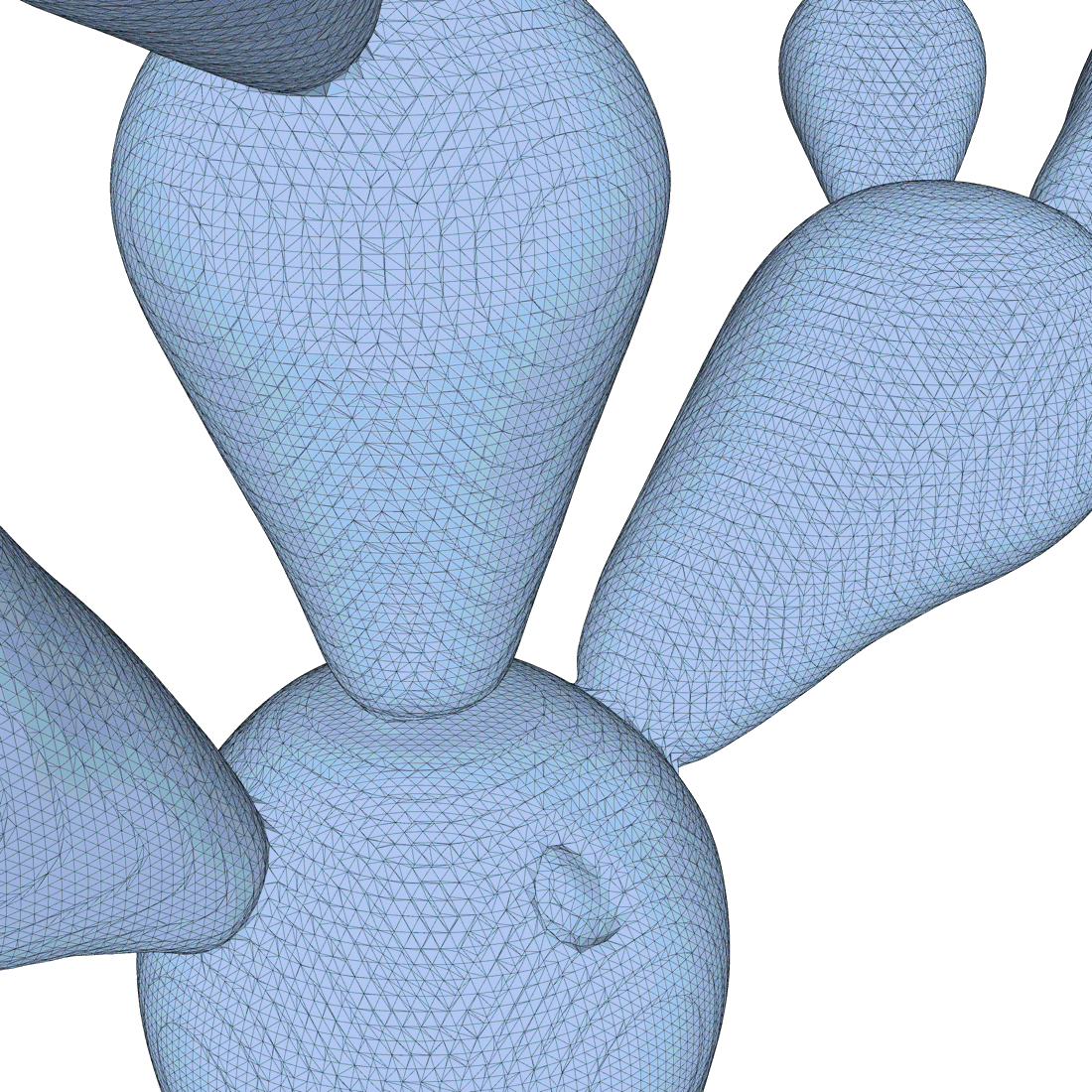} \\
{\scriptsize (d) SpVox wireframe, global} & {\scriptsize (e) SpVox wireframe, detail} & {\scriptsize (f) VecSet wireframe, detail} \\
\end{tabular}
\caption{Mesh quality of the SpVox backend. \emph{Slices} ((a)--(c)): (a)~shows where the plane cuts the object; (b)~and~(c)~are the resulting cross-sections for the two backends, drawn over the same coordinate range so that stroke positions are directly comparable. \textbf{Note on how the slices are drawn:} VecSet contours are shown filled, SpVox contours as unfilled dashed curves. This is deliberate and is not a geometric difference---a raw, winding-inconsistent mesh has no well-defined interior, so filling it, or distinguishing solid from dashed strokes, would assert more than the mesh supports. Compare stroke positions, not shading. On the outer wall the two agree closely (median offset $0.4\%$ of the object extent); the strokes SpVox draws \emph{inside} the outline are the interior sheets discussed in the text, which a continuous-field iso-surface cannot produce. \emph{Wireframes} ((d)--(f)): the raw triangulation, unshaded, at a shared camera. The tessellation is locally uniform at both scales, with no scrambling at grid-aligned positions; SpVox is denser and less regular than VecSet, which is the trace of its per-voxel construction rather than a layout artifact. Warm-toned faces are back-facing and light blue front-facing under double-sided rendering---in the raw output about four fifths of SpVox's visible faces are back-facing, the direct signature of its per-voxel orientation, which the face-orientation post-process resolves without moving any vertex. Blank regions are open boundaries, not culled back-faces.}
\label{fig:spvox_mesh_quality}
\Description{Six panels arranged in two rows examine a branching cactus toy. The upper row shows the cutting plane, SpVox cross-sections as unfilled dashed contours, and VecSet cross-sections as filled blue regions. The outer contours largely coincide, but SpVox includes additional strokes inside some outlines; the different fill styles are drawing conventions. The lower row shows a global SpVox wireframe and matched SpVox and VecSet close-ups. SpVox has a denser, less regular triangulation than VecSet. Warm faces indicate back-facing triangles and light blue faces indicate front-facing triangles; blank regions show open boundaries rather than culled faces.}
\end{figure*}

\highlight{Overlapping Structures and Layout Artifacts.}
\emph{Layout artifacts are absent.} The wireframes in Fig.~\ref{fig:spvox_mesh_quality}(d)--(f) show a locally uniform tessellation across the whole surface, at both the global and the zoomed scale, with no scrambled or degenerate neighbourhoods and no break in regularity at grid-aligned positions. The SpVox tessellation is visibly denser and more irregular than VecSet's---the direct trace of its per-voxel construction---but it is uniform in exactly the sense the concern is about: nothing in the output is traceable to the predetermined spatial layout as such, and the triangle density does not modulate with the grid.

\emph{Overlapping structures are minor but not entirely absent.} On the outer surface---the part the reconstruction is actually claiming---the two backends agree closely: overlaying the two cross-sections of Fig.~\ref{fig:spvox_mesh_quality}(b,~c), every point of the VecSet wall lies a median $0.4\%$ of the object's extent from an SpVox stroke, so SpVox places the visible surface where VecSet's iso-surface places it. What SpVox additionally draws are strokes in the object's \emph{interior}, which a continuous-field iso-surface cannot produce. Part of these follow internal structure of the object itself; the remainder are the doubled walls we report here, and they are concentrated on smooth, nearly untextured surfaces of near-constant depth---the setting in which adjacent dual cells can settle on slightly different iso-levels. Such offset sheets are cheap under a latent-space regression loss and nearly free under Chamfer Distance, which is precisely why an aggregate metric could not have surfaced them and why the direct inspection was warranted. They are also what the skewed self-intersection counts in Tab.~\ref{tab:mesh_topology} reflect: the objects on which interior sheets appear are the same ones that carry most of the self-intersecting faces, so the two diagnostics point at a single phenomenon rather than two. We record this as a genuine, if limited, property of the sparse-voxel backend rather than describing the output as artifact-free. It is worth being clear about the scope of the consequence: it does not displace the reconstructed surface, and it matters only for consumers that reason about the mesh interior.

\highlight{Which Backend to Use.}
Taken together with the metric comparison above, this analysis makes the backend choice concrete rather than a matter of taste---and it is the reason a decoder-agnostic interface matters in practice. \emph{Use VecSet when downstream consumers require guaranteed topology}: volume computation, physical simulation, 3D printing, boolean operations, or any pipeline that assumes a manifold, consistently oriented, closed-where-observed surface. Marching Cubes on a continuous field supplies these by construction, and VecSet also attains the better aggregate metrics. \emph{Use SpVox when the priority is geometric detail or inference cost}: it recovers sharp edges, thin handles and high-frequency variation more faithfully and avoids VecSet's dense grid evaluation, at the price of a triangulation that needs a cleanup pass before topology-sensitive use. That pass is cheap and standard---face-orientation unification, which we already apply, plus component pruning and self-intersection resolution where required---and it addresses the statistics in Tab.~\ref{tab:mesh_topology} without moving the reconstructed surface.

Neither option is strictly better, and that is the point. The sparse-voxel decoder we build on does not guarantee watertight output, VecSet does but over-smooths, and which of the two properties a user needs is determined by their downstream pipeline rather than by the reconstruction problem. Because NGTs form a decoder-independent interface, this trade-off is resolved at the decoder, by swapping backends, with the upstream encoder left untouched---so multi-backend support is a substantive capability of the design rather than an incidental demonstration of generality.

We close by stating what this does and does not say about the method's core claim. The NGT interface determines \emph{where the surface is}, and Tab.~\ref{tab:cross_decoder} shows it improves both backends there. Topological cleanliness is a property of the frozen pretrained decoder and its mesh-extraction rule---dual contouring on a sparse grid versus Marching Cubes on a continuous field---and is orthogonal to what NGTs contribute. A reader weighing the two backends should treat detail-versus-topology as the operative trade-off, not as evidence about NGTs either way.

\section{Positioning: Reconstruction vs. Generation}
\label{sec:supp:recon_vs_gen}

Section~5.3 makes the visual case that state-of-the-art generative 3D models produce surfaces with substantial disagreement to input observations, despite their visual plausibility. This appendix extends that argument by expanding on why the reconstruction--generation fidelity gap matters in practice.

Reconstruction and generation serve fundamentally different needs. Generative 3D models produce plausible and diverse shapes, but they hallucinate geometry even when conditioned on dense observations---a property that is desirable when the goal is creative synthesis but disqualifying when downstream tasks demand geometric fidelity, \eg{} 3D scanning, physical simulation, robotic manipulation, or any measurement-grade pipeline. This reconstruction--generation gap is increasingly recognized in the broader 3D community: RnG~\cite{xiang2026rng} documents an order-of-magnitude depth-accuracy gap between sparse-view reconstruction and the leading generative pipelines under matched supervision, and ReconViaGen~\cite{chang2025reconviagen} analyses why diffusion-based 3D generation achieves degraded multi-view consistency even with strong image priors. The gap has been documented for point-cloud, depth, and 3DGS reconstruction; \name{} shows that it persists, and is in fact widened, in feed-forward \emph{mesh} reconstruction (Sec.~5.1 reports $\sim\!30\times$ CD reduction over TRELLIS2 and Hunyuan3D).

\name{}'s positioning is therefore complementary rather than adversarial. It provides a high-fidelity baseline for tasks that cannot tolerate confident-but-unsupported geometry, while generative pipelines remain the right tool when the goal is shape \emph{generation}---a coherent extrapolation from minimal observations rather than a faithful recovery of an observed one.

\highlight{Where Hybrid Methods Sit.}
A growing line of work attempts to obtain fidelity from within the generative paradigm, and it is worth stating precisely how \name{} differs from it, since the distinction is architectural rather than a matter of degree. Two strategies are visible in recent work. The first couples a reconstruction front-end to a generative back-end, using multi-view reconstruction features to condition a generative mesh decoder and compensating for fidelity during sampling. The second enforces pixel alignment natively inside 3D generation, back-projecting 2D image features into the volume that a latent diffusion model denoises. Both meaningfully improve input fidelity over unconditioned generation.

What both retain, however, is a generative predictor as the sole producer of \emph{all} output geometry. Fidelity is therefore a soft signal: the sampler must rediscover the observed surface through its conditioning channel on every run, and nothing in the architecture prevents it from departing from that surface where the prior is strong. \name{} makes a different trade. It regresses observed geometry directly---no diffusion sampler, no VAE-based completion---so observed surfaces are recovered by construction rather than recovered by sampling. The cost of this choice is equally explicit: \name{} reconstructs only what the input views observe, and unobserved regions remain unresolved (Appendix~\ref{sec:supp:failures}). The two designs are thus not competitors on a single axis but occupy opposite ends of a fidelity-versus-flexibility trade-off, and a combined pipeline that anchors visible surfaces with \name{} and completes the remainder generatively is a natural direction that neither design forecloses.

\section{Failure Cases}
\label{sec:supp:failures}

The conclusion (Sec.~\ref{sec:conclusion}) notes that \name{} inherits several limitations from its feed-forward, photometry-driven design. This appendix makes those limitations concrete with four failure categories. For each, we describe representative cases, explain the underlying mechanism, give an honest account of whether the limitation is unique to \name{} or shared across the field, and outline a corresponding mitigation direction.

\subsection{Heavy Self-Occlusion / Cavity-Rich Geometry}

\highlight{Representative Cases.}
Cups, deep folds in clothing, the interior of bottles, and concave shells. In all of these, large portions of the surface are simultaneously hidden from every input view, so the reconstructed mesh exhibits open boundaries or missing interior walls.

\highlight{Mechanism.}
\name{} only reconstructs surfaces observable from at least one of the input views. Heavily self-occluded regions---deep cavities and interior surfaces of hollow objects---receive no per-pixel supervision and no test-time observation, so the NGT-conditioned aggregator has no signal with which to recover them. This is the intended behaviour of a strict-reconstruction system rather than a defect: closing such a surface would require inventing geometry that no view supports. It does, however, mean that \name{} does not deliver a complete watertight mesh, and applications requiring one need a separate completion stage.

\begin{figure*}[t]
\centering
\setlength{\tabcolsep}{1pt}
\renewcommand{\arraystretch}{0.6}
\resizebox{0.92\textwidth}{!}{%
\begin{tabular}{@{} c c cc cc c @{}}
    & \textbf{Input views}
    & \multicolumn{2}{c}{\textbf{\name{} (VecSet)}}
    & \multicolumn{2}{c}{\textbf{\name{} (SpVox)}}
    & \textbf{GT} \\[2pt]
    & & {\small visibility} & {\small reconstruction} & {\small visibility} & {\small reconstruction} & {\small ground truth} \\[2pt]
    \rotatebox{90}{\quad\textsc{Circo Fish}}
    & \includegraphics[width=2.35cm]{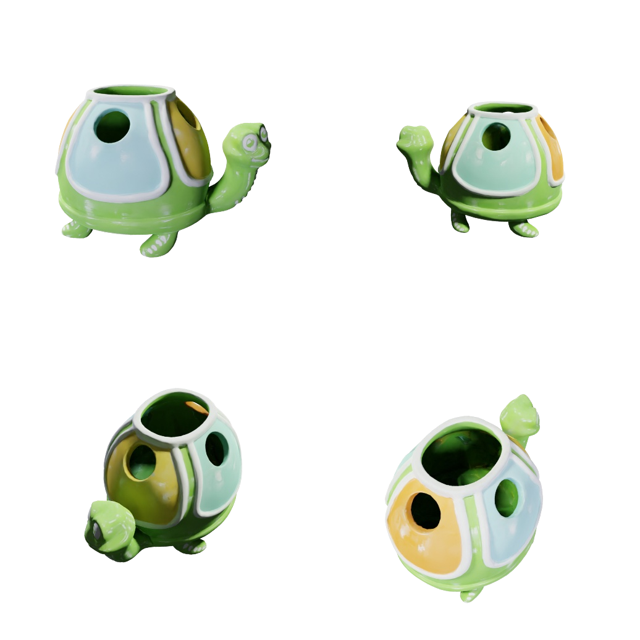} & \includegraphics[width=2.35cm]{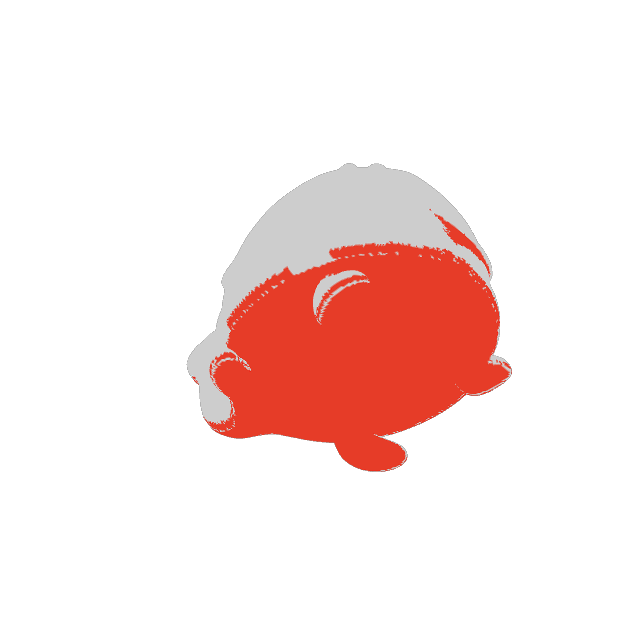} & \includegraphics[width=2.35cm]{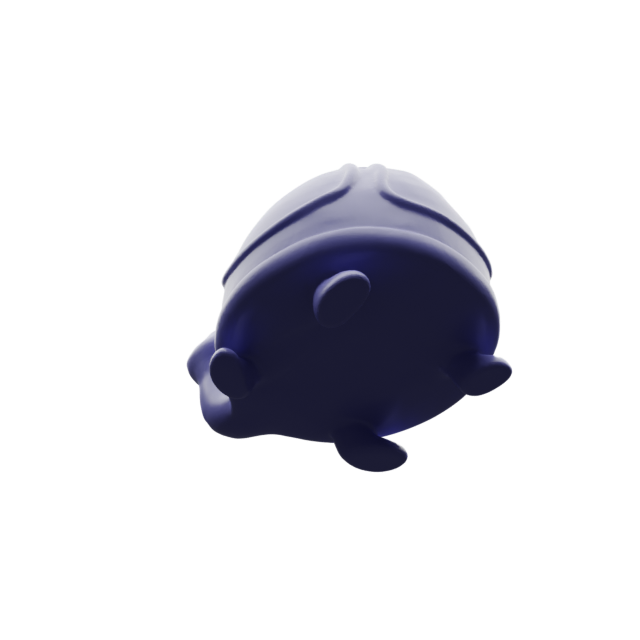}
    & \includegraphics[width=2.35cm]{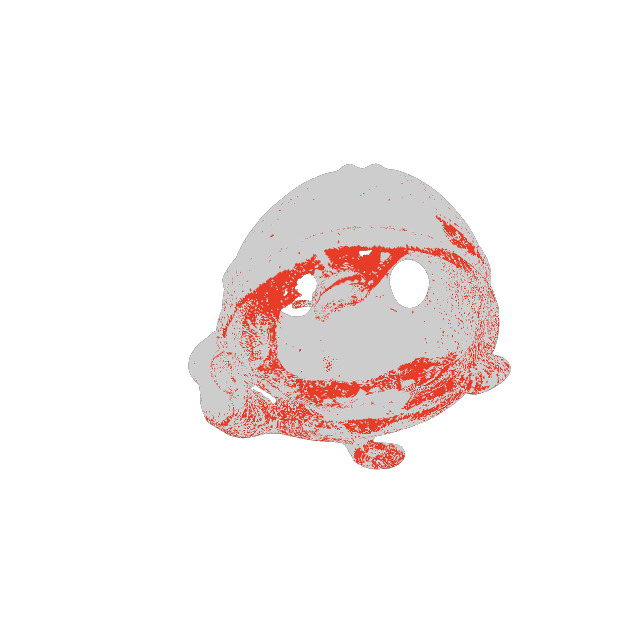} & \includegraphics[width=2.35cm]{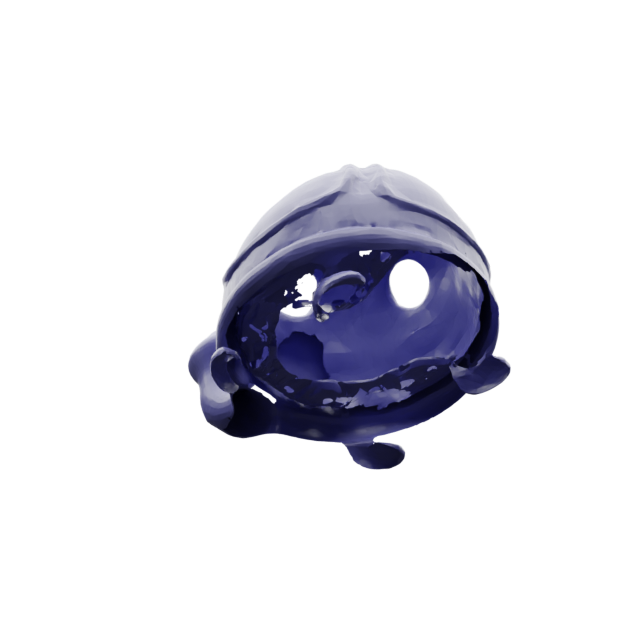} & \includegraphics[width=2.35cm]{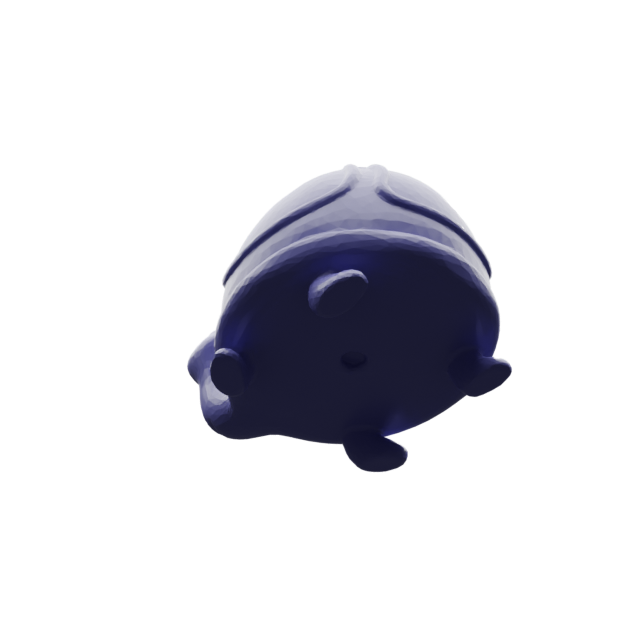} \\[1pt]
    \rotatebox{90}{\quad\textsc{LEGO 5887}}
    & \includegraphics[width=2.35cm]{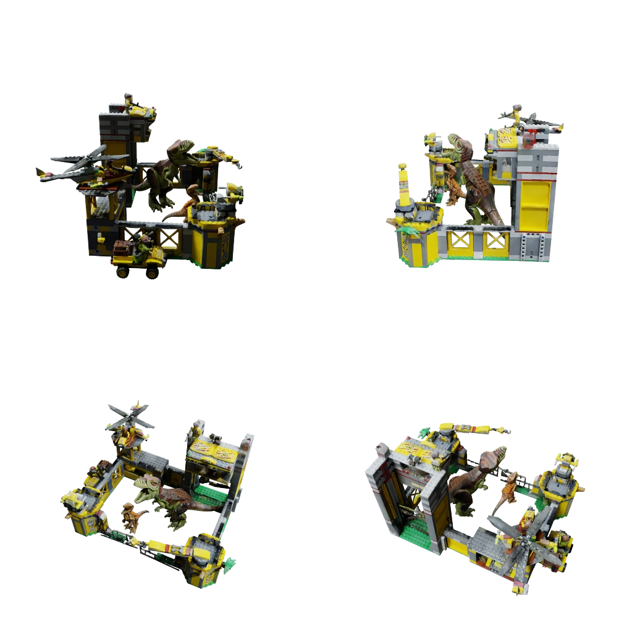} & \includegraphics[width=2.35cm]{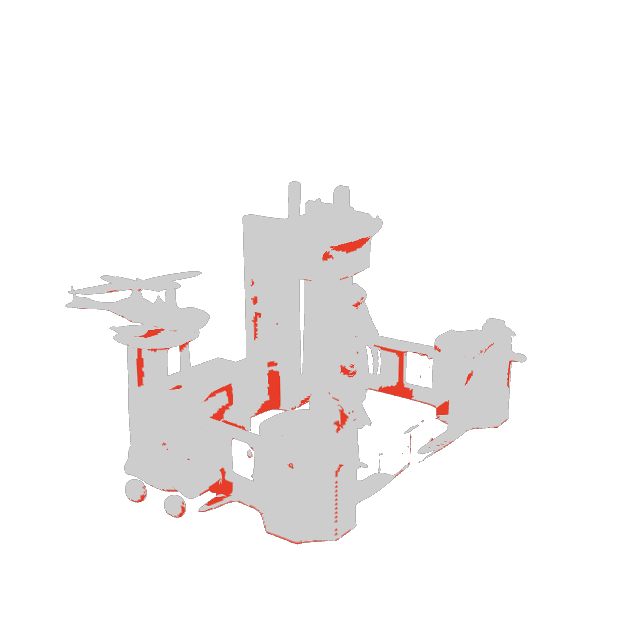} & \includegraphics[width=2.35cm]{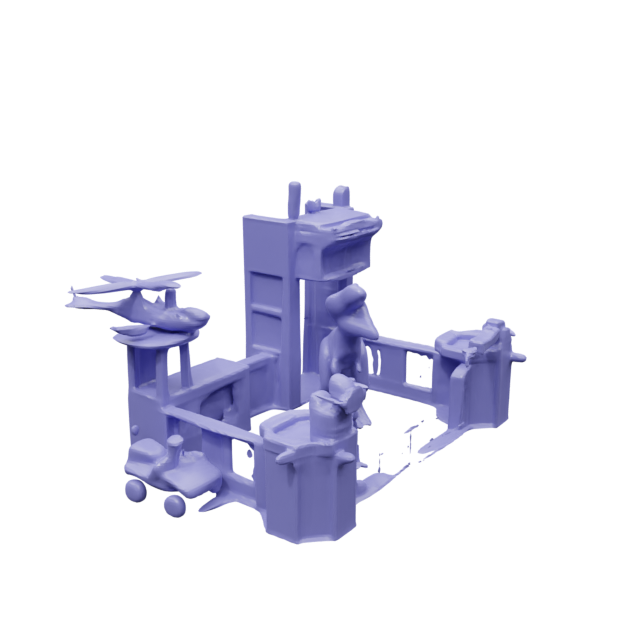}
    & \includegraphics[width=2.35cm]{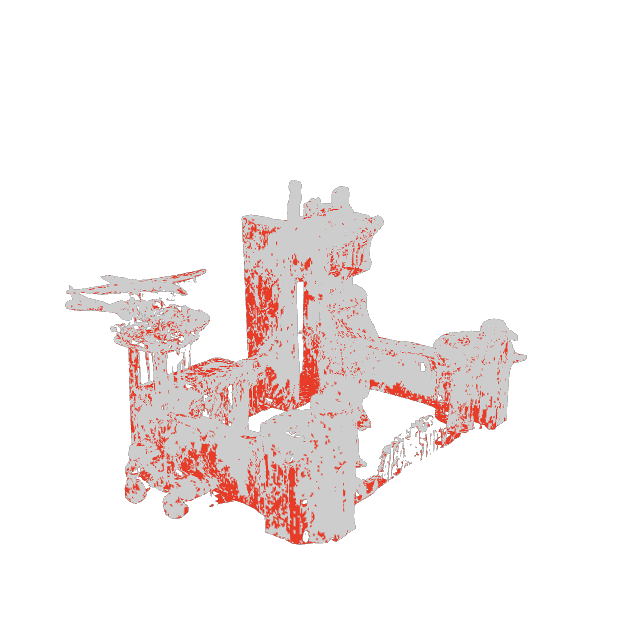} & \includegraphics[width=2.35cm]{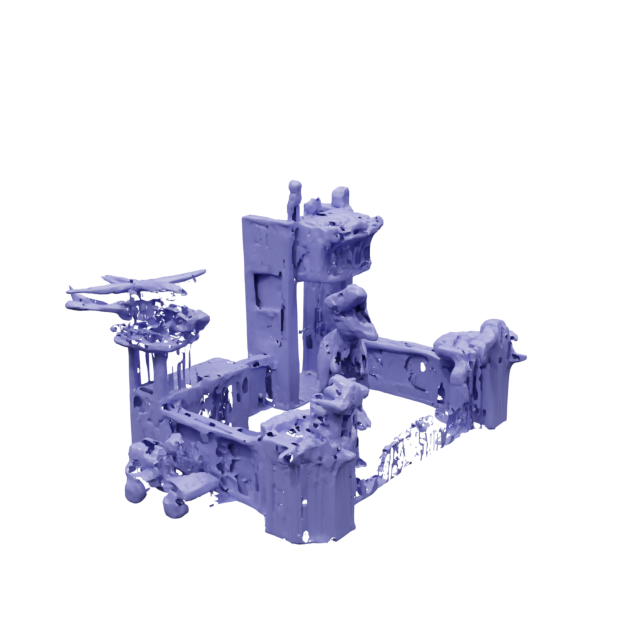} & \includegraphics[width=2.35cm]{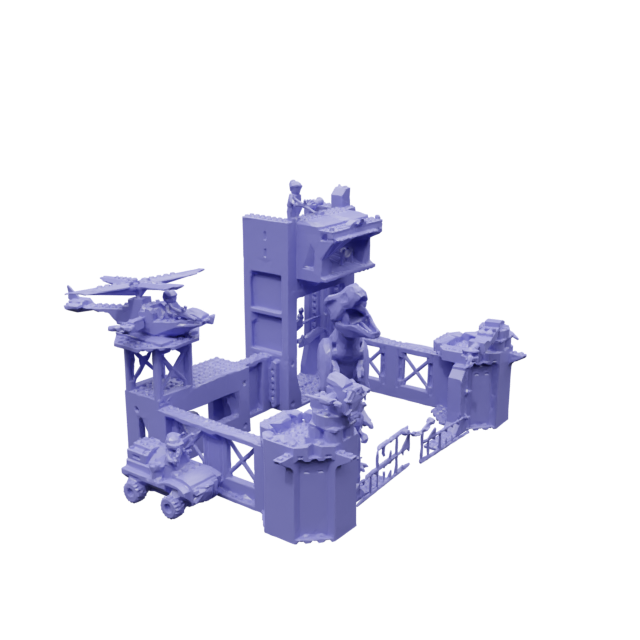} \\[2pt]
    & & \multicolumn{2}{c}{\small closes the cavity} & \multicolumn{2}{c}{\small truncates at the boundary} & {\small ground truth} \\
\end{tabular}%
}
\caption{Reconstruction behaviour in unobserved regions, for the two decoder backends on the same object and camera. The camera is placed below the object to expose the underside, which the four upper-hemisphere input views (left column) never see. \emph{Visibility:} the predicted mesh with vertices unobserved from every input view in \textcolor{red}{red} and observed vertices in grey. \emph{Reconstruction:} the same mesh under the paper's matte shader. The two backends behave oppositely where evidence runs out. VecSet closes the unobserved cavity into a smooth, watertight continuation of the observed shell---on \textsc{Circo Fish} it has no enclosed holes and matches the GT silhouette area to $0.6\%$, despite $69\%$ of that silhouette being unobserved. SpVox terminates the surface instead, leaving open apertures ($3.4\%$ of its rendered area, silhouette at $96\%$ of GT). This follows from the mesh-extraction rules of Appendix~\ref{sec:supp:cross_decoder:meshquality} rather than from the reconstruction: Marching Cubes on a continuous field cannot emit an open boundary, dual contouring on a sparse grid can. Visibility masks are computed per predicted mesh and so the red fractions are not directly comparable between the two backends; what is comparable is each backend's behaviour at its own boundary. On \textsc{LEGO 5887} the object's own construction contains genuine gaps, visible in GT as well, so apertures there should not be read as unobserved geometry.}
\label{fig:unobserved}
\Description{Two rows show Circo Fish and a LEGO model from below. Columns contain four input-view thumbnails, a visibility-colored mesh and shaded reconstruction for VecSet, the same pair for SpVox, and ground truth. Red vertices are unobserved from all input views and grey vertices are observed. VecSet forms a continuous underside across the fish's unseen cavity, while SpVox leaves openings. The LEGO model contains gaps in the ground truth as well as in the reconstructions. Visibility coloring is computed separately for each predicted mesh.}
\end{figure*}

\highlight{What the Unobserved Regions Actually Contain.}
Saying that unobserved geometry is ``not recovered'' understates what a downstream user needs to know: the decoder does not leave a hole wherever evidence runs out, it produces \emph{something}, and the practical question is what. Fig.~\ref{fig:unobserved} answers this by rendering reconstructions alongside the per-view visibility mask, so the observed and unobserved parts of the same surface can be read against each other.

What the renders show is that the answer is not a property of \name{} as a whole, but of the decoder backend---and the two backends resolve it in opposite directions. Neither leaves a hole where the evidence runs out. The VecSet backend closes the unobserved region completely: on \textsc{Circo Fish} in Fig.~\ref{fig:unobserved}, where $69\%$ of the rendered silhouette is unobserved from any input view, its reconstruction has no enclosed holes at all and its silhouette area matches GT to within $0.6\%$. It extrapolates the belly as a smooth continuation of the observed shell---plausible, unsupported, and visually indistinguishable from the measured part of the same surface. The SpVox backend instead terminates the surface near the visibility boundary, leaving open apertures that occupy $3.4\%$ of its rendered area on the same object and reduce its silhouette to $96\%$ of GT. The mechanism is the one analysed in Appendix~\ref{sec:supp:cross_decoder:meshquality}: Marching Cubes on VecSet's continuous field cannot produce an open boundary, whereas dual contouring on the sparse voxel grid emits a boundary wherever predicted occupancy runs out. Extrapolation-versus-truncation in unobserved regions is thus the same decoder property that Appendix~\ref{sec:supp:cross_decoder} discusses as watertight-versus-open topology, seen from the side that matters for fidelity.

Two caveats on reading the figure. First, the visibility masks are computed per predicted mesh, so the unobserved fraction is not comparable across backends as a number---the two columns differ partly because the meshes differ. What is comparable is the behaviour at the boundary within each column. Second, an open boundary is not the same as a missing region: SpVox does not truncate exactly at the visibility line, and on \textsc{LEGO 5887}, where the object's own construction leaves genuine gaps, both backends and GT show apertures, so an aperture there cannot be attributed to unobserved geometry alone. The clean reading of the extrapolation-versus-truncation contrast is the \textsc{Circo Fish} pair, where the unobserved region is a single large connected cavity.

Neither behaviour should be read as the model failing to try. Both are the visible signature of the fidelity-versus-flexibility trade-off of Appendix~\ref{sec:supp:recon_vs_gen}, and they differ in how legible the deficit is left. SpVox's open boundary is self-reporting: a downstream consumer can see where the evidence stopped. VecSet's smooth closure is not, and this is the sharper practical point---a strict-reconstruction method can still hand back an unsupported surface that looks exactly like a measured one, because the decoder's iso-surface has no way to abstain. A generative decoder would close both cases confidently and be wrong in a way no render exposes; \name{} narrows that failure to the unobserved region alone, but does not eliminate it. For downstream use the consequence is concrete: the Stage-1 visibility mask should be carried alongside the mesh, since with the VecSet backend it is the \emph{only} channel that distinguishes measured surface from extrapolated surface.

\highlight{Shared with the Field.}
Yes. All photometry-driven feed-forward methods exhibit this limitation, and even per-scene optimization with $128$ dense views struggles to recover deep interior geometry. The visibility filter of Appendix~\ref{sec:supp:eval} is partly designed to make the comparison fair across methods on the visible portion.

\highlight{Mitigation.}
Hybrid recon-plus-generation pipelines are the natural path: use NGTs to anchor visible surfaces faithfully, and use a generative prior (\eg, a 3D shape diffusion model) conditioned on the NGT-anchored geometry to inpaint the occluded portion. A lighter alternative is to expose the visibility mask predicted by Stage~1 to a post-process completion module, so that hallucinated interior geometry is clearly demarcated from the faithfully reconstructed visible surface.

\subsection{Very Thin Structures}

\highlight{Representative Cases.}
Lamp wires, mesh fabric, antennas, and plant stems. The thin elements are either smoothed into the surrounding background or disappear entirely from the output mesh, while the bulk geometry around them is reconstructed correctly.

\highlight{Mechanism.}
The SpVox backend operates at a coarse $R{=}16^3$ voxel grid before hierarchical subdivision; thin structures whose cross-section is comparable to or smaller than the post-subdivision voxel cell size are either smoothed out or pruned by the occupancy filter. The VecSet backend localizes capacity more flexibly but still inherits the spatial frequency limits of the pretrained 3DShape2VecSet decoder.

\highlight{Shared with the Field.}
Partially. Voxel-based methods (PRM, MeshFormer, optimization-based 2DGS/GOF) all share this limitation. Generative methods sometimes hallucinate plausible thin structure, but at the cost of fidelity (Appendix~\ref{sec:supp:recon_vs_gen}).

\highlight{Mitigation.}
Adaptive-resolution decoding---octree-style subdivision driven by local geometric-detail estimates derived from the NGTs themselves---is the most direct route. A complementary approach is NGT-aware post-process mesh refinement that re-queries the decoder along thin-feature candidate locations identified from per-view foreground masks, lifting the effective resolution where it is needed without globally inflating compute.

\subsection{Transparent / Reflective Surfaces}

\highlight{Representative Cases.}
Glass jars, mirrors, and polished metal vases. The reconstructed surfaces are visibly noisy, often exhibiting bumps where specular highlights are localized, and outright holes where the background leaks through transparent regions.

\highlight{Mechanism.}
Photometric appearance is view-inconsistent for transparent and specular materials: specular highlights move with camera, and transparent objects leak background. The Stage~1 depth/normal predictions become unreliable in these regions, so NGTs encode noisy or contradictory geometric cues, and the downstream mesh prediction degrades accordingly.

\highlight{Shared with the Field.}
Universal among photometry-based methods.

\highlight{Mitigation.}
Multi-modal inputs (\eg, polarization) and learned material priors can help. A natural in-architecture extension is to add an auxiliary BRDF-prediction head in Stage~1 pretraining, so the encoder learns to separate geometry from photometric appearance rather than fusing them; the predicted material map can also serve as an explicit uncertainty signal that downweights unreliable regions during NGT aggregation.

\subsection{Out-of-Distribution Camera Configuration}

\highlight{Representative Cases.}
Bottom-only views, telephoto-narrow FoV captures taken from extreme distance, and lower-hemisphere camera clusters. The reconstructions in these cases exhibit globally inconsistent geometry rather than localized artifacts, indicating a systematic failure of the aggregation stage rather than a missing-detail problem.

\highlight{Mechanism.}
\plucker embeddings are learned within the training camera distribution; configurations that fall far outside that distribution produce attention patterns that the encoder has not been trained to interpret.

\highlight{Shared with the Field.}
Partly. Methods conditioned on absolute pose embeddings (including most LRM-style methods) share this limitation; methods using only relative poses (\eg, SLAM3R-style) may generalize better, at the cost of additional architectural complexity.

\highlight{Mitigation.}
Pose-agnostic or canonical-pose-normalized embeddings, together with broader train-time camera distributions, are the most direct routes. The current training distribution covers the upper hemisphere only, which suffices for typical photography but explicitly excludes ground-perspective captures; augmenting the camera distribution at training time, or canonicalizing input poses against a learned reference frame, would extend \name{} to such configurations without altering the core architecture.

\end{document}